\documentclass[12pt,a4paper]{article}
\usepackage{jheppub}
\usepackage{amssymb}
\usepackage{amsmath}
\usepackage{mathtools}
\usepackage{bbm}
\usepackage{amsfonts}
\usepackage{dsfont}
\usepackage{pdfpages}
\usepackage{verbatim}
\usepackage{tensor}
\usepackage{tikz}
\usetikzlibrary{arrows,shapes,positioning}
\usetikzlibrary{decorations.markings}
\usetikzlibrary{decorations.pathreplacing}
\usetikzlibrary{shapes.misc}

\tikzset{cross/.style={cross out, draw=black, fill=none, minimum size=2*(#1-\pgflinewidth), inner sep=0pt, outer sep=0pt}, cross/.default={2pt}}

\usetikzlibrary{calc}
\tikzstyle arrowstyle=[scale=1]
\tikzstyle directed=[postaction={decorate,decoration={markings,
    mark=at position .55 with {\arrow[arrowstyle]{stealth}}}}]
\usepackage{ifthen}
\usepackage{multirow}
\usepackage{longtable}
\usepackage[boxsize=0.7em,centertableaux]{ytableau}
\usepackage{afterpage}

 \newcommand{\ba}{\begin{align}}

\newcommand{\be}{\begin{equation}}
\newcommand{\ee}{\end{equation}}
\newcommand{\tr}{\text{tr}}
\newcommand{\abs}[1]{\left| #1 \right|}
\newcommand{\ket}[1]{| #1 \rangle}
\newcommand*{\tran}{^{\mkern-1.5mu\mathsf{T}}}

\newcommand{\vacbubbletwo}[2]{ 
\begin{tikzpicture}[baseline={([yshift=-.5ex]current bounding box.center)}]
\draw[thick,->] (0,0) arc (5:355:.7) node[pos=.25, fill=white] {#1} node[pos=.75, fill=white] {#2};
\end{tikzpicture}
}
\newcommand{\vacbubblethree}[3]{ 
\begin{tikzpicture}[baseline={([yshift=-.5ex]current bounding box.center)}]
\draw[thick,->] (0,0) arc (5:355:.7) node[pos=.17, fill=white] {#1} node[pos=.5, fill=white] {#2} node[pos=.83, fill=white] {#3};
\end{tikzpicture}
}
\newcommand{\vacbubblefour}[4]{ 
\begin{tikzpicture}[baseline={([yshift=-.5ex]current bounding box.center)}]
\draw[thick,->] (0,0) arc (5:355:.7) node[pos=.125, fill=white] {#1} node[pos=.375, fill=white] {#2} node[pos=.625, fill=white] {#3} node[pos=.875, fill=white] {#4};
\end{tikzpicture}
}
\newcommand{\vacbubblefive}[5]{ 
\begin{tikzpicture}[baseline={([yshift=-.5ex]current bounding box.center)}]
\draw[thick,->] (0,0) arc (5:355:.7) node[pos=.1, fill=white] {#1} node[pos=.3, fill=white] {#2} node[pos=.5, fill=white] {#3} node[pos=.7, fill=white] {#4} node[pos=.9, fill=white] {#5};
\end{tikzpicture}
}

\newcommand{\branethree}[3]{ 
\begin{tikzpicture}[baseline={([yshift=-.5ex]current bounding box.center)}]
\draw[ultra thick] (-1.5,0) -- (1.5,0);
\draw[thick,->] (-1.3,0) arc (185:355:1.3 and .6) node[pos=.25, fill=white] {#1} node[pos=.5, fill=white] {#2} node[pos=.75, fill=white] {#3};
\end{tikzpicture}
}

\newcommand{\twobranetwo}[4]{ 
\begin{tikzpicture}[baseline={([yshift=-.5ex]current bounding box.center)}]
\draw[ultra thick] (-2,0) -- (2,0);
\draw[thick,->] (-1.3,0) arc (185:355:1.3 and .6) node[pos=.33, fill=white] {#1} node[pos=.67, fill=white] {#2};
\draw[thick,->] (-1.8,0) arc (185:355:1.8 and 1.2) node[pos=.33, fill=white] {#3} node[pos=.67, fill=white] {#4};
\end{tikzpicture}
}

\newcommand{\twobranetwocross}[4]{ 
\begin{tikzpicture}[baseline={([yshift=-.5ex]current bounding box.center)}]
\draw[ultra thick] (-2,0) -- (2,0);
\draw[thick,->] (-1.3,0) arc (185:355:1.55 and .6) node[pos=.3, fill=white] {#1} node[pos=.5, fill=white] {#2} node[pos=.7, fill=white]{};
\draw[thick,->] (-1.8,0) arc (185:355:1.55 and 1.2) node[pos=.33, fill=white] {#3} node[pos=.67, fill=white] {#4};
\end{tikzpicture}
}

\allowdisplaybreaks[1]

\title{The Grassmannian VOA}

\author[a]{Lorenz Eberhardt}
\author[b]{and Tom\'{a}\v{s} Proch\'{a}zka}
\affiliation[a]{School of Natural Sciences, Institute for Advanced Study, \\
\hspace*{0.3cm}Princeton, NJ 08540, USA}
\affiliation[b]{Arnold Sommerfeld Center for Theoretical Physics, \\
\hspace*{0.3cm}Ludwig Maximilian University of Munich, \\
\hspace*{0.3cm}Theresienstr.~37, D-80333 M\"unchen, Germany}
\emailAdd{elorenz@ias.edu}
\emailAdd{Tomas.Prochazka@lmu.de}

\abstract{
We study the 3-parametric family of vertex operator algebras based on the unitary Grassmannian coset CFT $\mathfrak{u}(M+N)_k/(\mathfrak{u}(M)_k \times \mathfrak{u}(N)_k)$. This VOA 
serves as a basic building block for a large class of cosets and
generalizes the $\mathcal{W}_\infty$ algebra.
We analyze representations and their characters in detail and find surprisingly simple character formulas for the representations in the generic parameter regime that admit an elegant combinatorial formulation. We also discuss truncations of the algebra and give a conjectural formula for the complete set of truncation curves.
We develop a theory of gluing for these algebras in order to build more complicated coset and non-coset algebras. We demonstrate the power of this technology with some examples and show in particular that the $\mathcal{N}=2$ supersymmetric Grassmannian can be obtained by gluing three bosonic Grassmannian algebras in a loop.
We finally speculate about the tantalizing possibility that this algebra is a specialization of an even larger 4-parametric family of algebras exhibiting pentality symmetry. Specialization of this conjectural family should include both the unitary Grassmannian family as well as the Lagrangian Grassmannian family of VOAs which interpolates between the unitary and the orthosymplectic cosets.

}

\begin{document}
\maketitle

%make math in all titles bold
\makeatletter
\g@addto@macro\bfseries{\boldmath}
\makeatother
%end code

\section{Introduction}

Recently there has been a renewed interest in $\mathcal{W}$-algebras, which are symmetry algebras of two-dimensional conformal field theories, in particular due to their connections to higher dimensional field theories. The AGT correspondence relates partition functions of 4d $\mathcal{N}=2$ supersymmetric gauge theories to correlation functions of Liouville and Toda theory in two dimensions \cite{Alday:2009aq,Wyllard:2009hg}. The mathematical reason behind this correspondence is the action of $\mathcal{W}$-algebras on the equivariant cohomology of instanton moduli spaces \cite{Maulik:2012wi,schiffmann2013cherednik}. In \cite{Beem:2013sza} $\mathcal{W}$-algebras were found to control operator product expansions of a certain class of local operators in 4d $\mathcal{N}=2$ superconformal theories. The index associated to these operators agrees with the vacuum character of the corresponding $\mathcal{W}$-algebras. $\mathcal{W}$-algebras were also found to describe the degrees of freedom associated to junctions of co-dimension $1$ defects in twisted $\mathcal{N}=4$ super Yang-Mills theory \cite{Gaiotto:2017euk}. $\mathcal{W}$-algebras play also an important role in $\text{AdS}_3/\text{CFT}_2$ holography \cite{Gaberdiel:2010pz,Gaberdiel:2012ku,Candu:2012ne, Gaberdiel:2014cha, Eberhardt:2018plx} where they organize the states of the conformal field theory side. Last but not least, the related algebraic structures play role in recent studies of $M$-theory in $\Omega$-background \cite{Costello:2017fbo,Gaiotto:2019wcc}

As an unexpected consequence of these developments, a certain family of $\mathcal{W}$-algebras was found to be equivalent to Yangian symmetries associated to affine Lie algebras \cite{Maulik:2012wi,Smirnov:2013hh,Tsymbaliuk:2014fvq,Zhu:2015nha,Prochazka:2019dvu,Negut:2020npc}. Using this description, one can use the tools of integrability to study these vertex operator algebras. In particular, the representation theory simplifies significantly in this Yangian picture and in many cases one finds natural combinatorial interpretation of the representation spaces. In particular, the algebra $\mathcal{W}_{1+\infty}$ which is a unifying algebra of $\mathcal{W}_N$ series of $\mathcal{W}$-algebras is closely connected to combinatorics of plane partitions (3d Young diagrams) encountered in the topological string \cite{Aganagic:2003db}.

It is clear that the space of vertex operator algebras is very large. In order to explore it, one can look for techniques which construct directly new vertex operator algebras out of old ones. Well-known examples of these constructions are GKO coset construction \cite{Goddard:1984vk}, Drinfe\v{l}d-Sokolov reduction \cite{drinfeld1985lie,Feigin:1990pn} or taking orbifolds \cite{Dixon:1985jw, Dixon:1986jc, Dixon:1986qv}. Alternatively, one can try to study $\mathcal{W}$-algebras by building them from smaller building blocks \cite{Bowcock:1999uy,Feigin:2017edu}. The idea is to describe a vertex operator algebra as an extension of a vertex operator subalgebra (typically of a product form) by a set of representations of the subalgebra. One particular realization of this construction is the gluing construction discussed in \cite{Prochazka:2017qum} (this construction was discussed in the Yangian language in series of papers \cite{Gaberdiel:2017hcn, Gaberdiel:2018nbs, Li:2019nna, Li:2019lgd} and recently generalized in \cite{Li:2020rij}). Here one uses the fact that a certain class of $\mathcal{W}$-algebras describes degrees of freedom associated to junctions of co-dimension $1$ defects in four-dimensional twisted $\mathcal{N}=4$ super Yang-Mills theory. In particular, once one understands the basic junction where three co-dimension $1$ defects meet, it is natural to consider more complicated configurations of these defects. Following this procedure, a large class of $\mathcal{W}$-algebras was constructed in \cite{Prochazka:2017qum}. In the $\mathcal{W}$-algebra language the building block is the $\mathcal{W}_{1+\infty}$ algebra or its truncations and the algebras resulting from this gluing procedure include the unitary affine Lie algebras, $\mathcal{W}$-algebras associated to non-principal Drinfe\v{l}d-Sokolov reductions or the $\mathcal{N}=2$ superconformal algebra and its extension to $\mathcal{N}=2$ $\mathcal{W}_\infty$. Despite some important differences, the whole procedure largely parallels the calculation of topological string partition functions on toric Calabi-Yau manifolds using the formalism of topological vertex \cite{Aganagic:2003db}.

Although the class of algebras constructible following the procedure described in the last paragraph is large, it clearly does not exhaust all the known algebras that can be constructed in another way. Perhaps the simplest example are the unitary Grassmannian cosets of the form \cite{Bais:1987zk}
\be 
\frac{\mathfrak{u}(\mu_1+\mu_2)_k}{\mathfrak{u}(\mu_1)_k \times \mathfrak{u}(\mu_2)_k}\ . \label{eq:coset intro}
\ee
Setting $\mu_2 = 1$ and $k$ to a positive integer these cosets give rise to $\mathcal{W}_k$ algebras, but if both $\mu_1$ and $\mu_2$ are different than one, the algebra that one gets is larger. It still only has one unique stress-energy tensor (and no spin-1 currents) and thus cannot be decomposed it into smaller pieces. This therefore indicates that the building block used in \cite{Prochazka:2017qum} should be generalized to include the Grassmannian cosets. The main subject of this work are these Grassmannian $\mathcal{W}$-algebras as well as their generalizations.

\subsection{Overview, summary of results}

In section~\ref{sec:grassmannian} we introduce the Grassmannian cosets as well as the associated universal algebra parametrized by three complex parameters which truncates to Grassmannian cosets for special values of parameters. The coset description has a manifest $\mathds{Z}_2$ symmetry exchanging the two parameters $\mu_1$ and $\mu_2$.
The first surprising feature of the algebra is that the $\mathds{Z}_2$ duality symmetry manifest in coset description \eqref{eq:coset intro} is enhanced to an $\mathcal{S}_3 \times \mathds{Z}_2$ symmetry, see also \cite{Creutzig:2019kro}. One way to see this enhancement is to observe that if we perform the coset \eqref{coset1} in two steps, the intermediate algebra is the matrix-valued $\mathcal{W}_{1+\infty}$ algebra (or the affine $\mathfrak{gl}(M)$ Yangian), which was studied previously in \cite{Arakawa:2016fbi,Creutzig:2018pts,Creutzig:2019qos,Eberhardt:2019xmf}. These algebras possess a conjugation duality which combined with the $\mathds{Z}_2$ duality of the coset \eqref{coset1} generates the triality symmetry $\mathcal{S}_3$. We should already stress at this point that this triality symmetry is not the generalization of the triality symmetry present in $\mathcal{W}_\infty$ \cite{Gaberdiel:2012ku}, but rather a different independent triality symmetry. The additional $\mathds{Z}_2$ symmetry changes sign of all rank and level parameters and is known already at the level of simple Lie algebras \cite{Cvitanovic:2008zz}.
We also introduce a generalization of the Grassmannian cosets to generalized flag manifolds, which have $n-1$ unitary groups in the denominator in \eqref{eq:coset intro}. In this case there is also an enhancement of the manifest $\mathcal{S}_{n-1}$ symmetry to $\mathcal{S}_n \times \mathds{Z}_2$ that acts on the $n$ parameters of the coset. We conclude the section by discussing representations that can be constructed using the coset description of the algebra as well as various free field constructions.

In the following section \ref{sec:characters} we study the characters of representations of the algebra. Working in the universal algebra with generic values of the parameters significantly simplifies the character calculations and we are able to determine not only the vacuum character of the algebra but any character which is visible from the coset description. They all take a very simple, but somewhat unusual form, for which we have no microscopic explanation. For instance, the vacuum character of the Grassmannian coset just reads
\be 
\prod_{n=1}^\infty \frac{(1-q^n)^2}{1-2q^n}\ . \label{eq:intro vac char}
\ee
 We introduce a convenient combinatorial picture of string diagrams decorated by beads stretched between branes. This picture can be also easily generalized to algebras associated to generalized flag manifolds.

In the following section we analyze the structure of the generalized flag VOAs with respect to gluing. The coset description of these generalized flag VOAs immediately provides certain decompositions of these algebras analogous to pair of pants decompositions of genus zero topological surfaces with labeled punctures. Different ways of decomposing a given VOA are related by an analogue of crossing symmetry. The basic Grassmannian algebra corresponds to a three-punctured sphere. We can attach to each of the punctures a copy of an affine Lie algebra. Attaching it to one of the punctures we find a matrix-valued $\mathcal{W}_\infty$ \cite{Arakawa:2016fbi,Creutzig:2018pts,Creutzig:2019qos,Eberhardt:2019xmf}. Attaching two affine Lie algebras to a Grassmannian results in an affine Lie algebra in the numerator \eqref{coset1}. Finally, gluing a Grassmannian algebra with three affine Lie algebras results in a special affine Lie algebra at a level which is minus one half of the rank. This affine Lie algebra is the maximal algebra out of which all generalized flag VOAs can be constructed using the coset construction. This description also manifests the maximal $n$-ality symmetry of this class of VOAs.

The remaining part of section~\ref{sec:gluing} discusses generalizations of the gluing procedure used in generalized flag VOAs. Analogously to \cite{Prochazka:2017qum}, there is actually a one-parametric family of gluings where the parameter controls the conformal dimension of the lightest gluing field. At present we do not know any alternative construction of this family of gluings, but thanks to detailed information that we have about characters of the Grassmannian VOA we can make non-trivial predictions for characters of these glued algebras. As an illustration, equation \eqref{eq:treegluingchar} gives a general conjecture for the vacuum character of any algebra obtained by the most general genus 0 gluing, i.e. described by a tree graph with no loops. This formula passes many non-trivial checks and is surprisingly simple.

The next section discusses various extensions and applications of the Grassmannian gluing. First of all, one can repeat the analysis of the unitary Grassmannian cosets in the orthosymplectic situation. We discuss in detail the properties of characters in this orthosymplectic setting. We also consider the hybrid cosets of unitary Lie algebras with respect to their orthogonal and symplectic subalgebras as well as the level-rank dual of these. The geometric analogue of these cosets are the Lagrangian Grassmannians. We find dualities mapping between these different cosets, so these Lagrangian cosets have their associated two-parametric universal algebra. To our surprise, the field content of this algebra agrees with the Grassmannian universal VOA up to spin $6$.

The next generalization that we consider is the introduction of supersymmetry. We discuss the $\mathcal{N}=2$ version of the Grassmannian VOA and from the gluing perspective it is found to be glued from three bosonic Grassmannians in a triangular loop, together with a $\mathfrak{u}(1)$-current. Topologically the corresponding gluing diagram is a genus 1 three-punctured surface and it serves as the first example of higher genus gluing. This reduces to the previously known result that the $\mathcal{N}=2$ $\mathcal{W}_{1+\infty}$ algebra can be obtained by gluing two bosonic $\mathcal{W}_{\infty}$ algebras, together with a $\mathfrak{u}(1)$-current \cite{Gaberdiel:2017hcn, Prochazka:2017qum, Gaberdiel:2018nbs}.
 The final example is the large $\mathcal{N}=4$ superconformal algebra, which can be obtained by gluing two $\mathfrak{su}(2)$ algebras to a (truncation of the) Grassmannian algebra. The gluing procedure in this case is slightly different than the general gluings considered so far.

In section~\ref{sec:OPEs} we discuss the structure of operator product expansions of the algebra in the primary basis, assuming the field content as predicted by the character calculations of section~\ref{sec:characters}. We start with the most general ansatz and impose associativity of the OPE.
Due to the very quickly growing number of primary fields in the algebra, it is very hard to solve the associativity conditions on OPEs for spins higher than $\sim 6$. Surprisingly, the results of this bootstrap calculation to the order we were able to reach are compatible with the existence of a \emph{four-parametric} family of algebras instead of the three-parametric Grassmannian family that we started with. We return to discussion of this possibility in the later part of the article.

In section~\ref{sec:truncations} we collect various pieces of information about truncations of the Grassmannian algebras. While none of the approaches to determine these truncations curves gave us complete information, we could combine these various sources to make a simple conjecture for the most general conjecture, which is eq.~\eqref{eq:grasstruncfull1}. We will return to this in section \ref{sec:4parameters} in the context of the conjectural four-parametric family of algebras.

Section~\ref{sec:4parameters} is of speculative character. We collect the hints that seem to indicate that both the unitary Grassmannian as well as the Lagrangian family of algebras could be one-parametric specializations of a four-parametric family of algebras possessing an $\mathcal{S}_5$ symmetry. We refer to this symmetry as pentality symmetry and it enhances the $\mathcal{S}_3 \times \mathds{Z}_2$ symmetry present in the Grassmannian.
 Since we do not know any direct way for constructing such an algebra, we list the circumstantial evidence that seems to indicate the existence of such an object. First of all, the invariant expressions like the central charge \eqref{eq:bigcentralcharge} or the dimensions and charges of the minimal representations \eqref{eq:bifundbig} can be written in a universal form that applies to both unitary and Lagrangian families. Another strong hint comes from the analysis of the truncation curves. All the truncation curves that we are aware of can be written as a simple equation \eqref{eq:bigtruncations} which is of the same form as the corresponding equation in $\mathcal{W}_\infty$ \cite{Prochazka:2014gqa,Prochazka:2017qum}. All the operator product expansions that we were able to calculate explicitly both in the Grassmannian algebra as well as in the Lagrangian coset are compatible with the existence of such unifying a object. Finally, also all the OPEs that we were able to compute by imposing just the associativity constraints in section~\ref{sec:OPEs} are compatible with the existence of such a four-parametric family. 
 
We end in section~\ref{sec:discussion} with a general discussion of our results. We discuss the intriguing possibility that the Grassmannian cosets have a stringy holographic dual. All their qualitative properties point to this scenario: they enjoy large-N factorization, the partition function has Hagedorn growth, etc. This would be an ideal laboratory to explore the stringy $\text{AdS}_3/\text{CFT}_2$ correspondence beyond its `free' point given by the symmetric product orbifold \cite{Maldacena:1997re, Seiberg:1999xz, Eberhardt:2018ouy, Eberhardt:2019ywk, Eberhardt:2020akk}. We finally mention the many open questions that remain and possible future directions.

\section{The unitary Grassmannian} \label{sec:grassmannian}
\subsection{Definition}
The main object of our study is the three-parametric family of vertex operator algebras associated to the family of GKO cosets
\begin{equation}
\label{coset1}
\frac{\mathfrak{u}(\mu_1+\mu_2)_k}{\mathfrak{u}(\mu_1)_k \times \mathfrak{u}(\mu_2)_k}
\end{equation}
or to the dual cosets
\begin{equation}
\label{coset2}
\frac{\mathfrak{su}(k)_{\mu_1} \times \mathfrak{su}(k)_{\mu_2}}{\mathfrak{su}(k)_{\mu_1+\mu_2}} \ .
\end{equation}
By the level-rank duality, for integer values of parameters $\mu_1, \mu_2$ and $k$ these VOAs (thought of as the respective simple quotients) are isomorphic. As a simplest non-trivial check, we can calculate the central charges and find
\begin{equation}
c = \frac{k(\mu_1+\mu_2-1)(\mu_1+\mu_2+1)}{k+\mu_1+\mu_2} - \frac{k(\mu_1-1)(\mu_1+1)}{k+\mu_1} - \frac{k(\mu_2-1)(\mu_2+1)}{k+\mu_2} - 1
\end{equation}
in the first case and
\begin{equation}
c = \frac{\mu_1(k-1)(k+1)}{k+\mu_1} + \frac{\mu_2(k-1)(k+1)}{k+\mu_2} - \frac{(\mu_1+\mu_2)(k-1)(k+1)}{k+\mu_1+\mu_2}
\end{equation}
in the second case and it is easy to see that both of these expressions are equal to
\begin{equation}
c = \frac{\mu_1 \mu_2(k-1)(k+1)(\mu_1+\mu_2+2k)}{(\mu_1+k)(\mu_2+k)(\mu_1+\mu_2+k)}.
\end{equation}
Both of these coset descriptions have manifest $\mathds{Z}_2$ duality exchanging $\mu_1 \leftrightarrow \mu_2$.

\paragraph{Universal algebra}
If we specialize one of the $\mu_j$ parameters to $1$, we get a standard coset description of $\mathcal{W}_N$ algebras. Instead of thinking of these algebras as a family parametrized by integers, it is very useful to study the universal algebra parametrized by arbitrary complex parameters. In the case of $\mathcal{W}_N$ family, the universal algebra is $\mathcal{W}_\infty$ \cite{Gaberdiel:2012ku,Prochazka:2014gqa,Linshaw:2017tvv}.

One way to define the universal algebra is using the operator product expansions of the generating fields: choosing a normalization of the generating fields appropriately, one finds that the OPE coefficients depend rationally (of even polynomially, \cite{Prochazka:2014gqa,Linshaw:2017tvv}) on the parameters of the algebra. One can thus define the universal algebra using these operator product expansions without any restrictions on integrality of the parameters. Once one has the universal algebra, it is easy to reconstruct back the coset algebras with integer values of parameters: all that we need to do is to specialize the parameters to be integers and take the simple quotient.

We can think of this procedure as effectively working in the limit of large values of $\mu_1, \mu_2$ and $k$ but without putting these parameters strictly equal to infinity. Another possible approach would be to replace the $\mathfrak{su}(N)$ algebra in the coset by the associated interpolating family $\mathfrak{hs}(\lambda)$ (as used in higher spin theories \cite{Gaberdiel:2012ku}). Either way, in the following we will mostly work with the universal algebra parametrized by three complex numbers $\mu_1, \mu_2$ and $k$ and only when needed we will discuss the specializations of parameters and the associated truncations.

\subsection{Symmetries} \label{subsec:symmetries}

\paragraph{$\boldsymbol{\mathds{Z}_2}$ duality}
Once we work with the universal Grassmannian algebra so that we do not need to restrict $\mu_1$ and $\mu_2$ to be integers, there is additional symmetry flipping signs of $\mu_1, \mu_2$ and $k$ at the same time \cite{Cvitanovic:2008zz}. We can see that this map indeed preserves the central charge and later we will see that also all other structure constants of the algebra that we calculated are invariant under this transformation.

\paragraph{Matrix $\mathcal{W}_{1+\infty}$} Instead of taking the double quotient as in (\ref{coset1}) at once, one can first study the coset
\begin{equation}
\label{matrixcoset}
\frac{\mathfrak{u}(L+M)_K}{\mathfrak{u}(L)_K}.
\end{equation}
It is clear that the resulting algebra contains an affine $\mathfrak{u}(M)_K$ as a subalgebra, but looking at characters (or studying the operator product expansions) we see that it is actually much larger. One can choose a set of generators of the corresponding universal algebra such that for each conformal dimension there is a $M \times M$ matrix of generators transforming in the adjoint representation of the global $\mathfrak{u}(M)$ symmetry. This algebra was studied from a different perspective in \cite{Arakawa:2016fbi,Creutzig:2019qos,Eberhardt:2019xmf, Rapcak:2019wzw} and one can think of it as a matrix extension of $\mathcal{W}_{1+\infty}$. Just like $\mathcal{W}_{1+\infty}$ can be understood as the Yangian of affine $\mathfrak{gl}(1)$, we can think of the matrix $\mathcal{W}_{1+\infty}$ as of the Yangian of affine $\mathfrak{gl}(M)$. Passage from (\ref{matrixcoset}) to (\ref{coset1}) is then equivalent to decoupling of spin $1$ generators of $\mathfrak{u}(M)_K$ subalgebra analogously to the $M=1$ situation where one decouples the spin $1$ current in $\mathcal{W}_{1+\infty}$ to get $\mathcal{W}_\infty$.

In \cite{Eberhardt:2019xmf}, the matrix-valued $\mathcal{W}_{1+\infty}$ was parametrized by the parameters $(M,N,\kappa)$. The parameter $M$ is the rank of the $\mathfrak{u}(M)$ subalgebra. The parameter $N$ is the analogue of $N$ in $\mathcal{W}_N$ algebras, i.e. it determines the spin of the generator of the highest spin (from the Yangian perspective it is the level). The algebra has a representation in terms of $N$ commuting $\mathfrak{u}(M)_\kappa$ algebras and $\kappa$ enters as their level. The map between $(M,N,\kappa)$ parameters and the parameters appearing in the coset (\ref{matrixcoset}) is
\begin{subequations}
\begin{align}
M & \leftrightarrow M \\
L & \leftrightarrow -\kappa(N-1) \\
K & \leftrightarrow N\kappa.
\end{align}
\end{subequations}
The matrix algebra (\ref{matrixcoset}) has a duality symmetry generalizing the conjugation symmetry of $\mathfrak{u}(M)_K$ \cite{Eberhardt:2019xmf}. It acts on the parameters of \cite{Eberhardt:2019xmf} as
\begin{subequations}
\begin{align}
M & \leftrightarrow M \\
N & \leftrightarrow -\frac{N\kappa}{M+\kappa} \\
\kappa & \leftrightarrow -M-\kappa.
\end{align}
\end{subequations}
It is easy to see that it is involutive, i.e. that it is a duality symmetry. Since this symmetry does not touch the $\mathfrak{u}(M)_K$ subalgebra, it survives the quotient and gives rise to a duality symmetry of (\ref{coset1}). In terms of parameters used there, we find an additional duality symmetry of the Grassmannian algebra under which
\begin{equation}
\mu_1 \to -\mu_1-\mu_2-2k, \quad\quad \mu_2 \to \mu_2, \quad\quad k \to k.
\end{equation}

\paragraph{Triality} The duality symmetries discussed previously do not commute so we can use them to generate more symmetry operations. To parametrize the algebra in a symmetric way, let us trade the level $k$ for another rank-like parameter $\mu_3$,
\begin{equation}
k = -\frac{1}{2}(\mu_1+\mu_2+\mu_3).
\end{equation}
In terms of these parameters the symmetries of the algebra act in a very simple way: there is a triality symmetry $\mathcal{S}_3$ permuting the $\mu_j$ and there is a $\mathds{Z}_2$ duality which changes simultaneously the sign of all $\mu_j$. The combined symmetry of the Grassmannian algebra that we found is thus $\mathcal{S}_3 \times \mathds{Z}_2$. Later we will verify that this is indeed a symmetry of the OPE coefficients of the algebra. The central charge can be written as
\begin{equation}
c = -\frac{2\mu_1 \mu_2 \mu_3(\mu_1+\mu_2+\mu_3-2)(\mu_1+\mu_2+\mu_3+2)}{(\mu_1+\mu_2-\mu_3)(\mu_1-\mu_2+\mu_3)(-\mu_1+\mu_2+\mu_3)}. \label{eq:central charge}
\end{equation}
which manifests all the discrete symmetries that we discussed. For later purposes it will be convenient useful to introduce one additional set of parameters, $\nu_j = \mu_j + k$ or more explicitly
\begin{equation}
\label{eq:mutonu}
\nu_1 = \frac{1}{2} \left( \mu_1 - \mu_2 - \mu_3 \right), \quad \nu_2 = \frac{1}{2} \left( -\mu_1 + \mu_2 - \mu_3 \right), \quad \nu_3 = \frac{1}{2} \left( -\mu_1 - \mu_2 + \mu_3 \right).
\end{equation}
The inverse transformation is simply
\begin{equation}
\mu_1 = -\nu_2-\nu_3, \quad \mu_2 = -\nu_1-\nu_3, \quad \mu_3 = -\nu_1-\nu_2, \quad k = \nu_1+\nu_2+\nu_3.
\end{equation}
The central charge written in terms of these parameters takes the form
\begin{equation}
\label{eq:grasscnu}
c = -\frac{(\nu_1+\nu_2)(\nu_1+\nu_3)(\nu_2+\nu_3)(\nu_1+\nu_2+\nu_3+1)(\nu_1+\nu_2+\nu_3-1)}{\nu_1\nu_2\nu_3}.
\end{equation}
We can write the coset (\ref{coset1}) slightly differently which makes the triality symmetry manifest. First of all, notice that the Grassmannian
\begin{equation}
\frac{\mathfrak{u}(\mu_1+\mu_2+\mu_3)_{-\frac{\mu_1+\mu_2+\mu_3}{2}}}{\mathfrak{u}(\mu_1+\mu_2)_{-\frac{\mu_1+\mu_2+\mu_3}{2}} \times \mathfrak{u}(\mu_3)_{-\frac{\mu_1+\mu_2+\mu_3}{2}}}
\end{equation}
is trivial (since it has parameters $(\mu_1+\mu_2,\mu_3,0)$). This means that we can write
\begin{equation}
\mathfrak{u}(\mu_1+\mu_2)_{-\frac{\mu_1+\mu_2+\mu_3}{2}} \simeq \frac{\mathfrak{u}(\mu_1+\mu_2+\mu_3)_{-\frac{\mu_1+\mu_2+\mu_3}{2}}}{\mathfrak{u}(\mu_3)_{-\frac{\mu_1+\mu_2+\mu_3}{2}}}
\end{equation}
and using this, the coset (\ref{coset1}) can be represented as
\begin{equation}
\label{eq:grassfromtop}
\frac{\mathfrak{u}(\mu_1+\mu_2+\mu_3)_{-\frac{\mu_1+\mu_2+\mu_3}{2}}}{\mathfrak{u}(\mu_1)_{-\frac{\mu_1+\mu_2+\mu_3}{2}} \times \mathfrak{u}(\mu_2)_{-\frac{\mu_1+\mu_2+\mu_3}{2}} \times \mathfrak{u}(\mu_3)_{-\frac{\mu_1+\mu_2+\mu_3}{2}}}
\end{equation}
which manifests the triality symmetry permuting $\mu_j$. The triality invariant description of associated level-rank dual coset (\ref{coset2}) is
\begin{equation}
\frac{\mathfrak{su}(-\frac{\mu_1+\mu_2+\mu_3}{2})_{\mu_1} \times \mathfrak{su}(-\frac{\mu_1+\mu_2+\mu_3}{2})_{\mu_2} \times \mathfrak{su}(-\frac{\mu_1+\mu_2+\mu_3}{2})_{\mu_3}}{\mathfrak{su}(-\frac{\mu_1+\mu_2+\mu_3}{2})_{\mu_1+\mu_2+\mu_3}}.
\end{equation}
%Note that the level of the affine Lie algebra in the denominator is exactly minus twice the dual Coxeter number which is precisely the condition for the standard coset BRST charge of \cite{Karabali:1989dk} to be nilpotent. In the context of $4d$ $\mathcal{N}=2$ theories this corresponds to gauging of a non-anomalous global symmetry \cite{Beem:2013sza}.

\paragraph{Conjugation symmetry} The algebra possesses one more automorphism that acts non-trivially on the fields in the algebra. This symmetry is the analogue of the parity or conjugation symmetry in $\mathcal{W}_\infty$ under which every (non-composite) even-spin primary field is even and every (non-composite) odd-spin primary is odd. 
The Grassmannian algebra possesses a unique primary spin-3 field $W_3$, which transforms under a similar conjugation symmetry as $W_3 \to -W_3$. This symmetry can be extended to the complete Grassmannian. From the coset perspective, it is induced from the conjugation symmetry of both the numerator and denominator algebras. We emphasize that in the Grassmannian, it is \emph{not} true that every non-composite even-spin field is even under this symmetry and every non-composite odd-spin field is odd under this symmetry. This starts to break down at spin 6, where we find a conjugation-odd field. The number of even and odd fields at a given spin can be found in table~\ref{tab:spin parity}.

\paragraph{Specialization to $\mathcal{W}_\infty$} A much better explored family of $\mathcal{W}$-algebras is $\mathcal{W}_\infty$. Its coset realizations are given by (\ref{coset1}) or (\ref{coset2}) with $\mu_1=1$.

\paragraph{Generalized Flag Manifolds} Sometimes we will consider the following direct generalization of the cosets \eqref{coset1} and \eqref{coset2},
\be 
\frac{\mathfrak{u}(\mu_1+\mu_2+\cdots+\mu_{n-1})_k}{\mathfrak{u}(\mu_1)_k \times  \cdots \mathfrak{u}(\mu_{n-1})_k}\cong 
\frac{\mathfrak{su}(k)_{\mu_1} \times \cdots \times\mathfrak{su}(k)_{\mu_{n-1}}}{\mathfrak{su}(k)_{\mu_1+\cdots+\mu_{n-1}}} \ .
\ee
Since the first coset describes a generalized flag manifold,\footnote{i.e.~the space of all embeddings of hermitian vectorspaces $\{0\} \subset V_1 \subset V_2 \subset \cdots \subset V_{n-1}$, where $V_i$ has complex dimension $\mu_1 +\cdots +\mu_i$.} we will refer to the corresponding vertex operator algebra as the generalized flag manifold algebra. The same arguments above show that one can rewrite the algebra as
\be 
\frac{\mathfrak{u}(\mu_1+\cdots+\mu_n)_k}{\mathfrak{u}(\mu_1)_k \times \mathfrak{u}(\mu_2)_k \times \cdots \mathfrak{u}(\mu_n)_k}\ ,
\ee
where
\be 
k=-\frac{1}{2}(\mu_1+ \cdots +\mu_n)\ .
\ee
Introducing the parameter $\mu_n$ makes the $n$-ality symmetry of the algebra manifest that permutes the parameters $\mu_j$. It is again useful to introduce the parameters $\nu_j=\mu_j+k$.

\subsection{Primary fields} \label{subsec:primary fields}
Coset primaries of the coset realization \eqref{coset2} are labeled by three $\mathfrak{su}(k)$ representations. We shall assume that $k$ is very large (or analytically continued) so that no truncations occur. Thus, we are effectively looking at $\mathfrak{su}(\infty)$ representations.  Such representations can be described by a pair of (finite) Young diagrams $\boldsymbol{\Lambda}=(\Lambda,\bar{\Lambda})$ to which we shall refer to as `boxes' and `antiboxes'. Here and in the following, $\Lambda$ and $\bar{\Lambda}$ are assumed to be independent. We will need several operations on these pairs of Young diagrams. We denote by
\begin{subequations}
\begin{align}
\lVert \boldsymbol{\Lambda} \rVert&=\abs{\Lambda}+\abs{\bar{\Lambda}}\ , \label{eq:number of boxes plus antiboxes}\\
\abs{ \boldsymbol{\Lambda} }&=\abs{\Lambda}-\abs{\bar{\Lambda}}\ , \label{eq:number of boxes minus antiboxes}\\
\bar{\boldsymbol{\Lambda} }&=(\bar{\Lambda},\Lambda)\ ,\label{eq:conjugate definition} \\
\boldsymbol{\Lambda} \tran&=(\Lambda\tran,\bar{\Lambda}\tran)\ ,\label{eq:transpose definition}
\end{align}
\end{subequations}
the total number of boxes and anti-boxes, the number of boxes minus the number of anti-boxes, the conjugate representation and the transposed representation. By transpose, we mean the representation that is given by the transposed Young-diagram.

 For instance, the fundamental representation, the antifundamental representation and the adjoint representation correspond to $(\ydiagram{1},\bullet)$, $(\bullet,\ydiagram{1})$ and $(\ydiagram{1},\ydiagram{1})$, respectively. Primary fields of the coset \eqref{coset2} are then labeled by three such pairs $(\boldsymbol{\Lambda}_1,\boldsymbol{\Lambda}_2,\boldsymbol{\Lambda}_3)$. The first two representations correspond to the numerator algebras and the last one is the \emph{conjugated} denominator representation.
This additional conjugation will make triality symmetry manifest.
The tensor product of $\mathfrak{su}(\infty)$ has a $\mathds{Z}$ selection rule -- upon taking tensor products the number of boxes minus number of antiboxes (i.e.~$\abs{\boldsymbol{\Lambda}}$) is conserved. This leads to the following selection rule of the coset representations
\be 
\abs{\boldsymbol{\Lambda}_1}+\abs{\boldsymbol{\Lambda}_2}+\abs{\boldsymbol{\Lambda}_3}=0\ . \label{eq:selection rule}
\ee
We will find a nice interpretation of this fact in section~\ref{sec:characters}. The conformal weight of a primary is $h_0(\boldsymbol{\Lambda}_1,\boldsymbol{\Lambda}_2,\boldsymbol{\Lambda}_3)+\text{integer}$, where
\be 
h_0(\boldsymbol{\Lambda}_1,\boldsymbol{\Lambda}_2,\boldsymbol{\Lambda}_3)=\sum_{i=1}^3 \frac{\mathcal{C}_k(\boldsymbol{\Lambda}_i)}{2(\mu_i+k)} = \sum_{i=1}^3 \frac{\mathcal{C}_k(\boldsymbol{\Lambda}_i)}{2\nu_i} \ . \label{eq:primary conformal weight}
\ee
Here, $\mathcal{C}_k$ is the $\mathfrak{su}(k)$ Casimir of the representation. This formula is manifestly triality invariant. The integer shift is caused by the fact that a coset representation might not appear on the level of the affine primary fields, but only further down in the module. For reference, we recall the following formula for the quadratic Casimir:
\begin{multline} 
\mathcal{C}_k(\Lambda,\bar{\Lambda})=k \lVert \boldsymbol{\Lambda} \rVert-\frac{\abs{\boldsymbol{\Lambda}}^2}{k}+\sum_i \text{row}_i(\Lambda)^2
+\sum_i \text{row}_i(\bar{\Lambda})^2\\
-\sum_i \text{col}_i(\Lambda)^2-\sum_i \text{col}_i(\bar{\Lambda})^2\ . \label{eq:suinfty Casimir}
\end{multline}
By $\text{row}_i(\Lambda)$, we mean the length of the $i$-th row of $\Lambda$ and by $\text{column}_i(\Lambda)$, we mean the length of the $i$-th column.
Under the $\mathds{Z}_2$ duality that sends $\mu_i \to -\mu_i$, the representation labels transform non-trivially. From the explicit form of the Casimir operator, it follows that the conformal weight is invariant if we accompany the sign flip of the $\mu_i$'s with the transformation $\boldsymbol{\Lambda}_i\to \boldsymbol{\Lambda}_i\tran$, where $\tran$ denotes the transpose diagram, see eq.~\eqref{eq:transpose definition}. 

To summarize, the symmetry group $\mathcal{S}_3 \times \mathds{Z}_2$ acts on the three representation labels $(\boldsymbol{\Lambda}_1,\boldsymbol{\Lambda}_2,\boldsymbol{\Lambda}_3)$ by permutations and by the transpose operation. Finally the conjugation symmetry acts by $(\boldsymbol{\Lambda}_1,\boldsymbol{\Lambda}_2,\boldsymbol{\Lambda}_3) \to (\bar{\boldsymbol{\Lambda}}_1,\bar{\boldsymbol{\Lambda}}_2,\bar{\boldsymbol{\Lambda}}_3)$

There are two classes of representations that can be considered minimal. They play a special role because other representations can be obtained by fusing these. They also play an important role when identifying the duality symmetries in $\mathcal{W}$-algebras \cite{Gaberdiel:2012ku,Prochazka:2019yrm}. The first of these, which we call the bifundamental representation, has labels $((\ydiagram{1},\bullet),(\bullet,\ydiagram{1}),(\bullet,\bullet))$. It is associated to the choice of two $\nu_j$ parameters and it also has a complex conjugate representation. The second representation is the adjoint representation with labels $((\ydiagram{1},\ydiagram{1}),(\bullet,\bullet),(\bullet,\bullet))$ and there are three of these representations associated to three $\nu_j$ parameters. The conformal weights of these representations are
\begin{align} 
h\big((\ydiagram{1},\bullet),(\bullet,\ydiagram{1}),(\bullet,\bullet)\big)& = \frac{(k^2-1)}{2k}\left(\frac{1}{\nu_1}+\frac{1}{\nu_2}\right) \ ,\label{eq:conformal dimension minimal representation} \\
h\big((\ydiagram{1},\ydiagram{1}),(\bullet,\bullet),(\bullet,\bullet)\big)&=\frac{k}{\nu_1}+1 = \frac{2\nu_1+\nu_2+\nu_3}{\nu_1} \ . \label{eq:conformal dimension minimal adjoint representation}
\end{align}
The conformal weight of the adjoint representation receives an integer shift of $1$, which we have included in the formula. The reason for this shift is that the singlet representation of the denominator in \eqref{coset2} does not appear at the highest level of the numerator representation but rather at level $1$.

It is useful to describe these primaries also from a level-rank dual perspective in the coset description \eqref{coset1}. In this description, the representations are also labeled by three pairs of Young diagrams $(\boldsymbol{\Gamma}_1,\boldsymbol{\Gamma}_2,\boldsymbol{\Gamma}_3)$. The third pair of Young diagrams labels the \emph{conjugated} numerator representation, whereas the first and second labels correspond to the denominator representations. There is again a selection rule on these representations, which takes the same form as before,
\be 
\abs{\boldsymbol{\Gamma}_1}+\abs{\boldsymbol{\Gamma}_2}+\abs{\boldsymbol{\Gamma}_3}=0\ . \label{eq:selection rule2}
\ee
We are now describing $\mathfrak{u}(\infty)$ representations, so we should in principle also include three $\mathfrak{u}(1)$ charges. However, two $\mathfrak{u}(1)$ charges are actually redundant, since we could divide out the overall $\mathfrak{u}(1)$ in both the numerator and denominator in \eqref{coset1}. The final $\mathfrak{u}(1)$ charge in the denominator is fully determined by selection rules. To write symmetric formulas, we will keep all three $\mathfrak{u}(1)$ charges and normalize the $\mathfrak{u}(1)$ charge of the numerator in a natural (but arbitrary) way. We will declare that the third $\mathfrak{u}(1)$-charge is $u_3=-\abs{\boldsymbol{\Gamma}_3}$ (so that the fundamental representation has charge 1). The $\mathfrak{u}(1)$ charges of the denominator are then fixed to $u_i=\abs{\boldsymbol{\Gamma}_i}$ for $i=1$, $2$. The conformal weight of a primary equals then $\tilde{h}_0(\boldsymbol{\Gamma}_1,\boldsymbol{\Gamma}_2,\boldsymbol{\Gamma}_3)+\text{integer}$, where
\be 
\tilde{h}_0(\boldsymbol{\Gamma}_1,\boldsymbol{\Gamma}_2,\boldsymbol{\Gamma}_3)=\frac{\mathcal{C}_{\mu_1+\mu_2}(\boldsymbol{\Gamma}_3)}{2(\mu_1+\mu_2+k)}+\frac{\abs{\boldsymbol{\Gamma}_3}^2}{2k(\mu_1+\mu_2)}-\sum_{i=1}^2 \left(\frac{\mathcal{C}_{\mu_i}(\boldsymbol{\Gamma}_i)}{2(\mu_i+k)}+\frac{\abs{\boldsymbol{\Gamma}_i}^2}{2\mu_i k}\right)\ .
\ee
This does not look triality invariant, but using the explicit formula for the Casimir \eqref{eq:suinfty Casimir}, we can rewrite it in terms of the conformal weight of the primary \eqref{eq:primary conformal weight} as
\be 
\tilde{h}_0(\boldsymbol{\Gamma}_1,\boldsymbol{\Gamma}_2,\boldsymbol{\Gamma}_3)=h_0(\boldsymbol{\Gamma}_1\tran,\boldsymbol{\Gamma}_2\tran,\boldsymbol{\Gamma}_3\tran)
+\frac{1}{2}(\lVert\boldsymbol{\Gamma}_3\rVert-\lVert\boldsymbol{\Gamma}_1\rVert-\lVert\boldsymbol{\Gamma}_2\rVert)\ .
\ee
Here, $\tran$ denotes the transposed Young diagram, see eq.~\eqref{eq:transpose definition}. The last term is always an integer thanks to the selection rule \eqref{eq:selection rule2}. Since the integer can be absorbed in the integer shift of the primary, we are hence motivated to identify the coset representations by $\boldsymbol{\Lambda}_i=\boldsymbol{\Gamma}_i\tran$. We have checked in various examples that the integer shift also works out correctly. Since the first description in terms of $\boldsymbol{\Lambda}_i$ does not involve subtleties with $\mathfrak{u}(1)$ charges and is manifestly symmetric under triality and the additional $\mathds{Z}_2$ duality, we shall in the following work with this representation.

\subsection{Limits and free field realizations} \label{subsec:free fields}
For various special choices, the Grassmannian algebra can be realized via free field constructions. In the parametrization of the $\mu_i$'s, they usually require that some parameters are sent to infinity. In the following, we list all free field realizations known to us. The free field realizations are also special in that for these choices of parameters, the algebra possesses either a $\mathcal{W}_{\infty}$ or $\mathcal{W}^\text{even}_\infty$ subalgebra. As discussed in section~\ref{sec:OPEs}, this is \emph{not} the case for generic choices of the parameters.
\paragraph{Adjoint boson}
Consider free scalars in the adjoint representation of $\mathfrak{su}(k)$:
\be 
\partial \tensor{\Phi}{^a_b} (z) \partial \tensor{\Phi}{^c_d}(w) \sim \frac{k\tensor{\delta}{^c_b}\tensor{\delta}{^a_d}- \tensor{\delta}{^a_b}\tensor{\delta}{^c_d}}{(z-w)^2}\ .
\ee
Then the singlet part
\be 
\frac{\text{adjoint scalar}}{\mathrm{SU}(k)}
\ee
gives a free field realization for the Grassmannian. The parameters are
\be 
\mu_1 \to \infty\ ,\qquad \mu_2 \to \infty
\ee
with $\mu_1+\mu_2+\mu_3=-2k$ kept fixed. This limit should be well-defined and independent of the order in which we take the limit. This realization comes from taking both levels to infinity in the coset \eqref{coset2}.
This free-field realization has a $\mathcal{W}_\infty^\text{even}$ subalgebra generated by all bilinears of the free boson.
\paragraph{Bifundamental boson}
We can similarly take the large-level limit in the level-rank dual coset \eqref{coset1}, which leads to the free-field realization in terms of a bifundamental boson with defining OPE
\be 
\partial \tensor{\Phi}{^a_i}(z) \partial \tensor{\bar{\Phi}}{_b^j}(w) \sim \frac{\tensor{\delta}{^a_b} \tensor{\delta}{_i^j}}{(z-w)^2}\ ,
\ee
where $a$, $b$ are indices of $\mathrm{U}(\mu_1)$ and $i$, $j$ are indices of $\mathrm{U}(\mu_2)$. This yields the free-field realization
\be 
\frac{\text{complex scalar in bifundamental}}{\mathrm{U}(\mu_1)\times \mathrm{U}(\mu_2)}\ .
\ee
This realization has parameters
\be 
\mu_1=\mu_1\ , \qquad \mu_2=\mu_2\ , \qquad \mu_3 \to \infty\ .
\ee
This free field realization has a $\mathcal{W}_\infty$ subalgebra generated by all bilinears.
\paragraph{Adjoint fermion}
We can also give a fermionic free-field realization. For this, consider again the large-level limit of the coset \eqref{coset2} for one of the levels. Consider then setting the level equal to the rank, i.e.~
\be 
\frac{\mathfrak{su}(k)_k}{\mathrm{SU}(k)}\ .
\ee
But $\mathfrak{su}(k)_k$ has a free-field realization in terms of fermions in the adjoint representation,
\be 
\tensor{\psi}{^a_b} (z) \tensor{\psi}{^c_d}(w) \sim \frac{k\tensor{\delta}{^c_b}\tensor{\delta}{^a_d}- \tensor{\delta}{^a_b}\tensor{\delta}{^c_d}}{z-w}\ .
 \ee
More precisely, $\mathfrak{su}(k)_k$ has a conformal embedding into fermions in the adjoint representation. Hence, we obtain the realization
\be 
\frac{\text{adjoint fermion}}{\mathrm{SU}(k)}\ .
\ee
Because the algebra of adjoint fermions contains $\mathfrak{su}(k)_k$ as a subalgebra, we actually obtain a conformal extension of the algebra. 
The parameters for this realization are
\be 
\mu_1 \to \infty\ , \qquad \mu_2=k
\ee
with $\mu_1+\mu_3=-3k$ kept fixed.

This free field realization has a $\mathcal{W}_{\infty}$ subalgebra generated by the bilinears. This subalgebra does not lie inside the actual Grassmannian, but only its conformal extension.
\paragraph{Symmetric and antisymmetric fermions}
We can give two more examples of a similar flavor.
We can repeat the same trick using the free-field realizations \cite{Bais:1986zs}
\begin{subequations}
\begin{align}
\mathfrak{u}(k)_{k+2}&=\text{complex free fermions in symmetric representation}\ , \\
\mathfrak{u}(k)_{k-2}&=\text{complex free fermions in antisymmetric representation}\ .
\end{align}
\end{subequations}
The defining OPEs read
\begin{subequations}
\begin{align}
\tensor{\psi}{^{ab}}(z)\tensor{\bar{\psi}}{_{cd}}(w) &\sim \frac{\tensor{\delta}{^a_c}\tensor{\delta}{^b_d}+\tensor{\delta}{^a_d}\tensor{\delta}{^b_c}}{z-w}\ , \\
\tensor{\psi}{^{ab}}(z)\tensor{\bar{\psi}}{_{cd}}(w) &\sim \frac{\tensor{\delta}{^a_c}\tensor{\delta}{^b_d}-\tensor{\delta}{^a_d}\tensor{\delta}{^b_c}}{z-w}\ ,
\end{align}
\end{subequations}
where in the first case $\psi^{ab}=\psi^{ba}$ and in the second case $\psi^{ab}=-\psi^{ba}$.
Thus, we obtain the free field realizations
\begin{align}
&\frac{\text{complex free fermions in symmetric representation}}{\mathrm{SU}(k)}\ , \\
&\frac{\text{complex free fermions in antisymmetric representation}}{\mathrm{SU}(k)}\ .
\end{align}
These are again conformal extensions of the Grassmannian algebra, which both contain $\mathcal{W}_\infty$ subalgebras.\footnote{As before, the $\mathcal{W}_\infty$ subalgebras are \emph{not} subalgebras of the Grassmannian, but only of the conformal extension.}

\section{Characters} \label{sec:characters}
\subsection{Characters from the coset}
We will now study characters of the algebra starting from the Grassmannian coset \eqref{coset2}. Partial results in this direction already have been obtained in \cite{Kumar:2018dso}.
We will be mainly interested in characters for all parameters $\mu_1, \mu_2$ and $\mu_3$ generic. For $\mu_1$ and $\mu_2$ generic and $k$ a positive integer there are no null states associated to special values of the level, i.e. we can use the Weyl character formula where the summation is over the Weyl group of $\mathfrak{su}(k)$ rather than over the full affine Weyl group. In this situation, the characters of representations of the coset \eqref{coset2} can be written as
\be 
\mathrm{ch}[\boldsymbol{\Lambda}_1,\boldsymbol{\Lambda}_2,\boldsymbol{\Lambda}_3](q)=\frac{1}{k!}\int \prod_{i=1}^k \frac{\mathrm{d}z_i}{2\pi i z_i} \Delta(z) \prod_{l=1}^3 \mathrm{ch}[\boldsymbol{\Lambda}_l](z) \prod_{n=1}^\infty \frac{ (1-q^n)}{\prod_{i,j=1}^k (1-z_i z_j^{-1} q^n)}\ . \label{eq:character projection formula}
\ee
In this expression, $z$ stands collectively for $(z_1,\dots,z_k)$.
The last factor is the vacuum character of affine $\mathfrak{su}(k)$. There are two such factors coming from the numerator algebra of which one is canceled by the denominator. We denoted by $\text{ch}[\boldsymbol{\Lambda}_l](z)$ the characters of the global $\mathfrak{su}(k)$ algebras, as given by the Weyl character formula:
\be 
\text{ch}[\boldsymbol{\Lambda}](z)=\frac{\sum_{\sigma \in \mathcal{S}_k} \text{sgn}(\sigma) z^{\sigma(\boldsymbol{\Lambda}+\rho)}}{\sum_{\sigma \in \mathcal{S}_k} \text{sgn}(\sigma) z^{\sigma(\rho)}}\ .
\ee
The three labels $\boldsymbol{\Lambda}_i$ were discussed in section~\ref{subsec:primary fields}.
The symmetric group $\mathcal{S}_k$ arises as the Weyl group of $\mathrm{SU}(k)$. $\rho$ is the Weyl vector (half-sum of all positive roots) and we denote the highest weight vector of the representation $\boldsymbol{\Lambda}$ by the same symbol.

Turning back to \eqref{eq:character projection formula}, we should think of $\boldsymbol{\Lambda}_1$ and $\boldsymbol{\Lambda}_2$ as describing the numerator representations. Since we want to project onto singlets of the denominator algebra, which is diagonally embedded, we only have to keep track of the chemical potentials associated to this diagonal subalgebra and hence use always $z$ for this set of fugacity variables. We want to analyze the branching of this numerator representation into the denominator representation, whose representation we denote by $\bar{\boldsymbol{\Lambda}}_3$. This is however the same problem as searching for singlets in the triple tensor product $\boldsymbol{\Lambda}_1 \otimes \boldsymbol{\Lambda}_2 \otimes \boldsymbol{\Lambda}_3$ and thus the character of $\boldsymbol{\Lambda}_3$ appears symmetrically with the other two representations. This is consistent with the triality symmetry.

Finally, the prefactor in \eqref{eq:character projection formula} achieves the projection on singlets. The integral is an integral over the Cartan torus, with the Vandermonde determinant giving the correct Haar measure,
\be 
\Delta(z)=\prod_{i\ne j} (1-z_i z_j^{-1})\ .
\ee
The prefactor $\frac{1}{k!}$ accounts for the remaining Weyl group symmetry.

We should note that we have not included the conformal weight of the primary state in \eqref{eq:character projection formula}. To get the complete character, one should also include the prefactor $q^{h_0(\boldsymbol{\Lambda}_1,\boldsymbol{\Lambda}_2,\boldsymbol{\Lambda}_3)}$, where $h_0(\boldsymbol{\Lambda}_1,\boldsymbol{\Lambda}_2,\boldsymbol{\Lambda}_3)$ is given by \eqref{eq:primary conformal weight}. We also do not include a factor $q^{-\frac{c}{24}}$, where the central charge is given by \eqref{eq:central charge} that is usually included to improve modular properties of the characters.
%\footnote{Since there are infinitely many characters, we do not expect them to have good modular properties.}

In principle, this formula gives the Grassmannian characters, but it is computationally very inefficient because $k$ was still assumed to be positive integer. In the following we will develop a formalism that computes the characters in the limit $k \to \infty$, i.e. in the regime where all three $\mu_j$ parameters are generic, and no truncation of the algebra occurs.

From \eqref{eq:character projection formula}, it is clear that the resulting expression of the character only depends on the tensor product $\boldsymbol{\Lambda}_1 \otimes \boldsymbol{\Lambda}_2 \otimes \boldsymbol{\Lambda}_3$, i.e.
\be 
\text{ch}[\boldsymbol{\Lambda}_1,\boldsymbol{\Lambda}_2,\boldsymbol{\Lambda}_3](q)=\sum_{\boldsymbol{\Lambda} \in \boldsymbol{\Lambda}_1\otimes \boldsymbol{\Lambda}_2\otimes \boldsymbol{\Lambda}_3} \text{ch}[\boldsymbol{\Lambda}](q)\ ,
\ee
where the sum runs over all irreducible representations in the tensor product $\boldsymbol{\Lambda}_1 \otimes \boldsymbol{\Lambda}_2 \otimes \boldsymbol{\Lambda}_3$ and $\text{ch}[\boldsymbol{\Lambda}](q)$ is the character with only one non-trivial representation label. The tensor product is taken in $\mathfrak{su}(\infty)$, so the relevant representations of $\mathfrak{su}(\infty)$ have a finite number of boxes and anti-boxes, i.e. they have only a finite number of non-zero Dynkin labels. 
\subsection{Vacuum character}\label{subsec:vacuum character}
We start by determining the vacuum character. We do this combinatorially. It is convenient to consider the limit of the coset \eqref{coset2}, where $\mu_1 \to \infty$, in which case one is left with a coset
\be 
\frac{\mathfrak{su}(k)_{\mu_2}}{\mathrm{SU}(k)}\ .
\ee
In other words, the vacuum character is given by all $\mathfrak{su}(k)$ singlets in the vacuum representation of the affine algebra $\mathfrak{su}(k)_{\mu_2}$. This is indeed also what \eqref{eq:character projection formula} tells us.

We can count these singlets as follows. Let us denote the $\mathfrak{su}(k)$-valued currents by the matrix $J_m$, where $m$ denotes as usual the mode number. The matrices are traceless, i.e.~$\tr (J_m)=0$. A general $\mathfrak{su}(k)$ singlet has then the form
\be 
\tr(J_{-m_{1,1}} \dots J_{-m_{1,\ell_1}}) \cdots \tr(J_{-m_{n,1}} \dots J_{-m_{n,\ell_n}})\ket{0}\ . \label{eq:generic state vacuum rep}
\ee
We should note that a trace of one current vanishes and the trace is cyclically symmetric, which gives some identifications. Reordering of the traces would produce commutators and the resulting expression has fewer currents $J$. Thus, for the purpose of counting the number of singlets, we can treat the traces as commutative. We have thus reduced problem to counting single-trace contributions, since multi-trace contributions can be accounted for by taking the plethystic exponential of the single-trace contribution. 

It is useful to introduce a pictorial way of writing these traces that makes cyclicity manifest. We denote e.g.
\be 
\vacbubblethree{1}{4}{2}\qquad =\qquad \tr( J_{-1} J_{-4} J_{-2})\ .
\ee
At low lying levels, one can easily list the possible traces. We have listed them explicitly in table~\ref{tab:low-lying necklaces}.
\begin{table}
\begin{tabular}{c|l}
Level 2 & \vacbubbletwo{1}{1} \\
Level 3 & \vacbubbletwo{1}{2},\ \vacbubblethree{1}{1}{1}  \\
Level 4 & \vacbubbletwo{1}{3},\  \vacbubbletwo{2}{2},\ \vacbubblethree{1}{1}{2}, \ \vacbubblefour{1}{1}{1}{1}  \\
Level 5 & \vacbubbletwo{1}{4},\  \vacbubbletwo{2}{3},\ \vacbubblethree{1}{2}{2},\ \vacbubblethree{1}{1}{3}, \ \vacbubblefour{1}{1}{1}{2}, \ \vacbubblefive{1}{1}{1}{1}{1}  \\
\end{tabular}
\caption{The low-lying traces.}  \label{tab:low-lying necklaces}
\end{table}

In the mathematics literature, these combinatorial objects are known as \emph{necklaces}. They can be counted using the P\'olya enumeration theorem. We review the necessary ingredients in appendix~\ref{app:combinatorics} and state here the result. We introduce the generating function
\be 
Z(q)=\sum_{n=1}^\infty q^n \times \text{(number of necklaces with total mode $n$)}\ .
\ee
Then \cite{Flajolet}
\be 
Z(q)=\sum_{m=1}^\infty \frac{\phi(m)}{m} \log \left(\frac{(1-q^m)^2}{1-2q^m}\right)\ , \label{eq:single necklaces generating function}
\ee
where $\phi(m)$ is the Euler totient function. 

It is now straightforward to calculate the plethystic exponential and hence the vacuum character
\be 
\text{ch}[\text{vac}](q)=\exp\left(\sum_{n=1}^\infty \frac{Z(q^n)}{n}\right)=\prod_{n=1}^\infty \frac{(1-q^n)^2}{1-2q^n}\ . \label{eq:Grassmannian vacuum character}
\ee
We note that the result has a surprisingly simple form. From a CFT point of view the occurrence of the factor 2 in the denominator is quite unusual and evades a simple oscillator interpretation.

\subsection{The adjoint representation}
We now analyze the simplest non-trivial representation, which is given by $\boldsymbol{\Lambda}=(\ydiagram{1},\ydiagram{1})$.\footnote{Recall that a single box is not allowed, because of the selection rule \eqref{eq:selection rule}.} We can again work out the character combinatorially. We can for instance assume that $(\ydiagram{1},\ydiagram{1})$ is the representation of one of the numerator algebras. In the same large level limit, we were considering in the previous subsection, we hence build states on a ground state that transforms itself in the adjoint representation and hence can be identified with a traceless matrix, which we can call $\Omega$. Thus, the states in this representation look exactly the same as in \eqref{eq:generic state vacuum rep}, but have one further special trace of the form 
\be 
 \tr(J_{-m_1} \dots J_{-m_\ell}\Omega)\ .
\ee
Pictorially, we will represent this trace by a `brane' with an `open string' attached to it, e.g.
\be 
\branethree{2}{3}{1} \qquad = \qquad \tr (J_{-2}J_{-3}J_{-1}\Omega)\ .
\ee 
Thus, counting states in this representation is essentially reduced to counting the number of open string configurations. Of course, such an open string configuration can always be multiplied by additional vacuum bubbles, e.g.
\be 
\vacbubblethree{3}{5}{1}\ \vacbubblefour{2}{2}{2}{2} \ \branethree{3}{7}{1} 
\ee
would represent a state at level 28. It is clear that the character factorizes,
\be 
\text{ch}[(\ydiagram{1},\ydiagram{1})](q)=\text{ch}[\text{vac}](q)\, \Phi(q)\ ,
\ee
where the `wedge character' $\Phi(q)$ only counts the connected open string configurations. Let us write\footnote{There is no empty string without numbers, since $\Omega$ is traceless.}
\be 
\Phi(q)=\sum_{n=1}^\infty c(n)q^n\ .
\ee
Then $c(n)$ satisfies a simple recursion relation. Consider the first number $m$ on an open string configuration, which can be any integer from $1$ to $n$. Removing it leads to a valid open string configuration, where the total level is $n-m$. Thus, $c(n)$ satisfies the recursion
\be 
c(n)=1+\sum_{m=1}^{n-1} c(m)\ , \label{eq:minimal wedge character recursion relation}
\ee
which is solved by $c(n)=2^{n-1}$. Thus, we conclude that
\be 
\Phi(q)=\sum_{n=1}^\infty 2^{n-1}q^n=\frac{q}{1-2q}\ . \label{eq:minimal wedge character}
\ee
Hence the full character of the minimal representation is given by
\be 
\text{ch}[(\ydiagram{1},\ydiagram{1})](q)=\frac{q}{1-2q}\prod_{n=1}^\infty \frac{(1-q^n)^2}{1-2q^n}\ .
\ee
We should mention that this character starts at order $q+\mathcal{O}(q^2)$ and hence incorporates the integer shift that we have discussed in section~\ref{subsec:primary fields}.
\subsection{Higher representations}
We now generalize the previous analysis to obtain the characters of any other representation systematically.
We begin by illustrating the method with the help of the representations with Dynkin labels
\be 
[2,0,\dots,0,2]\ , \quad [0,1,0,\dots,0,2]\ , \quad [2,0,\dots,0,1,0]\ , \quad [0,1,0,\dots,0,1,0]\ ,
\ee
which appear in the tensor product of the adjoint representation with itself. We will in the following denote them by their box and anti-box labels, i.e.~by
\be 
\big( \ydiagram{2},\ydiagram{2}\big)\ , \quad \big( \ydiagram{1,1},\ydiagram{2}\big)\ , \quad \big( \ydiagram{2},\ydiagram{1,1}\big)\ , \quad\big( \ydiagram{1,1},\ydiagram{1,1}\big)\ .
\ee
Written in terms of tensors, they are tensors $\tensor{\Omega}{_{ab}^{cd}}$ that are (anti)symmetrized in the indices $ab$ and $cd$ and that have all traces removed. Hence we can think of them in terms of a brane on which two open strings end. Due to tracelessness condition, every open string starting and ending on the same brane must carry at least one mode number. The endpoint of the strings are (anti)symmetrized depending on the representation. We only determine the wedge character $\Phi_{\boldsymbol{\Lambda}}(q)$ of the corresponding representation, the full character is obtained by multiplying with the vacuum character. A generic open string configuration with two strings is given by
\be 
\twobranetwo{2}{3}{1}{5}\ , \qquad \twobranetwocross{2}{3}{1}{5}\ .
\ee
These two expressions stand for the contractions
\be 
\tensor{\Omega}{_{ab}^{cd}} \tensor{J}{_{-2}^b_e} \tensor{J}{_{-3}^e_c} \tensor{J}{_{-1}^a_f} \tensor{J}{_{-5}^f_d}\ , \qquad \tensor{\Omega}{_{ab}^{cd}} \tensor{J}{_{-2}^b_e} \tensor{J}{_{-3}^e_d} \tensor{J}{_{-1}^a_f} \tensor{J}{_{-5}^f_c}\ .
\ee
The order of the end-points of the open strings is hence important, since it specifies the way in which indices are contracted. A general state in the representation specified by the labels $[2,0,\dots,0,2]$ has to be correctly symmetrized, which means that the actual state is given by
\begin{multline} 
\frac{1}{4}\Bigg(\ \twobranetwo{2}{3}{1}{5}+\twobranetwocross{2}{3}{1}{5}\\
+\twobranetwo{1}{5}{2}{3}+\twobranetwocross{1}{5}{2}{3}\ \Bigg)\ .
\end{multline}
We hence see that the number of states is almost given by $\tfrac{1}{2} \Phi(q)^2$, where $\Phi(q)$ is the wedge character of the single string we determined in \eqref{eq:minimal wedge character}. The only states we are not correctly accounting with this are the diagonal states of the form
\be 
 \twobranetwo{2}{3}{2}{3}\ ,
\ee
since they survive in the symmetrization, but not in the antisymmetrization. To correct for it, one adds the term $\tfrac{1}{2} \Phi(q^2)$ and hence
\be 
\Phi_{\text{\tiny \ydiagram{2}, \ydiagram{2}}}(q)=\frac{1}{2}\big(\Phi(q)^2+\Phi(q^2)\big)=\frac{q^2(1-q)^2}{(1-2q)^2(1-2q^2)}\ .
\ee
Similarly, one determines the wedge character in the other cases, which yields
\begin{subequations}
\begin{align}
\Phi_{\text{\tiny \ydiagram{1,1}, \ydiagram{1,1}}}(q)&=\Phi_{\text{\tiny \ydiagram{2}, \ydiagram{2}}}(q)=\frac{1}{2}\big(\Phi(q)^2+\Phi(q^2)\big)=\frac{q^2(1-q)^2}{(1-2q)^2(1-2q^2)}\ , \\
\Phi_{\text{\tiny \ydiagram{2}, \ydiagram{1,1}}}(q)&=\Phi_{\text{\tiny \ydiagram{1,1}, \ydiagram{2}}}(q)=\frac{1}{2}\big(\Phi(q)^2-\Phi(q^2)\big)=\frac{q^3(2-3q)}{(1-2q)^2(1-2q^2)}\ .
\end{align}
\end{subequations}

We notice that we essentially only used representation theory of the diagonal symmetric subgroup $\mathcal{S}_2 \subset \mathcal{S}_2 \times \mathcal{S}_2$ of the two symmetric groups that permute the end of the strings. The representation under the symmetric group permuting the ingoing strings alone is not important, since one can always (anti)symmetrize
\be 
\twobranetwo{1}{5}{2}{3}\pm \twobranetwocross{1}{5}{2}{3}\ .
\ee
There can never be any cancellations, since the pictures will always be distinct. We only care about the representation under \emph{simultaneous} exchange of outgoing and incoming strings and hence about the subgroup $\mathcal{S}_2 \subset \mathcal{S}_2 \times \mathcal{S}_2$. This explains why $\Phi_{\text{\tiny \ydiagram{1,1}, \ydiagram{1,1}}}(q)=\Phi_{\text{\tiny \ydiagram{2}, \ydiagram{2}}}(q)$ and $\Phi_{\text{\tiny \ydiagram{2}, \ydiagram{1,1}}}(q)=\Phi_{\text{\tiny \ydiagram{1,1}, \ydiagram{2}}}(q)$. The former two representations are associated to the representations $(\ydiagram{2},\, \ydiagram{2})$ and $(\ydiagram{1,1},\, \ydiagram{1,1})$ under the group $\mathcal{S}_2 \times \mathcal{S}_2$. Here and in the following, we often view Young diagrams as specifying irreducible representations of the unitary group, but also of the symmetric group. However, under the diagonal subgroup, they transform always in the trivial representation $\ydiagram{2}$, which leads to the same character.

Thus, one can convince oneself that
\be 
\Phi_{\Lambda,\, \bar{\Lambda}}(q)=\Phi_{\Lambda \otimes \bar{\Lambda}}(q)\ ,
\ee
where the tensor product is taken as $\mathcal{S}_n$ representations with $n$ being the number of (anti)boxes. For only one non-trivial representation label $\boldsymbol{\Lambda}$, the selection rule \eqref{eq:selection rule} guarantees that there are always equally many boxes and anti-boxes. This reduces our task to determining $\Phi_\Lambda(q)$ in general for any irreducible representation $\Lambda$ of $\mathcal{S}_n$. The formula extends to reducible representations by linearity. 

Determining the remaining wedge-characters proceeds very much as in $\mathcal{W}_{1+\infty}$ \cite{Gaberdiel:2015wpo}. We have
\be 
\Phi_\Lambda(q)=\frac{1}{n!} \sum_{\sigma \in \mathcal{S}_n} \chi_\Lambda(\sigma) \Phi_\sigma(q)\ , \label{eq:general wedge character}
\ee
where for $\sigma$ with cycle structure $(1)^{m_1}(2)^{m_2}(3)^{m_3} \cdots$, we define
\be 
\Phi_\sigma(q)= \prod_{i \ge 1} \Phi(q^i)^{m_i}\ , \label{eq:Phi permutation}
\ee
and $\chi_\Lambda(\sigma)$ denotes the character of the symmetric group in the respective irreducible representation.
For instance,
\begin{subequations}
\begin{align}
\Phi_\text{\tiny \ydiagram{1}}(q)&=\Phi(q)\ , \\
\Phi_\text{\tiny \ydiagram{2}}(q)&=\frac{1}{2}\left(\Phi(q)^2+\Phi(q^2)\right)\ , \\
\Phi_\text{\tiny \ydiagram{1,1}}(q)&=\frac{1}{2}\left(\Phi(q)^2-\Phi(q^2)\right)\ , \\
\Phi_\text{\tiny \ydiagram{3}}(q)&=\frac{1}{6}\left(\Phi(q)^3+3\Phi(q)\Phi(q^2)+2\Phi(q^3)\right)\ , \\
\Phi_\text{\tiny \ydiagram{2,1}}(q)&=\frac{1}{3}\left(\Phi(q)^3-\Phi(q^3)\right)\ , \\
\Phi_\text{\tiny \ydiagram{1,1,1}}(q)&=\frac{1}{6}\left(\Phi(q)^3-3\Phi(q)\Phi(q^2)+2\Phi(q^3)\right)\ .
\end{align}
\end{subequations}
Note that this is of the form of the transformation between Schur polynomials (LHS) and the Newton power sum polynomials (RHS). The wedge character associated to the representation $[1,1,0,\dots,0,1,1]$ is then for example
\be 
\Phi_\text{\tiny \ydiagram{2,1}, \ydiagram{2,1}}(q)= \Phi_{\text{\tiny \ydiagram{2,1}} \,\otimes \,\text{\tiny \ydiagram{2,1}}}(q)=\Phi_{\text{\tiny \ydiagram{3}}}+\Phi_{\text{\tiny \ydiagram{2,1}}}(q)+\Phi_{\text{\tiny \ydiagram{1,1,1}}}(q)=\frac{1}{3} \left(2\Phi(q)^3+\Phi(q^3)\right)\ .
\ee
We should mention that we have exposed another symmetry that was not visible on the level of the formula \eqref{eq:character projection formula}. Namely, the branching to the diagonal symmetric group $\mathcal{S}_n \times \mathcal{S}_n \to \mathcal{S}_n$ is invariant under sending $(\Lambda,\bar{\Lambda}) \to (\Lambda \otimes \text{sgn}, \bar{\Lambda} \otimes \text{sgn})$, where sgn denotes the alternating representation of the symmetric group. On the level of the Young diagram, this is exactly the transposition $\boldsymbol{\Lambda} \to \boldsymbol{\Lambda}\tran$ that was discussed in section~\ref{subsec:primary fields}. Thus, we have shown that the characters are invariant under $(\boldsymbol{\Lambda}_1,\boldsymbol{\Lambda}_2,\boldsymbol{\Lambda}_3)\to (\boldsymbol{\Lambda}_1,\tran\boldsymbol{\Lambda}_2\tran,\boldsymbol{\Lambda}_3\tran)$. This should be expected, since we have argued that this corresponds to the action of the $\mathds{Z}_2$ duality.
\subsection{Conjugation symmetry}
As we have discussed in section~\ref{subsec:symmetries}, the Grassmannian algebra possesses a $\mathds{Z}_2$ automorphism that is induced from the involution $J \mapsto -J\tran$ in the coset. Thus, we want to refine the vacuum character by including a $\mathds{Z}_2$ fugacity $x$, that satisfies $x^2=1$. Let us first consider the effect on a trace:
\be 
\tr(J_{-m_1} \cdots J_{-m_\ell})\longmapsto (-1)^{\ell}\tr(J_{-m_1}\tran \cdots J_{-m_\ell}\tran)= (-1)^{\ell}\tr(J_{-m_\ell} \cdots J_{-m_1})\ .
\ee
Thus, the parity reflects the necklace and weights it with a sign, depending on whether the length is even or odd. Such necklaces can be again counted by the P\'olya enumeration theorem (although the relevant group is now the dihedral group). We derive in appendix~\ref{app:characters} the result for the single-trace generating function, which takes the form
\be 
Z(q)=\frac{1+x}{2}\sum_{m=1}^\infty \frac{\phi(m)}{m} \log \left(\frac{(1-q^m)^2}{1-2q^m}\right)+\frac{(1-x)q^2}{2(1-2q^2)}\ .
\ee
Upon taking the plethystic exponential, we obtain the refined vacuum character, which takes the form
\be 
\text{ch}[\text{vac}](q,x)%=\exp\left(\sum_{n=1}^\infty \frac{Z(q^n)}{n}\right)
=\prod_{n=1}^\infty \left(\frac{(1-q^{2n-1})^2}{1-2q^{2n-1}}\right)^{\frac{1+x}{2}}\left(\frac{(1-q^{2n})^2}{1-2q^{2n}}\right)^{\frac{3+x}{4}}\left(\frac{1+q^{2n}}{1-q^{2n}}\right)^{(1-x)2^{n-3}} .
\ee
From this character, we can read off how many fields there are of a given spin and parity. We write the vacuum character in the form
\be 
\text{ch}[\text{vac}](q,x)=\prod_{s=1}^\infty \prod_{n=s}^\infty \frac{1}{(1-q^n)^{N_+(s)}(1-x q^n)^{N_-(s)}}\ ,
\ee
where $N_+(s)$ is the number of parity even fields and $N_-(s)$ the number of parity odd fields of a given spin. For low values of spin, we obtain table~\ref{tab:spin parity}. While at spins $\le 5$, even (odd) spin fields are even (odd) under parity, this pattern quickly breaks at higher spins. This is one of the main differences of the Grassmannian to the $\mathcal{W}_\infty$ algebra.
\begin{table}
\begin{center}
\begin{tabular}{c|ccccccccccccccc}
spin $s$  & 1 & 2 & 3 & 4 & 5 & 6 & 7 & 8 & 9 & 10 & 11 & 12 & 13 & 14 & 15 \\
\hline
$N_+(s)$ & 0 & 1 & 0 & 2 & 0 & 5 & 1 & 12 & 8 & 32 & 32 & 98 & 124 & 307 & 473 \\
$N_-(s)$ & 0 & 0 & 1 & 0 & 2 & 1 & 5 & 4 & 16 & 16 & 48 & 66 & 156 & 243 & 537 \\
\end{tabular}
\end{center}
\caption{Number of strong generators of the Grassmannian at a low spin.}
\label{tab:spin parity}
\end{table}

One can also include the $\mathds{Z}_2$ fugacity $x$ in the wedge character. On representations, conjugation acts by $(\boldsymbol{\Lambda}_1,\boldsymbol{\Lambda}_2,\boldsymbol{\Lambda}_3) \to (\bar{\boldsymbol{\Lambda}}_1,\bar{\boldsymbol{\Lambda}}_2,\bar{\boldsymbol{\Lambda}}_3)$. Thus, it makes only sense to talk about the $\mathds{Z}_2$ fugacity in characters that are self-conjugate.
 On the string/brane picture, the parity reverses the direction of the strings and weights the picture with the sign $(-1)^L$, where $L$ is the sum of lengths of all strings.
 
The minimal wedge character is self-conjugate and one can repeat the above discussion to arrive at a recursion relation analogous to \eqref{eq:minimal wedge character recursion relation}. Here, one sums over the first and last number in the string, giving
\be 
c(n)=x+\sum_{r,s=1}^n c(m-r-s) \times \begin{cases}
1\ , & r=s\,,\\
\frac{1+x}{2}\ , & r \ne s\,.
\end{cases}
\ee
Here, we put conventionally $c(0)=1$ and $c(m)=0$ for $m<0$ to simplify the notation. The solution is given by
\be 
c(n)=\begin{cases}
2^{n-2}(1+x)\ , & n\text{ even}\,, \\
2^{n-2}(1+x)+2^{\frac{n-3}{2}}(x-1)\ , & n\text{ odd}\, ,
\end{cases}
\ee
and hence the refined wedge character reads
\be
\Phi(q,x)=\sum_{n=1}^\infty c(n)=\frac{q \left(x+q-xq-q^2-xq^2\right)}{(1-2 q) \left(1-2 q^2\right)}\ .
\ee
The above discussion of how to obtain the higher wedge characters still goes through except that in \eqref{eq:general wedge character}, also $x$ has to be raised to the same power as $q$, i.e.~\eqref{eq:Phi permutation} gets modified to
\be 
\Phi_\sigma(q,x)=\prod_{i \ge 1} \Phi(q^i,x^i)^{m_i}\ . 
\ee
\subsection{Summary and generalization to higher Grassmannians}
Let us summarize the result for the characters, that we have obtained. We focus here again on the unrefined characters. We immediately state the generalization to a generalized flag manifold with $N$ parameters, whose representations are labeled by $N$ $\mathrm{SU}(\infty)$ representations $\boldsymbol{\Lambda}_1,\dots, \boldsymbol{\Lambda}_N$. Above, we have discussed the minimal case of $N=3$. All arguments carry through directly to the higher Grassmannians. The character of a representation $(\boldsymbol{\Lambda}_1,\dots,\boldsymbol{\Lambda}_N)$ factorizes according to
\be 
\text{ch}[\boldsymbol{\Lambda}_1,\dots,\boldsymbol{\Lambda}_N](q)=\text{ch}[\text{vac}](q)\Phi_{\boldsymbol{\Lambda}_1,\dots,\boldsymbol{\Lambda}_N}(q)\ ,
\ee
where the vacuum character is given by
\be 
\text{ch}[\text{vac}](q)=\prod_{n=1}^\infty\frac{(1-q^n)^{N-1}}{1-(N-1) q^n}
\ee
The wedge character $\Phi_{\boldsymbol{\Lambda}_1,\dots,\boldsymbol{\Lambda}_N}(q)$ counts the number of open string configurations ending on $N$ branes. For example, the Grassmannian representation described by the label $\big(( \ydiagram{1,1},\, \ydiagram{2}),\, (\bullet,\, \bullet),\, ( \ydiagram{1},\, \bullet),\, ( \bullet,\, \ydiagram{1})\big)$ would count open string configurations of the form
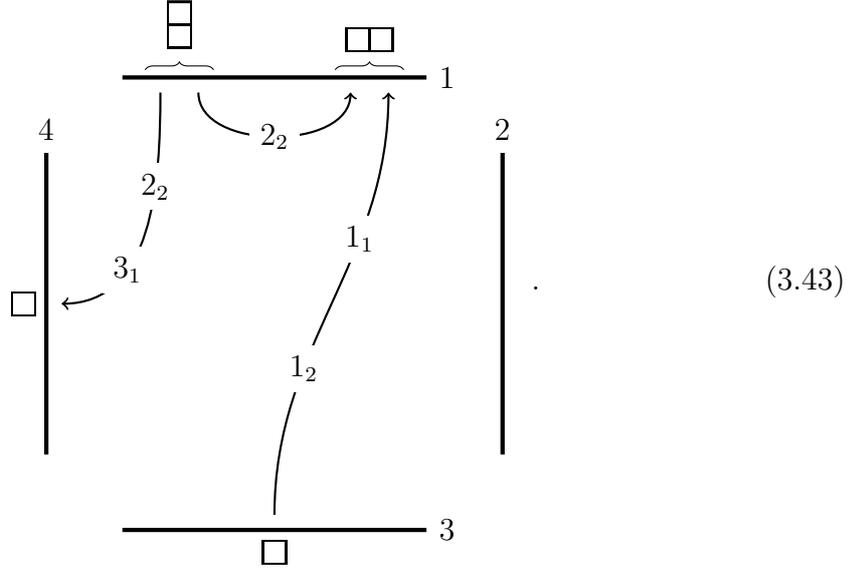
\be 
\begin{tikzpicture}[baseline={([yshift=-.5ex]current bounding box.center)}]
\draw[ultra thick] (-3,-2) -- (-3,2) node[above] {4};
\draw[ultra thick] (3,-2) -- (3,2) node[above] {2};
\draw[ultra thick] (-2,3) -- (2,3) node[right] {1};
\draw[ultra thick] (-2,-3) -- (2,-3) node[right] {3};
\draw[thick,<-,out=-90,in=90] (1.5,2.8) to node[pos=.33, fill=white] {$1_1$} node[pos=.67, fill=white] {$1_2$} (0,-2.8);
\draw[thick,<-,out=-90,in=-90] (1,2.8) to node[pos=.5, fill=white] {$2_2$} (-1,2.8);
\draw[thick,->,out=-90,in=0] (-1.5,2.8) to node[pos=.33, fill=white] {$2_2$} node[pos=.67, fill=white] {$3_1$} (-2.8,0);
\draw [decorate,decoration={brace,amplitude=3pt}]
(-1.7,3.1) -- (-.8,3.1) node [black,midway,yshift=.6cm] {$ \ydiagram{1,1}$};
\draw [decorate,decoration={brace,amplitude=3pt}]
(.8,3.1) -- (1.7,3.1) node [black,midway,yshift=.4cm] {$ \ydiagram{2}$};
\node at (-3.3,0) {$ \ydiagram{1}$};
\node at (0,-3.3) {$ \ydiagram{1}$};
\end{tikzpicture}\ . \label{eq:four brane example state}
\ee
The picture shows a state at level 9. Multiple strings ending on the same brane are (anti)symmetrized according to the representation labeling the specific brane. With respect to the basic Grassmannian, there is one novelty. The numbers we put on the strings now have $N-2$ different species and we distinguish them by the corresponding subscript. This can be seen as before in the large level limit, which now counts singlets in the vacuum representation of the product algebra $\mathfrak{su}(k)_{\mu_1} \times \cdots \times \mathfrak{su}(k)_{\mu_{N-2}}$.

The wedge character depends only on the tensor product in $\mathfrak{su}(\infty)$
\be
\label{eq:wedgecharacter}
\Phi_{\boldsymbol{\Lambda}_1,\dots,\boldsymbol{\Lambda}_N}(q)=\sum_{\boldsymbol{\Lambda} \in \boldsymbol{\Lambda}_1 \otimes \boldsymbol{\Lambda}_2 \otimes \dots \otimes \boldsymbol{\Lambda}_N} \Phi_{\boldsymbol{\Lambda}}(q)\ .
\ee
Writing $\Phi_{\boldsymbol{\Lambda}}(q)=\Phi_{\Lambda,\, \bar{\Lambda}}(q)$ in terms of box and anti-box labels, this character in turn only depends on the $\mathcal{S}_n$ tensor product $\Lambda\otimes \bar{\Lambda}$. Finally, $\Phi_{\Lambda \otimes \bar{\Lambda}}(q)$ can always be reduced to the wedge character of a single open string involving only one brane using \eqref{eq:general wedge character}. The wedge character involving only a single string and a single brane (the adjoint representation) is in general given by
\be 
\Phi_{\mathrm{adj}}(q)=\frac{(N-2)q}{1-(N-1)q}\ . \label{eq:minimal wedge character flag manifold}
\ee

\paragraph{Bifundamental representation}
It is easy to find directly the wedge character of the bifundamental representation. Combinatorially this corresponds to an open string stretched between two branes. We can find a similar recursion relation as in the case of adjoint representation: introducing a counting function $c(j)$ such that
\begin{equation}
\Phi_{\mathrm{bif}}(q) = \sum_{j=0}^\infty c(j) q^j\ ,
\end{equation}
we have first of all $c(0)=1$ because the configuration with no mode numbers labeling the string is allowed and unique and it corresponds to the highest weight state. The recursion relation satisfied by $c(j)$ is
\begin{equation}
c(j) = (N-2) \sum_{k=1}^j c(j-k)
\end{equation}
which says simply that removing the left-most mode index (whose value is $k$ and can be of one of $N-2$ species) we obtain another open string with total mode decreased by $k$. The solution of this recursion relation is
\begin{equation}
c(j) = (N-2) (N-1)^{j-1}
\end{equation}
and the corresponding wedge character is
\begin{equation}
\Phi_{\mathrm{bif}}(q) = 1 + (N-2) \sum_{j=1}^\infty (N-1)^{j-1} q^j = \frac{1-q}{1-(N-1)q}=1+\Phi_\text{adj}(q)\ .
\end{equation}

\paragraph{More complicated example}
To illustrate formula \eqref{eq:wedgecharacter} on a more complicated example, we explicitly compute the wedge character of the representation $\big(( \ydiagram{1,1},\, \ydiagram{2}),\, (\bullet,\, \bullet),\, ( \ydiagram{1},\, \bullet),\, ( \bullet,\, \ydiagram{1})\big)$. First, taking the $\mathrm{SU}(\infty)$ tensor product leads to
\begin{multline} 
\Phi_\text{\tiny \ydiagram{1}, \ydiagram{1}}(q)+\Phi_\text{\tiny \ydiagram{2}, \ydiagram{2}}(q)+\Phi_\text{\tiny \ydiagram{1,1}, \ydiagram{1,1}}(q)+3\Phi_\text{\tiny \ydiagram{1,1}, \ydiagram{2}}(q)+\Phi_\text{\tiny \ydiagram{1,1,1}, \ydiagram{3}}(q)\\
+\Phi_\text{\tiny \ydiagram{1,1,1}, \ydiagram{2,1}}(q)+\Phi_\text{\tiny \ydiagram{2,1}, \ydiagram{2,1}}(q)+\Phi_\text{\tiny \ydiagram{2,1}, \ydiagram{3}}(q)\ .
\end{multline}
Next, we perform the tensor product of the box representations with the anti-box representations inside the symmetric group, which reduces the expression to
\be 
\Phi_\text{\tiny \ydiagram{1}}(q)+2 \Phi_\text{\tiny \ydiagram{2}}(q)+3 \Phi_\text{\tiny \ydiagram{1,1}}(q)+3 \Phi_\text{\tiny \ydiagram{1,1}}(q)+\Phi_\text{\tiny \ydiagram{3}}(q)+3 \Phi_\text{\tiny \ydiagram{2,1}}(q)+2 \Phi_\text{\tiny \ydiagram{1,1,1}}(q)\ .
\ee
Finally, we use the symmetrisation formula \eqref{eq:general wedge character} to express this in terms of the minimal wedge character \eqref{eq:minimal wedge character flag manifold} . This finally yields
\begin{multline} 
\Phi(q)+\tfrac{5}{2}\Phi(q)^2-\tfrac{1}{2}\Phi(q^2)+\tfrac{3}{2} \Phi(q)^3-\tfrac{1}{2} \Phi(q)\Phi(q)^2\\
=\frac{q(1-q) \left(2-q-9 q^3\right)}{(1-3 q)^3 \left(1-3 q^2\right)}=2 q+15 q^2+88 q^3+423 q^4+1866 q^5+\mathcal{O}\left(q^6\right)\ .
\end{multline}
From this we see that there is no way to draw a picture like \eqref{eq:four brane example state} with no numbers put on the strings that respects the (anti)symmetrizations and tracelessness.\footnote{Remember that there cannot be an empty string that connects the same brane with itself.} There are exactly two ways to put one number on the strings and they are given by
\begin{multline} 
\begin{tikzpicture}[baseline={([yshift=-.5ex]current bounding box.center)}]
\draw[ultra thick] (-2,-1.5) -- (-2,1.5);
\draw[ultra thick] (2,-1.5) -- (2,1.5);
\draw[ultra thick] (-1.5,2) -- (1.5,2);
\draw[ultra thick] (-1.5,-2) -- (1.5,-2);
\draw[thick,<-,out=-90,in=90] (1.2,1.8) to(0,-1.8);
\draw[thick,<-,out=-90,in=-90] (.7,1.8) to node[pos=.5, fill=white] {$1_1$} (-.7,1.8);
\draw[thick,->,out=-90,in=0] (-1.2,1.8) to (-1.8,0);
\end{tikzpicture}\  -\ \begin{tikzpicture}[baseline={([yshift=-.5ex]current bounding box.center)}]
\draw[ultra thick] (-2,-1.5) -- (-2,1.5);
\draw[ultra thick] (2,-1.5) -- (2,1.5);
\draw[ultra thick] (-1.5,2) -- (1.5,2);
\draw[ultra thick] (-1.5,-2) -- (1.5,-2);
\draw[thick,<-,out=-90,in=90] (1.2,1.8) to (0,-1.8);
\draw[thick,->,out=-90,in=0] (-.7,1.8) to node[pos=.2, fill=white] {}  (-1.8,0);
\draw[thick,<-,out=-90,in=-90] (.7,1.8) to node[pos=.5, fill=white] {$1_1$} (-1.2,1.8);
\end{tikzpicture}\\
+\ 
\begin{tikzpicture}[baseline={([yshift=-.5ex]current bounding box.center)}]
\draw[ultra thick] (-2,-1.5) -- (-2,1.5);
\draw[ultra thick] (2,-1.5) -- (2,1.5);
\draw[ultra thick] (-1.5,2) -- (1.5,2);
\draw[ultra thick] (-1.5,-2) -- (1.5,-2);
\draw[thick,<-,out=-90,in=90] (.7,1.8) to node[pos=.13,fill=white] {} (0,-1.8);
\draw[thick,<-,out=-90,in=-90] (1.2,1.8) to node[pos=.5, fill=white] {$1_1$} (-.7,1.8);
\draw[thick,->,out=-90,in=0] (-1.2,1.8) to (-1.8,0);
\end{tikzpicture}\ -\ 
\begin{tikzpicture}[baseline={([yshift=-.5ex]current bounding box.center)}]
\draw[ultra thick] (-2,-1.5) -- (-2,1.5);
\draw[ultra thick] (2,-1.5) -- (2,1.5);
\draw[ultra thick] (-1.5,2) -- (1.5,2);
\draw[ultra thick] (-1.5,-2) -- (1.5,-2);
\draw[thick,<-,out=-90,in=90] (.7,1.8) to node[pos=.13,fill=white] {} (0,-1.8);
\draw[thick,->,out=-90,in=0] (-.7,1.8) to node[pos=.2, fill=white] {}  (-1.8,0);
\draw[thick,<-,out=-90,in=-90] (1.2,1.8) to node[pos=.5, fill=white] {$1_1$} (-1.2,1.8);
\end{tikzpicture}
\end{multline}
and the same picture for the other species of numbers. Counting states in this way becomes quickly very cumbersome. As we have mentioned several times, the complete character of this representation is obtained by multiplying with the vacuum character and putting in the conformal weights of the primary field.%\footnote{We choose not to insert an additional $q^{-\frac{c}{24}}$ in the character.} 
We hence have
\be
\label{eq:examplecharacter}
\text{ch}\big[( \ydiagram{1,1},\, \ydiagram{2}),\, (\bullet,\, \bullet),\, ( \ydiagram{1},\, \bullet),\, ( \bullet,\, \ydiagram{1})\big](q)=\frac{q^{1+h_0}(1-q) \left(2-q-9 q^3\right)}{(1-3 q)^3 \left(1-3 q^2\right)}\prod_{n=1}^\infty \frac{(1-q^n)^3}{1-3q^n}\ ,
\ee
where $h_0$ is given by \eqref{eq:primary conformal weight}. Let us emphasize that the character formulas such as \eqref{eq:examplecharacter} are as simple as they are because we work under assumption that $\mu_j$ parameters are generic. For special values of parameters of the algebra, there would be singular vectors and the characters would depend on these special values of $\mu_j$ parameters.

\section{Gluing construction} \label{sec:gluing}
In this section, we explore the Grassmannian as fundamental building block to build bigger vertex operator algebras. The gluing construction we introduce is analogous to the gluing construction of $\mathcal{W}_\infty$, that was successfully used to build more complicated $\mathcal{W}$-algebras \cite{Gaiotto:2017euk, Prochazka:2017qum, Gaberdiel:2017hcn, Prochazka:2018tlo, Gaberdiel:2018nbs, Li:2019nna, Li:2019lgd}.

To denote these gluings, we find it convenient to employ a graph notation, where the Grassmannian itself is represented as a trivalent vertex. As we will discuss in section~\ref{sec:4parameters}, we suspect that the Grassmannian can be further extended to a four-parameter family of algebras. If true, we would only touch the tip of the iceberg and there is a far richer and more general story underlying this construction. We will restrict in this section to representations that are accessible to us through the coset construction. 

\subsection{The Grassmannian as a 3-punctured sphere}
The Grassmannian is parametrized by parameters $\nu_1$, $\nu_2$ and $\nu_3$ which are permuted by triality. See section~\ref{subsec:symmetries} for an overview of the different parametrizations of the Grasmannian algebra.
Moreover, representations accessible from the coset construction are parametrized by three labels, associated to these three parameters. We pictured them in section~\ref{sec:characters} by three branes on which open strings can end. For the purpose of gluing, it will be helpful to think of the Grassmannian as a trivalent vertex or a 3-punctured sphere, where the three legs (punctures) are associated to the three parameters.

We can of course freely permute the legs and thus triality becomes manifest in this picture.
\be 
\text{Gr}(\nu_1,\nu_2,\nu_3)=\begin{tikzpicture}[baseline={([yshift=-.5ex]current bounding box.center)}]
\draw[very thick] (0,0) to (120:1) node[left] {$\nu_1$};
\draw[very thick] (0,0) to (240:1) node[left] {$\nu_2$};
\draw[very thick] (0,0) to (0:1) node[right] {$\nu_3$};
\end{tikzpicture}=\begin{tikzpicture}[baseline={([yshift=-.5ex]current bounding box.center)}]
\draw[thick] (0,0) circle (1.3);
\fill[black!50!white] (-.5,.9) circle (.1) node[black, right] {$\nu_1$};
\fill[black!50!white] (-.9,0) circle (.1) node[black, right] {$\nu_2$};
\fill[black!50!white] (1,-.5) circle (.1) node[black, left] {$\nu_3$};
\end{tikzpicture}\ .
\ee
Representations are now associated to the three legs/punctures as well. We similarly identify the generalized flag algebra $\text{Gr}(\nu_1,\dots,\nu_n)$ with the $n$-punctured sphere. In terms of graphs we can associate any tree graph with $n$ external legs. We will discuss below that all choices are equivalent.
\subsection{Gluing Grassmannians}
This identification makes only sense if it is compatible with a suitable pair of pants decomposition. We can decompose the four-punctured sphere as follows
\be 
\label{eq:grass4glue}
\text{Gr}(\nu_1,\nu_2,\nu_3,\nu_4)=\begin{tikzpicture}[baseline={([yshift=-.5ex]current bounding box.center)}]
\draw[thick] (0,0) circle (1.3);
\fill[black!50!white] (-.5,.8) circle (.1) node[black, right] {$\nu_1$};
\fill[black!50!white] (-.5,-.8) circle (.1) node[black, right] {$\nu_2$};
\draw[thick] (4,0) circle (1.3);
\fill[black!50!white] (4.5,.8) circle (.1) node[black, left] {$\nu_3$};
\fill[black!50!white] (4.5,-.8) circle (.1) node[black, left] {$\nu_4$};
\fill[white] (1,.3) rectangle (3,-.3);
\draw[bend right=10, thick] (1,.2) to (3,.2);
\draw[bend left=10, thick] (1,-.2) to (3,-.2);
\end{tikzpicture}\ .
\ee
This would hence suggest that there is a conformal embedding
\be 
\text{Gr}(\nu_1,\nu_2,\nu) \times \text{Gr}(\nu',\nu_3,\nu_4)\subset \text{Gr}(\nu_1,\nu_2,\nu_3,\nu_4) \label{eq:Grassmannian times Grassmannian conformal embedding}
\ee
for some choice of $\nu$ and $\nu'$. From the coset description, we can see that this is indeed true. In terms of $\mu$-parameters, we have
\begin{align}
 \text{Gr}(\mu_1,\mu_2,\mu_3,\mu_4)&=\frac{\mathfrak{su}(k)_{\mu_1} \times \mathfrak{su}(k)_{\mu_2} \times \mathfrak{su}(k)_{\mu_3}}{\mathfrak{su}(k)_{\mu_1+\mu_2+\mu_3}} \\
 &\supset \frac{\mathfrak{su}(k)_{\mu_1} \times \mathfrak{su}(k)_{\mu_2} }{\mathfrak{su}(k)_{\mu_1+\mu_2}}\times \frac{\mathfrak{su}(k)_{\mu_1+\mu_2} \times \mathfrak{su}(k)_{\mu_3}}{\mathfrak{su}(k)_{\mu_1+\mu_2+\mu_3}}\\
\label{eq:basicgluing}
 &=\text{Gr}(\mu_1,\mu_2,\mu_3+\mu_4) \times \text{Gr}(\mu_1+\mu_2,\mu_3,\mu_4)\ .
\end{align}
Translating back to $\nu$-parameters, we hence obtain \eqref{eq:Grassmannian times Grassmannian conformal embedding}, where $\nu$ and $\nu'$ are chosen such that
\begin{subequations}
\begin{align}
\nu&=-\nu'\ , \\
\nu_1+\nu_2+\nu&=\nu_3+\nu_4+\nu'\ .
\end{align} \label{eq:basic gluing conditions}
\end{subequations}
While the coset description only gives us access to the parameter range $k \in \mathds{Z}_{\ge 0}$, everything depends through rational functions on the $\mu$- or $\nu$-parameters. Thus we can analytically continue the gluing construction and it should hence also hold for arbitrary values of parameters.
Thus, we can obtain the four-parameter flag manifold by gluing two Grassmannians along two punctures,
\be 
\begin{tikzpicture}[baseline={([yshift=-.5ex]current bounding box.center)}]
\draw[thick] (0,0) circle (1.3);
\fill[black!50!white] (-.5,.8) circle (.1) node[black, right] {$\nu_1$};
\fill[black!50!white] (-.5,-.8) circle (.1) node[black, right] {$\nu_2$};
\draw[thick] (4,0) circle (1.3);
\fill[black!50!white] (4.5,.8) circle (.1) node[black, left] {$\nu_3$};
\fill[black!50!white] (4.5,-.8) circle (.1) node[black, left] {$\nu_4$};
\fill[white] (1,.3) rectangle (3,-.3);
\draw[bend right=10, thick] (1,.2) to (3,.2);
\draw[bend left=10, thick] (1,-.2) to (3,-.2);
\fill[black!50!white] (1.1,0) circle (.1) node[black, left] {$\nu$};
\fill[black!50!white] (2.9,0) circle (.1) node[black, right] {$\nu'$};
\end{tikzpicture}\ =\ 
\begin{tikzpicture}[baseline={([yshift=-.5ex]current bounding box.center)}]
\draw[thick] (0,0) circle (1.3);
\fill[black!50!white] (-.5,.9) circle (.1) node[black, right] {$\nu_1$};
\fill[black!50!white] (-.9,0) circle (.1) node[black, right] {$\nu_2$};
\fill[black!50!white] (.8,.5) circle (.1) node[black, left] {$\nu_3$};
\fill[black!50!white] (1,-.5) circle (.1) node[black, left] {$\nu_4$};
\end{tikzpicture}\ .
\ee
Of course, we can decompose the four-punctured sphere in three inequivalent ways into pairs of pants. The quadriality of the $\text{Gr}(\nu_1,\nu_2,\nu_3,\nu_4)$ ensures that they are all equivalent, e.g.
\be 
\begin{tikzpicture}[baseline={([yshift=-.5ex]current bounding box.center)}]
\draw[thick] (0,0) circle (1.3);
\fill[black!50!white] (-.5,.8) circle (.1) node[black, right] {$\nu_1$};
\fill[black!50!white] (-.5,-.8) circle (.1) node[black, right] {$\nu_2$};
\draw[thick] (4,0) circle (1.3);
\fill[black!50!white] (4.5,.8) circle (.1) node[black, left] {$\nu_3$};
\fill[black!50!white] (4.5,-.8) circle (.1) node[black, left] {$\nu_4$};
\fill[white] (1,.3) rectangle (3,-.3);
\draw[bend right=10, thick] (1,.2) to (3,.2);
\draw[bend left=10, thick] (1,-.2) to (3,-.2);
\fill[black!50!white] (1.1,0) circle (.1);
\fill[black!50!white] (2.9,0) circle (.1);
\end{tikzpicture}\ =\ 
\begin{tikzpicture}[baseline={([yshift=-.5ex]current bounding box.center)}]
\draw[thick] (0,0) circle (1.3);
\fill[black!50!white] (.8,-.5) circle (.1) node[black, above] {$\nu_4$};
\fill[black!50!white] (-.8,-.5) circle (.1) node[black, above] {$\nu_2$};
\draw[thick] (0,4) circle (1.3);
\fill[black!50!white] (.8,4.5) circle (.1) node[black, below] {$\nu_3$};
\fill[black!50!white] (-.8,4.5) circle (.1) node[black, below] {$\nu_1$};
\fill[white] (.3,1) rectangle (-.3,3);
\draw[bend left=10, thick] (.2,1) to (.2,3);
\draw[bend right=10, thick] (-.2,1) to (-.2,3);
\fill[black!50!white] (0,1.1) circle (.1);
\fill[black!50!white] (0,2.9) circle (.1);
\end{tikzpicture}\ . \label{eq:crossing}
\ee
On the level of operator product expansions, we verified that the total stress-energy tensor of the four-punctured algebra admits exactly these three decompositions.

Of course, one can also iterate the same construction to obtain the higher generalized flag manifolds, i.e.
\be 
\text{Gr}(\nu_1,\nu_2,\nu_3,\nu_4,\nu_5)=\begin{tikzpicture}[baseline={([yshift=-.5ex]current bounding box.center)}]
\draw[thick] (0,0) circle (1.3);
\fill[black!50!white] (-.5,.8) circle (.1) node[black, right] {$\nu_1$};
\fill[black!50!white] (-.5,-.8) circle (.1) node[black, right] {$\nu_2$};
\draw[thick] (4,0) circle (1.3);
\fill[black!50!white] (4,.8) circle (.1) node[black, above] {$\nu_3$};
\fill[white] (1,.3) rectangle (3,-.3);
\draw[bend right=10, thick] (1,.2) to (3,.2);
\draw[bend left=10, thick] (1,-.2) to (3,-.2);
\draw[thick] (8,0) circle (1.3);
\fill[black!50!white] (8.5,.8) circle (.1) node[black, left] {$\nu_4$};
\fill[black!50!white] (8.5,-.8) circle (.1) node[black, left] {$\nu_5$};
\fill[white] (5,.3) rectangle (7,-.3);
\draw[bend right=10, thick] (5,.2) to (7,.2);
\draw[bend left=10, thick] (5,-.2) to (7,-.2);
\fill[black!50!white] (1.1,0) circle (.1) node[black, left] {$\nu_{\text{L}}$};
\fill[black!50!white] (2.9,0) circle (.1) node[black, right] {$\nu_{\text{L}}'$};
\fill[black!50!white] (5.1,0) circle (.1) node[black, left] {$\nu_{\text{R}}$};
\fill[black!50!white] (6.9,0) circle (.1) node[black, right] {$\nu_{\text{R}}'$};
\end{tikzpicture}\ ,
\ee
where $\nu_{\text{L}}$, $\nu_{\text{L}}'$, $\nu_{\text{R}}$ and $\nu_{\text{R}}'$ are chosen such that
\begin{subequations}
\begin{align}
\nu_\text{L}&=-\nu_{\text{L}}'\ , \\
\nu_\text{R}&=-\nu_{\text{R}}'\ , \\
\nu_1+\nu_2+\nu_\text{L}&=\nu_{\text{L}}'+\nu_3+\nu_\text{R}=\nu_{\text{R}}'+\nu_4+\nu_5\ .
\end{align}
\end{subequations} 

We finally observe that there is a very simple rule determining the $\nu$-parameters of the glued puncture. For two punctures to be gluable, the gluing $\nu$-parameters have to be opposite and the sum of all the $\nu$-parameters on every three-punctured algebra has to coincide. We will explore a generalization of this rule in section~\ref{subsec:more gluings}.

It should now be clear why the generalized flag manifolds can represented through an arbitrary trivalent tree graph. The different graphs correspond to the different pair of pants decompositions of the $n$-puncture sphere into three-punctured spheres. Because of the basic crossing relation \eqref{eq:crossing}, they are all equivalent.

\paragraph{Gluing fields} In the gluing, one has to add some ``gluing fields'' that are supported along the gluing tube. In other words, the four-parameter flag manifold decomposes into the product of two Grassmannians together with an infinite tower of non-trivial representations. In terms of the chiral algebras, we can write
\label{eq:gluing33to4}
\be 
\text{Gr}(\nu_1,\nu_2,\nu_3,\nu_4)=\bigoplus_{\boldsymbol{\Lambda},\, \abs{\boldsymbol{\Lambda}}=0}\big(\bullet,\bullet,\boldsymbol{\Lambda}\big)\times \big(\bar{\boldsymbol{\Lambda}}, \bullet,\bullet\big)\ .
\ee
Here the RHS is a sum of representations of $\text{Gr}(\nu_1,\nu_2,\nu) \times \text{Gr}(\nu',\nu_3,\nu_4)$. $\boldsymbol{\Lambda}$ runs over all possible representations of $\mathfrak{su}(\infty)$ with equal number of boxes and antiboxes exactly once and we claim that in this decomposition, every character appears indeed exactly once. For this to make sense, we should check that the dimension of these representations is integer with respect to the total stress-energy tensor (which is an algebraic sum of the partial stress-energy tensors). It takes the form
\begin{align}
h_0&=h_0^{\nu_1,\nu_2,\nu}\big( \bullet,\bullet,\boldsymbol{\Lambda}\big)+h_0^{\nu',\nu_3,\nu_4} \big(\bar{\boldsymbol{\Lambda}},\bullet,\bullet\big) \\
&= \frac{\mathcal{C}_{\nu_1+\nu_2+\nu}(\boldsymbol{\Lambda})}{2\nu}+\frac{\mathcal{C}_{\nu_3+\nu_4+\nu'}(\bar{\boldsymbol{\Lambda}})}{2\nu'}=0\ .
\end{align}
Here, $\mathcal{C}_N$ is the Casimir of $\mathfrak{su}(N)$ of the respective representation. The two conditions \eqref{eq:basic gluing conditions} hence ensure cancellation of the two terms.
Thus, the dimension at which the gluing fields appear is entirely determined by the integer shift of the representation that we discussed on the level of the wedge character in section~\ref{sec:characters}.
From the brane picture, it is actually quite easy to see that this shift is at least equal to the number of boxes in $\boldsymbol{\Lambda}$. Taking into account both factors in the decomposition, a representation $\boldsymbol{\Lambda}$ only starts to contribute at conformal weight $\ge \lVert\boldsymbol{\Lambda}\rVert$.
%For a representation involving only brane as we are currently discussing, this integer shift is easy to determine. Let $\Lambda$ and $\bar{\Lambda}$ both have $\abs{\Lambda}$ many boxes or anti-boxes, respectively. In the brane-picture, there are hence $\abs{\Lambda}$ many outgoing and ingoing strings. Because of the traceless

The gluing \eqref{eq:gluing33to4} at the level of algebras requires the following character identity to hold:
\be 
\prod_{n=1}^\infty \frac{(1-q^n)^3}{1-3q^n}=\left(\prod_{n=1}^\infty \frac{(1-q^n)^2}{1-2q^n}\right)^2 \sum_{\boldsymbol{\Lambda},\, \abs{\boldsymbol{\Lambda}}=0} \Phi_{\boldsymbol{\Lambda}}(q) \Phi_{\bar{\boldsymbol{\Lambda}}}(q)\ .
\ee
We implemented the wedge characters in \texttt{Mathematica} and checked this identity up to $\mathcal{O}(q^{21})$. 

Similarly, we can glue higher analogues of Grassmannians, the generalized flag algebras. Gluing an $m_1$-punctured sphere to an $m_2$-punctured sphere leads to an $(m_1+m_2-2)$-punctured sphere. The corresponding character identity reads
\be 
\prod_{n=1}^\infty \frac{(1-q^n)^{m_1+m_2-3}}{1-(m_1+m_2-3)q^n}=\prod_{i=1,\, 2}\prod_{n=1}^\infty \frac{(1-q^n)^{m_i-1}}{1-(m_i-1)q^n} \sum_{\boldsymbol{\Lambda},\, \abs{\boldsymbol{\Lambda}}=0} \Phi^{m_1}_{\boldsymbol{\Lambda}}(q)\Phi^{m_2}_{\bar{\boldsymbol{\Lambda}}}(q)\ .
\ee
Here, the superscript in the wedge character indicates that these are wedge characters of the $m_1$-punctured or $m_2$-punctured algebra, respectively.

Apart from the verification of the gluing prescription at the level characters, we can also check the compatibility with representation-theoretic properties like the conformal dimensions. In particular, in the four-punctured algebra we have bifundamental representations associated to pairs of punctures. These representations should be permuted under the action of the quadrality symmetry, which is however not manifest after the decomposition of the algebra as in \eqref{eq:grass4glue}. Let us compare the dimension of bifundamental representation associated to punctures $(\nu_1,\nu_2)$ and $(\nu_1,\nu_3)$ in $\text{Gr}(\nu_1,\nu_2,\nu_3,\nu_4)$. In the first case, the dimension is given directly by the formula (\ref{eq:conformal dimension minimal representation}) applied to the left sphere. The dimension is
\be
\frac{k^2-1}{2k} \left(\frac{1}{\nu_1} + \frac{1}{\nu_2} \right)
\ee
where $k=\tfrac{1}{2}(\nu_1+\nu_2+\nu_3+\nu_4)$. On the other hand, the bifundamental representation associated to punctures $(\nu_1,\nu_3)$ is obtained by connecting the puncture $\nu_1$ to the gluing tube by bifundamental representation and by connecting the tube to the puncture $\nu_3$ by another bifundamental representation. The conformal dimension is now
\be
\frac{k^2-1}{2k} \left( \frac{1}{\nu_1} + \frac{2}{\nu_3+\nu_4-\nu_1-\nu_2} + \frac{1}{\nu_3} + \frac{2}{\nu_1+\nu_2-\nu_3-\nu_4} \right) = \frac{k^2-1}{2k} \left(\frac{1}{\nu_1} + \frac{1}{\nu_3} \right)
\ee
where the contributions associated to the gluing tube canceled and we find a result compatible with the quadrality of the algebra.

\subsection{Gluing affine algebras} 
We can also glue affine algebras to the punctured spheres/graphs. In this subsection, we will use the $\mu$-parameters, since they facilitate the notation.
 To see that this is possible, we can again use the coset description. In one level-rank dual frame, we see that
\be 
\frac{\mathfrak{u}(\mu_1+\mu_2)_k}{\mathfrak{u}(\mu_1)_k \times \mathfrak{u}(\mu_2)_k} \times \mathfrak{u}(\mu_1)_k \subset \frac{\mathfrak{u}(\mu_1+\mu_2)_k}{\mathfrak{u}(\mu_2)_k}
\ee
and thus there is a conformal extension that one can perform. There is a second way to attach an affine algebra to the Grassmannian algebra, which can be seen in the level-rank dual perspective:
\be 
\frac{\mathfrak{su}(k)_{\mu_1} \times \mathfrak{su}(k)_{\mu_2}}{\mathfrak{su}(k)_{\mu_1+\mu_2}} \times \mathfrak{su}(k)_{\mu_1+\mu_2} \subset \mathfrak{su}(k)_{\mu_1} \times \mathfrak{su}(k)_{\mu_2}\ .
\ee
This second way of gluing completely separates the algebra into two affine algebras. Thus, we will mostly be concerned with the first way of gluing.

We will denote affine algebra punctures by a cross, i.e.
\be 
\begin{tikzpicture}[baseline={([yshift=-.5ex]current bounding box.center)}]
\draw[thick] (0,0) circle (1.3);
\fill[black!50!white] (-.5,.9) circle (.1) node[black, right] {$\mu_1$};
\fill[black!50!white] (-.9,0) circle (.1) node[black, right] {$\mu_2$};
\fill[black!50!white] (1,-.5) circle (.1) node[black, left] {$\mu_3$};
\end{tikzpicture}\times \mathfrak{u}(\mu_1)_k \subset \begin{tikzpicture}[baseline={([yshift=-.5ex]current bounding box.center)}]
\draw[thick] (0,0) circle (1.3);
\draw[black!50!white] (-.5,.9) node[cross=3pt, thick] {};
\draw (-.5,.9) node[right] {$\mu_1$};
\fill[black!50!white] (-.9,0) circle (.1) node[black, right] {$\mu_2$};
\fill[black!50!white] (1,-.5) circle (.1) node[black, left] {$\mu_3$};
\end{tikzpicture}\ .
\ee
From the coset realization and from what we discussed in section~\ref{sec:grassmannian}, we can see that the sphere with two Grassmannian punctures and one affine algebra puncture is the matrix $\mathcal{W}_{1+\infty}$ algebra. Here are some identifications with various known algebras:
\begin{align}
\text{matrix $\mathcal{W}_{1+\infty}$} &= \begin{tikzpicture}[baseline={([yshift=-.5ex]current bounding box.center)}]
\draw[thick] (0,0) circle (1.3);
\draw[black!50!white] (-.5,.9) node[cross=3pt, thick] {};
\draw (-.5,.9) node[right] {$\mu_1$};
\fill[black!50!white] (-.9,0) circle (.1) node[black, right] {$\mu_2$};
\fill[black!50!white] (1,-.5) circle (.1) node[black, left] {$\mu_3$};
\end{tikzpicture}\ , \\
\mathfrak{u}(\mu_1+\mu_2)_{-\frac{1}{2}(\mu_1+\mu_2+\mu_3)} &= \begin{tikzpicture}[baseline={([yshift=-.5ex]current bounding box.center)}]
\draw[thick] (0,0) circle (1.3);
\draw[black!50!white] (-.5,.9) node[cross=3pt, thick] {};
\draw (-.5,.9) node[right] {$\mu_1$};
\draw[black!50!white] (-.9,0) node[cross=3pt, thick] {};
\draw (-.9,0) node[right] {$\mu_2$};
\fill[black!50!white] (1,-.5) circle (.1) node[black, left] {$\mu_3$};
\end{tikzpicture}\ , \\
\mathfrak{u}(\mu_1+\mu_2+\mu_3)_{-\frac{1}{2}(\mu_1+\mu_2+\mu_3)} &= \begin{tikzpicture}[baseline={([yshift=-.5ex]current bounding box.center)}]
\draw[thick] (0,0) circle (1.3);
\draw[black!50!white] (-.5,.9) node[cross=3pt, thick] {};
\draw (-.5,.9) node[right] {$\mu_1$};
\draw[black!50!white] (-.9,0) node[cross=3pt, thick] {};
\draw (-.9,0) node[right] {$\mu_2$};
\draw[black!50!white] (1,-.5) node[cross=3pt, thick] {};
\draw (1,-.5) node[left] {$\mu_3$};
\end{tikzpicture}\ .
\end{align}
While the first two identifications follow from the coset representation, the last one is more non-trivial, but was already discussed in section~\ref{sec:grassmannian}. For these gluings to make sense, we again have to check that the gluing fields have integer dimension. We claim that the gluing is summarized by the following branching of the glued algebra
\be 
\begin{tikzpicture}[baseline={([yshift=-.5ex]current bounding box.center)}]
\draw[thick] (0,0) circle (1.3);
\draw[black!50!white] (-.5,.9) node[cross=3pt, thick] {};
\draw (-.5,.9) node[right] {$\mu_1$};
\fill[black!50!white] (-.9,0) circle (.1) node[black, right] {$\mu_2$};
\fill[black!50!white] (1,-.5) circle (.1) node[black, left] {$\mu_3$};
\end{tikzpicture}=\bigoplus_{\boldsymbol{\Lambda},\, \abs{\boldsymbol{\Lambda}}=0} \big(\boldsymbol{\Lambda},\bullet,\bullet\big) \times \bar{\boldsymbol{\Lambda}}\tran\ . \label{eq:Winf decomposition}
\ee
Here, the second factor is an affine $\mathfrak{u}(\mu_1)_k$ representation with label $\boldsymbol{\Lambda}\tran$. Since $\abs{\boldsymbol{\Lambda}}=0$, the overall $\mathfrak{u}(1)$ charge vanishes. The conformal weight of the representations in this decomposition is
\be 
\frac{\mathcal{C}_k(\boldsymbol{\Lambda})}{2(\mu_1+k)} +\frac{\mathcal{C}_{\mu_1}(\bar{\boldsymbol{\Lambda}}\tran)}{2(\mu_1+k)}=\frac{1}{2}\lVert \boldsymbol{\Lambda}\rVert\ .
\ee
Here we used the explicit formula \eqref{eq:suinfty Casimir} for the Casimir of $\mathfrak{su}(\infty)$ representations. Thus, we predict the following character
identity
\be 
\prod_{n=1}^\infty \frac{1}{(1-q^n)^{n \mu^2}}=\prod_{n=1}^\infty \frac{(1-q^n)^2}{(1-2q^n)(1-q^n)^{\mu^2}} \sum_{\boldsymbol{\Lambda},\,\abs{\boldsymbol{\Lambda}}=0}q^{\frac{1}{2}\lVert\boldsymbol{\Lambda}\rVert} \Phi_{\boldsymbol{\Lambda}}(q) \dim_\mu (\bar{\boldsymbol{\Lambda}}\tran)\ . \label{eq:affine gluing character identity}
\ee
The LHS is the vacuum character of matrix $\mathcal{W}_{1+\infty}$ that we expect to obtain from this gluing. In the RHS, we factored out the vacuum characters of the Grassmannian and the affine algebra. We then sum over all wedge characters. Here, $\dim_\mu(\bar{\boldsymbol{\Lambda}}\tran)$ denotes the dimension of this $\mathfrak{su}(\mu)$ representation. We have again checked directly in \texttt{Mathematica} that this identity holds up to $\mathcal{O}(q^{21})$. We should also note that the transpose is actually not visible at the level of the wedge characters, since we have $\Phi_{\boldsymbol{\Lambda}\tran}(q)=\Phi_{\boldsymbol{\Lambda}}(q)$, as discussed in section~\ref{sec:characters}. This is a reflection that we could choose not to transpose $\boldsymbol{\Lambda}$, but send $\mu_i \to -\mu_i$ instead in the Grassmannian.

One can similarly check the other gluings that we identified above. To make things easier, we can actually note that the coset description makes it obvious that
\begin{align}
\frac{\mathfrak{u}(\mu_1+\cdots+\mu_{n-1})_k}{\mathfrak{u}(\mu_1)_k \times \cdots \times \mathfrak{u}(\mu_{n-3})_k} 
&\supset \frac{\mathfrak{u}(\mu_1+\cdots+\mu_{n-1})_k}{\mathfrak{u}(\mu_1)_k \times \cdots \mathfrak{u}(\mu_{n-2}+\mu_{n-1})_k} \times \mathfrak{u}(\mu_{n-2}+\mu_{n-1})_k\\
&\supset \frac{\mathfrak{u}(\mu_1+\cdots+\mu_{n-1})_k}{\mathfrak{u}(\mu_1)_k \times\cdots  \times \mathfrak{u}(\mu_{n-1})_k} \times \mathfrak{u}(\mu_{n-2})_k\times\mathfrak{u}(\mu_{n-1})_k\ .
\end{align}
In pictures,
\be 
\begin{tikzpicture}[baseline={([yshift=-.5ex]current bounding box.center)}]
\draw[thick] (0,0) circle (1.3);
\draw[black!50!white] (-.5,.9) node[cross=3pt, thick] {};
\draw (-.5,.9) node[right] {$\mu_1$};
\draw[black!50!white] (.3,.5) node[cross=3pt, thick] {};
\draw (.3,.5) node[right] {$\mu_2$};
\fill[black!50!white] (-.9,0) circle (.1) node[black, right] {$\mu_3$};
\fill[black!50!white] (0,-.6) circle (.1) node[black, left] {$\mu_4$};
\fill[black!50!white] (1,-.5) circle (.1) node[black, left] {$\mu_5$};
\node at (0,-.3) {$ \dots$};
\end{tikzpicture}\ =\ \begin{tikzpicture}[baseline={([yshift=-.5ex]current bounding box.center)}]
\draw[thick] (0,0) circle (1.3);
\draw[black!50!white] (-.6,.6) node[cross=3pt, thick] {};
\draw (-.6,.6) node[right] {$\mu_1+\mu_2$};
\fill[black!50!white] (-.9,0) circle (.1) node[black, right] {$\mu_3$};
\fill[black!50!white] (0,-.6) circle (.1) node[black, left] {$\mu_4$};
\fill[black!50!white] (1,-.5) circle (.1) node[black, left] {$\mu_5$};
\node at (0,-.3) {$ \dots$};
\end{tikzpicture}\ ,
\ee
and hence one can always merge two crosses into one. The manifest symmetry $\mathfrak{u}(\mu_1) \times \mathfrak{u}(\mu_2)$ always gets enhanced to $\mathfrak{u}(\mu_1+\mu_2)$. We can hence also write
\be 
\mathfrak{u}(\mu_1)_{-\frac{1}{2}(\mu_1+\mu_2)} = \ \begin{tikzpicture}[baseline={([yshift=-.5ex]current bounding box.center)}]
\draw[thick] (0,0) circle (1.3);
\draw[black!50!white] (-.5,.9) node[cross=3pt, thick] {};
\draw (-.5,.9) node[right] {$\mu_1$};
\fill[black!50!white] (1,-.5) circle (.1) node[black, left] {$\mu_2$};
\end{tikzpicture}\ , \qquad 
\mathfrak{u}(\mu)_{-\frac{\mu}{2}} =\  \begin{tikzpicture}[baseline={([yshift=-.5ex]current bounding box.center)}]
\draw[thick] (0,0) circle (1.3);
\draw[black!50!white] (-.5,0) node[cross=3pt, thick] {};
\draw (-.5,0) node[right] {$\mu$};
\end{tikzpicture}\ .
\ee
We have already motivated the last identification in section~\ref{sec:grassmannian}, but we can show this now on the level of the characters. Observe that we have the gluing
\be 
 \begin{tikzpicture}[baseline={([yshift=-.5ex]current bounding box.center)}]
\draw[thick] (0,0) circle (1.3);
\draw[black!50!white] (-.5,.9) node[cross=3pt, thick] {};
\draw (-.5,.9) node[right] {$\mu_1$};
\draw[black!50!white] (.5,0) node[cross=3pt, thick] {};
\draw (.5,0) node[left] {$\mu_2$};
\end{tikzpicture}\ =\ \begin{tikzpicture}[baseline={([yshift=-.5ex]current bounding box.center)}]
\draw[thick] (0,0) circle (1.3);
\draw[black!50!white] (-.5,.2) node[cross=3pt, thick] {};
\draw (-.5,.2) node[right] {$\mu_1$};
\draw[thick] (4,0) circle (1.3);
\draw[black!50!white] (4.5,.1) node[cross=3pt, thick] {};
\draw (4.5,.1) node[right] {$\mu_2$};
\fill[white] (1,.3) rectangle (3,-.3);
\draw[bend right=10, thick] (1,.2) to (3,.2);
\draw[bend left=10, thick] (1,-.2) to (3,-.2);
\fill[black!50!white] (1.1,0) circle (.1) node[black, left] {$\mu_2$};
\fill[black!50!white] (2.9,0) circle (.1) node[black, right] {$\mu_1$};
\end{tikzpicture}
\ee
Since this is a gluing of the type we have discussed above, we know the decomposition of the vacuum module of the LHS. We expect
\be 
\mathfrak{u}(\mu_1+\mu_2)_{-\frac{1}{2}(\mu_1+\mu_2)} = \bigoplus_{\boldsymbol{\Lambda}} \boldsymbol{\Lambda} \times \bar{\boldsymbol{\Lambda}}\ .
\ee
Here, the decomposition is in terms of affine $\mathfrak{u}(\mu_1)_{-\frac{1}{2}(\mu_1+\mu_2)} \times \mathfrak{u}(\mu_2)_{-\frac{1}{2}(\mu_1+\mu_2)}$ representations. The $\mathfrak{u}(1)$ charges are determined as explained in section~\ref{subsec:primary fields}. In this case, there is no restriction on $\abs{\boldsymbol{\Lambda}}$. The conformal weight of the gluing representations is 
\be
h(\boldsymbol{\Lambda})=\frac{\mathcal{C}_{\mu_1}(\boldsymbol{\Lambda})}{\mu_1-\mu_2}+\frac{\abs{\boldsymbol{\Lambda}}^2}{(\mu_1-\mu_2)(\mu_1+\mu_2)} +\frac{\mathcal{C}_{\mu_2}(\bar{\boldsymbol{\Lambda}})}{\mu_2-\mu_1}+\frac{\abs{\boldsymbol{\Lambda}}^2}{(\mu_2-\mu_1)(\mu_1+\mu_2)}=\lVert\boldsymbol{\Lambda}\rVert \ .
\ee
The second and fourth term are contributions to the conformal dimension from the non-zero $\mathfrak{u}(1)$ charges. To simplify, we used \eqref{eq:suinfty Casimir}. On the level of the character, we hence obtain the identity
\be 
\prod_{n=1}^\infty \frac{(1+q^n)^{(\mu_1+\mu_2)^2-1}}{1-q^n}=\prod_{n=1}^\infty \frac{1}{(1-q^n)^{\mu_1^2+\mu_2^2}}\sum_{\boldsymbol{\Lambda}} q^{\lVert\boldsymbol{\Lambda}\rVert} \dim_{\mu_1}(\boldsymbol{\Lambda})\dim_{\mu_2}(\bar{\boldsymbol{\Lambda}})\ . \label{eq:top algebra gluing}
\ee
The left hand side of this equation is the vacuum character of $\mathfrak{u}(\mu_1+\mu_2)_{-\frac{1}{2}(\mu_1+\mu_2)}$. We have checked this identity again up to $\mathcal{O}(q^{21})$.

One can generalize this gluing by gluing an affine algebra to an $m$-punctured sphere. While the resulting algebra has not been studied in the literature, we can predict its vacuum character from our gluing construction. For this, we simply compute the RHS of \eqref{eq:affine gluing character identity} for the algebra of the generalized flag manifold. We computed this character up to $\mathcal{O}(q^{21})$ and found that the result is consistent with the following vacuum character
\be 
\text{ch} \left[\begin{tikzpicture}[baseline={([yshift=-.5ex]current bounding box.center)}]
\draw[thick] (0,0) circle (1.3);
\draw[black!50!white] (-.5,.9) node[cross=3pt, thick] {};
\draw (-.5,.9) node[right] {$\mu_1$};
\fill[black!50!white] (-.9,0) circle (.1) node[black, right] {$\mu_2$};
\fill[black!50!white] (0,-.6) circle (.1) node[black, left] {$\mu_3$};
\fill[black!50!white] (1,-.5) circle (.1) node[black, left] {$\mu_m$};
\node at (0,-.3) {$ \dots$};
\end{tikzpicture}\right](q)=\prod_{n=1}^\infty \frac{(1-q^n)^{m-2}}{(1-(m-2)q^n)(1-q^n)^{\frac{(m-2)^n-1}{m-3} \mu_1^2}}\ .
\ee
This result is also valid for the original case $m=3$ we discussed above by taking the limit. Much more surprising, the formula gives the correct character for $m=2$ and even for $m=1$, even though our derivation of it breaks down. For $m=2$, we obtain indeed the vacuum character of an affine algebra. Finally for $m=1$, the character can be rewritten as
\be 
\prod_{n=1}^\infty \frac{(1+q^n)^{\mu_1^2-1}}{1-q^n}\ ,
\ee
which agrees with the LHS of \eqref{eq:top algebra gluing} and is the vacuum character of $\mathfrak{u}(\mu_1)_{-\frac{1}{2}\mu_1}$.

Finally, let us also briefly discuss the second way in which we can attach an affine algebra to the Grassmannian, in a way that completely separates the Grassmannian into a product of affine algebras. We are attaching $\mathfrak{su}(k)_{\mu_2+\mu_3}$ this time. The decomposition of the glued algebra is very similar to \eqref{eq:Winf decomposition}, except for a missing transpose,
\be 
\mathfrak{su}(k)_{\mu_2} \times \mathfrak{su}(k)_{\mu_3} = \bigoplus_{\boldsymbol{\Lambda},\, \abs{\boldsymbol{\Lambda}}=0} \big(\boldsymbol{\Lambda},\bullet,\bullet\big) \times \bar{\boldsymbol{\Lambda}}\ .
\ee
Here the RHS denotes modules of $\text{Gr}(\mu_1,\mu_2,\mu_3) \times \mathfrak{su}(k)_{\mu_2+\mu_3}$. The conformal weight of the gluing field can be seen to vanish (up to the integer shift that is implicitly contained in the wedge character). On the character level, we obtain the identity
\be 
\prod_{n=1}^\infty \frac{1}{(1-q^n)^{2k^2}}=\prod_{n=1}^\infty \frac{(1-q^n)^2}{(1-2q^n)(1-q^n)^{k^2}}\sum_{\boldsymbol{\Lambda},\, \abs{\boldsymbol{\Lambda}}=0} \Phi_{\boldsymbol{\Lambda}}(q) \dim_k(\bar{\boldsymbol{\Lambda}})\ .
\ee
We checked this identity up to $\mathcal{O}(q^{11})$.\footnote{Since gluing fields appear at a lower dimension, it is computationally more difficult to check this equality.} We also checked the corresponding identity for the higher punctured algebra. %Since this type of gluing does not lead to new algebras, we do not consider further.

\subsection{More gluings} \label{subsec:more gluings}
A natural question is whether these gluings exhaust the list of possible gluings. This is not the case, as can be seen already at the level of the coset. Restricting the attention to gluings of Grassmannians without gluing affine Lie algebras, we have for instance the following gluing,
\begin{multline}
\label{eq:mixedgluing}
\frac{\mathfrak{su}(k)_{\mu_2} \times \mathfrak{su}(k)_{\mu_3}}{\mathfrak{su}(k)_{\mu_2+\mu_3}}\times \frac{\mathfrak{su}(k)_{\mu_2+\mu_3}}{\mathfrak{s}(\mathfrak{u}(\mu_2')_{\mu_2+\mu_3}\times \mathfrak{u}(k-\mu_2')_{\mu_2+\mu_3})} \\
\subset \frac{\mathfrak{su}(k)_{\mu_2} \times \mathfrak{su}(k)_{\mu_3}}{\mathfrak{s}(\mathfrak{u}(\mu_2')_{\mu_2+\mu_3}\times \mathfrak{u}(k-\mu_2')_{\mu_2+\mu_3})}\ .
\end{multline}
Translating parameters, this gives a gluing of
\be 
\text{Gr}(\nu_1,\nu_2,\nu_3) \times \text{Gr}(\nu_1',\nu_2',\nu_3')
\ee
where the parameters are constrained by
\begin{subequations}
\begin{align} 
\nu_1&=\nu_1'\ , \\
-\nu_1&=\nu_1+\nu_2+\nu_3+\nu_1'+\nu_2'+\nu_3'\ .
\end{align}
\end{subequations}
In the usual gluing we have considered before, the parameters were instead constrained by\footnote{With respect to \eqref{eq:basic gluing conditions}, we sent $\nu_i'\to -\nu_i'$, but because of $\mathds{Z}_2$ duality this produces isomorphic algebras.}
\begin{subequations}
\begin{align} 
\nu_1&=\nu_1'\ , \\
0&=\nu_1+\nu_2+\nu_3+\nu_1'+\nu_2'+\nu_3'\ .
\end{align}
\end{subequations}

\paragraph{Gluing conditions}
In general, a necessary condition for a gluing to be possible is that all the gluing fields have (half-)integer conformal dimensions. In the following we will analyze this condition in some detail and find a family of possible gluings that resembles very much the structure found in $\mathcal{W}_\infty$ gluings \cite{Prochazka:2017qum}.
Assume that we are gluing the two Grassmannians $\text{Gr}(\nu_1,\nu_2,\nu_3)$ and $\text{Gr}(\nu_1',\nu_2',\nu_3')$ along the first puncture. Gluing fields have conformal dimension
\be 
\Delta=\frac{\mathcal{C}_{k}(\boldsymbol{\Lambda})}{2\nu_1}+\frac{\mathcal{C}_{k'}(\boldsymbol{\Lambda}')}{2\nu_1'}+\text{integer}\ ,
\ee
where as usual $k=\nu_1+\nu_2+\nu_3$. We can consider the two cases $\boldsymbol{\Lambda}'=\bar{\boldsymbol{\Lambda}}$ and $\boldsymbol{\Lambda}'=\bar{\boldsymbol{\Lambda}}\tran$. Without loss of generality, we can restrict ourselves to the second case, since the first is obtained from the second by applying the $\mathds{Z}_2$ duality to $\text{Gr}(\nu_1',\nu_2',\nu_3')$. In the second case, the conformal dimension of the gluing field becomes
\be 
\Delta=\frac{\mathcal{C}_{k}(\boldsymbol{\Lambda})}{2\nu_1}+\frac{\mathcal{C}_{k'}(\boldsymbol{\Lambda}\tran)}{2\nu_1'}+\text{integer}\ ,
\ee
where we used that the quadratic Casimir is invariant under conjugation. In light of the last terms in the explicit formula for Casimir \eqref{eq:suinfty Casimir}, we need $\nu_1'=\nu_1$ for $\Delta$ to have a chance to be an integer for all $\boldsymbol{\Lambda}$. We hence obtain
\be 
\Delta=\frac{k+k'}{2\nu_1} \lVert \boldsymbol{\Lambda} \rVert+\text{integer}\ .
\ee
We did not include the term proportional to $\abs{\boldsymbol{\Lambda}}$, since $\abs{\boldsymbol{\Lambda}}=0$ in the tree-level gluings we consider. 
%This restriction can be lifted by considering Grassmannians with non-trivial external representations, which change the selection rules. In this case, this term can always be compensated by the inclusion of an additional $\mathfrak{u}(1)$ charge. One explicit such example is given in section~\ref{subsec:N2 Grassmannian}. 
Hence, for $\Delta$ to be half-integer, we need that\footnote{One might wonder whether also half-integer $p$ is allowed. For $p \in \mathds{Z}$, the gluing fields have actually always integer conformal dimension, because $\abs{\boldsymbol{\Lambda}}=0$ implies that $\lVert  \boldsymbol{\Lambda} \rVert$ is even. However that restriction can be lifted by gluing Grassmannians in a loop. We will an explicit such example in section~\ref{subsec:N2 Grassmannian}.
However, it could very well be that gluings for half-integer $p$ are also possible if one restricts to tree-level gluings.}
\be 
\frac{k+k'}{\nu_1}=p \in \mathds{Z}\ .
\ee
Thus, we found two necessary conditions on the parameters of the Grassmannians for gluing to be possible. One can similarly treat the other case that is related by $\nu_i' \to -\nu_i'$, see the discussion in section~\ref{sec:grassmannian}. In summary, the following conditions are necessary for a gluing to be possible:
\begin{align}
\nu_1&=\varepsilon \nu_1'\ , \\
p \nu_1&=\nu_1+\nu_2+\nu_3+\varepsilon(\nu_1'+\nu_2'+\nu_3')\ .
\end{align}
Here, we allowed for the inclusion of $\varepsilon=\pm 1$, which is obtained by applying $\mathds{Z}_2$ duality on the second algebra.
Before taking into account integer shifts, $p$ is the minimal conformal dimension that appears in this gluing. The integer shifts contained in the wedge characters shift this dimension up. The integer shift is at least $\frac{1}{2} \lVert \boldsymbol{\Lambda} \rVert$ and thus the gluing fields actually have dimension
\be 
\Delta=\frac{p+2}{2} \lVert \boldsymbol{\Lambda} \rVert+\text{integer}\ .
\ee 
By construction, the additional integer vanishes for instance in the adjoint representation. Thus, positivity of the gluing field requires $p \ge -1$.\footnote{In more complicated gluings where several Grassmannians are glued together, this bound changes and in general we can only take some gluings to have $p=-1$, but not all of them.} The two gluings above \eqref{eq:basicgluing}, \eqref{eq:mixedgluing} clearly correspond to the cases $p=0$ and $p=-1$. For the case $p =1$, we will see one explicit example in section~\ref{subsec:N2 Grassmannian}. In general, we do not have a direct construction for these algebras, but we believe that such gluings exist nonetheless. Presumably, the list of possible gluings is still incomplete and we will meet one example of a gluing in section~\ref{subsec:N4 algebra} that does not fit in this framework.

\paragraph{Vacuum characters}
It is interesting to determine the vacuum characters of the glued algebras, since they exhibit a very simple structure. We will consider only gluings without loops, so that we can without loss of generality always choose $\varepsilon=1$.

Let us start by looking at the simplest gluing
\be 
\begin{tikzpicture}[baseline={([yshift=-.5ex]current bounding box.center)}]
\draw[very thick] (0,0) to (120:1) node[left] {$\nu_1$};
\draw[very thick] (0,0) to (240:1) node[left] {$\nu_2$};
\draw[very thick] (0,0) to node[above] {$p$} (1,0);
\draw[very thick] (1,0) to (1.5,.86) node[right] {$\nu_3$};
\draw[very thick] (1,0) to (1.5,-.86) node[right] {$\nu_4$};
\end{tikzpicture}
\ee
Using the character, technology, we can easily compute the vacuum character  and we have done so up to order $\mathcal{O}(q^{\min(5p+11,21)})$. For the first few cases, we find up to this order results that are consistent with
\begin{subequations}
\begin{align}
p&=-1: &  \prod_{n=1}^\infty \left(\frac{(1-q^n)^2}{1-2q^n}\right)^2\sum_{\boldsymbol{\Lambda},\, \abs{\boldsymbol{\Lambda}}=0} q^{-\frac{1}{2}\lVert{\boldsymbol{\Lambda}}\rVert}\Phi_{\boldsymbol{\Lambda}}(q)^2&=\prod_{n=1}^\infty \frac{(1-q^n)^3}{1-4q^n}\ , \\
p&=0: &  \prod_{n=1}^\infty\left(\frac{(1-q^n)^2}{1-2q^n}\right)^2\sum_{\boldsymbol{\Lambda},\, \abs{\boldsymbol{\Lambda}}=0} \Phi_{\boldsymbol{\Lambda}}(q)^2&=\prod_{n=1}^\infty \frac{(1-q^n)^3}{1-3q^n}\ , \\
p&=1: &  \prod_{n=1}^\infty\left(\frac{(1-q^n)^2}{1-2q^n}\right)^2\sum_{\boldsymbol{\Lambda},\, \abs{\boldsymbol{\Lambda}}=0} q^{\frac{1}{2}\lVert{\boldsymbol{\Lambda}}\rVert}\Phi_{\boldsymbol{\Lambda}}(q)^2&=\prod_{n=1}^\infty \frac{(1-q^n)^3}{1-3q^n+q^{2n}}\ .
\end{align}
\end{subequations}
In fact, one can guess a formula that is valid for any $p$,
\be 
\prod_{n=1}^\infty\left(\frac{(1-q^n)^2}{1-2q^n}\right)^2\sum_{\boldsymbol{\Lambda},\, \abs{\boldsymbol{\Lambda}}=0} q^{\frac{p}{2}\lVert{\boldsymbol{\Lambda}}\rVert}\Phi_{\boldsymbol{\Lambda}}(q)^2=\prod_{n=1}^\infty \frac{(1-q^n)^4}{1-4q^n+4q^{2n}-q^{n(p+2)}}\ ,
\ee
which we have checked for $p=-1,\dots,10$. Notice that the $p=0$ case reproduces as expected the result of the generalized flag manifold representing the four-punctured sphere. We find it surprising that the general result takes such a simple form.

A similar formula even holds true when gluing two generalized flag manifolds in the same manner. Gluing a $m_1$-punctured sphere and a $m_2$-punctured sphere for arbitrary $p$ leads to the following vacuum character,
\begin{multline} 
\prod_{i=1,\,2}\prod_{n=1}^\infty\left(\frac{(1-q^n)^{m_i-1}}{1-(m_i-1)q^n}\right)\sum_{\boldsymbol{\Lambda},\, \abs{\boldsymbol{\Lambda}}=0} q^{\frac{p}{2}\lVert{\boldsymbol{\Lambda}}\rVert}\Phi_{\boldsymbol{\Lambda}}^{m_1}(q)\Phi_{\boldsymbol{\Lambda}}^{m_2}(q)\\
=\prod_{n=1}^\infty \frac{(1-q^n)^{m_1+m_2-2}}{(1-(m_1-1)q^n)(1-(m_2-1)q^n)-(m_1-2)(m_2-2)q^{n(p+2)}}\ .
\end{multline}
Again, nothing guaranteed a priori that these vacuum characters admit such a simple form. Another surprise is the fact that the $p=-1$ gluing leads to a vacuum character that coincides with the $(m_1-1)(m_2-1)+1$-punctured sphere (up to $\mathfrak{u}(1)$ factors). We do not have an explanation for this coincidence.

We now look at the next possible gluing involving three generalized flag manifolds representing punctured spheres with $m_1$, $m_2$ or $m_3$ punctures. We keep $m_i$ general, since it will be easier to spot the pattern. Let $p_{12}$ be the parameter of gluing connecting the $m_1$ and $m_2$-punctured sphere and $p_{23}$ the parameter of gluing connecting the $m_2$ and $m_3$-punctured sphere, i.e.
\be 
\begin{tikzpicture}[baseline={([yshift=-.5ex]current bounding box.center)}]
\draw[very thick] (-1,0) to (-1.5,.86);
\draw[very thick] (-1,0) to (-1.5,-.86);
\draw[very thick] (-1,0) to (-1.86,-.5);
\draw[very thick] (-1,0) to (-1.5,-.86);
\node at (-1.7,.2) {$m_1\ \vdots$};
\draw[very thick] (-1,0) to node[above] {$p_{12}$} (0,0);
\draw[very thick] (0,0) to (0,-1);
\draw[very thick] (0,0) to (1,0);
\node at (.8,-.5) {\reflectbox{$\ddots$} $m_2$};
\draw[very thick] (0,0) to node[right] {$p_{23}$} (.5,.86);
\draw[very thick] (0,0) to (.5,.86) to (1.5,.86);
\draw[very thick] (0,0) to (.5,.86) to (0,1.73);
\draw[very thick] (.5,.86) to (.5,1.86);
\node at (1.3,1.5) {$\ddots\ m_3$};
\end{tikzpicture}
\ee
The $m_1$, $m_2$ and $m_3$ labels make us remember that the vertices $m_1$-, $m_2$- and $m_3$-valent. We again calculated the corresponding vacuum character and in all cases found it to be compatible with the following answer:
\begin{align}
\prod_{n=1}^\infty&\prod_{i=1}^3 \frac{(1-q^n)^{m_i-1}}{1-(m_i-1)q^n}\sum_{\boldsymbol{\Lambda}_1,\, \boldsymbol{\Lambda}_2,\, \abs{\boldsymbol{\Lambda}_1}=\abs{\boldsymbol{\Lambda}_2}=0} q^{\frac{p_{12}}{2}\lVert\boldsymbol{\Lambda}_1\rVert+\frac{p_{23}}{2}\lVert\boldsymbol{\Lambda}_2\rVert}\Phi_{\boldsymbol{\Lambda}_1}^{m_1}(q)\Phi_{\boldsymbol{\Lambda}_2}^{m_3}(q)\Phi_{\boldsymbol{\Lambda}_1,\boldsymbol{\Lambda}_2}^{m_2}(q)\nonumber\\
&=\prod_{n=1}^\infty (1-q^n)^{m_1+m_2+m_3-3} \Big((1-(m_1-1) q^n)(1-(m_2-1) q^n)(1-(m_3-1) q^n)\nonumber\\
&\qquad-(m_1-2)(m_2-2)q^{n(p_{12}+2)}(1-(m_3-1)q^n)\nonumber\\
&\qquad-(m_2-2)(m_3-2)q^{n(p_{23}+2)}(1-(m_1-1)q^n)\nonumber\\
&\qquad-(m_1-2)(m_3-2)q^{n(p_{12}+p_{23}+2)}(1+(m_2-3)q^n)\Big)^{-1}\ .
\end{align}
Here we observe another surprise. Taking in this formula $m_1=m_2=m_3=3$ and $p_{12}=p_{23}=1$, we get the vacuum character of the four-punctured algebra! In other words, it seems that one can obtain the four punctured algebra either in the standard $p=0$ gluing of two Grassmannians or by gluing \emph{three} Grassmannians with the $p=1$ gluing. In this new realization, we actually get an algebra that depends on \emph{five} parameters. Thus this gives a strong hint that the algebras we are considering actually depend on more parameters than are visible through a coset description. We will come back to this point in section~\ref{sec:4parameters}.

It is interesting to push these formulas even further to three gluings, where the emergent structure will become clear. Here we have two possibilities -- a star-shaped quiver or a linear quiver. We parametrize them by
\begin{align} 
\begin{tikzpicture}[baseline={([yshift=-.5ex]current bounding box.center)}]
\draw[very thick] (-1,0) to (-1.5,.86);
\draw[very thick] (-1,0) to (-1.5,-.86);
\draw[very thick] (-1,0) to (-1.5,-.86);
\draw[very thick] (-1,0) to node[above] {$p_{14}$} (0,0);
\draw[very thick] (0,0) to node[right] {$p_{24}$} (.5,-.86);
\draw[very thick] (0,0) to node[right] {$p_{34}$} (.5,.86);
\draw[very thick] (0,0) to (.5,.86) to (1.5,.86);
\draw[very thick] (0,0) to (.5,.86) to (0,1.73);
\draw[very thick] (0,0) to (.5,-.86) to (1.5,-.86);
\draw[very thick] (0,0) to (.5,-.86) to (0,-1.73);
\end{tikzpicture}
\qquad\text{and}\qquad
\begin{tikzpicture}[baseline={([yshift=-.5ex]current bounding box.center)}]
\draw[very thick] (-1,0) to (-1.5,.86);
\draw[very thick] (-1,0) to (-1.5,-.86);
\draw[very thick] (-1,0) to (-1.5,-.86);
\draw[very thick] (-1,0) to node[above] {$p_{12}$} (0,0);
\draw[very thick] (0,0) to (.5,-.86);
\draw[very thick] (0,0) to node[right] {$p_{23}$} (.5,.86);
\draw[very thick] (0,0) to (.5,.86) to node[above] {$p_{34}$} (1.5,.86);
\draw[very thick] (0,0) to (.5,.86) to (0,1.73);
\draw[very thick] (1.5,.86)  to (2,0);
\draw[very thick] (1.5,.86)  to (2,1.73);
\end{tikzpicture}\ .
\end{align}
For simplicity, we drew the graphs as trivalent, but in the following formulae we keep $m_1$, $m_2$, $m_3$ and $m_4$ general.

For the star-shaped quiver, we obtain
\begin{align}
\prod_{n=1}^\infty&\prod_{i=1}^4 \frac{(1-q^n)^{m_i-1}}{1-(m_i-1)q^n}\sum_{\boldsymbol{\Lambda}_1,\, \boldsymbol{\Lambda}_2,\, \boldsymbol{\Lambda}_3,\, \abs{\boldsymbol{\Lambda}_1}=\abs{\boldsymbol{\Lambda}_2}=\abs{\boldsymbol{\Lambda}_3}=0} q^{\frac{p_{14}}{2}\lVert\boldsymbol{\Lambda}_1\rVert+\frac{p_{24}}{2}\lVert\boldsymbol{\Lambda}_2\rVert+\frac{p_{34}}{2}\lVert\boldsymbol{\Lambda}_3\rVert}\nonumber\\
&\qquad\qquad\qquad\qquad\qquad\qquad\qquad\qquad\times\Phi_{\boldsymbol{\Lambda}_1}^{m_1}(q)\Phi_{\boldsymbol{\Lambda}_2}^{m_2}(q)\Phi_{\boldsymbol{\Lambda}_3}^{m_3}(q)\Phi_{\boldsymbol{\Lambda}_1,\boldsymbol{\Lambda}_2,\boldsymbol{\Lambda}_3}^{m_4}(q)\nonumber\\
&= \prod_{n=1}^\infty  (1-q^n)^{m_1+m_2+m_3+m_4-4}\Bigg(\prod_{i=1}^4 (1-(m_i-1)q^n)\nonumber\\
&\qquad-\sum_{i=1}^3 (m_i-2)(m_4-2) q^{(p_i+2)n}\prod_{j=1,\, j \ne i}^3 (1-(m_j-1)q^n)\nonumber\\
&\qquad-\!\!\sum_{1 \le i <j \le 3}\!\! (m_i-2)(m_j-2)q^{(p_i+p_j+2)n}(1+(m_4-3)q^n)\!\!\!\!\prod_{k=1,\, k \ne i,\, j}^3 \!\!(1-(m_k-1)q^n)\nonumber\\
&\qquad-\prod_{i=1}^3 (m_i-2)q^{(p_1+p_2+p_3+3)n}(2+(m_4-4)q^n)\Bigg)\ .
\end{align}
For the linear quiver, we obtain
\begin{align}
\prod_{n=1}^\infty&\prod_{i=1}^4 \frac{(1-q^n)^{m_i-1}}{1-(m_i-1)q^n}\sum_{\boldsymbol{\Lambda}_1,\, \boldsymbol{\Lambda}_2,\, \boldsymbol{\Lambda}_3,\, \abs{\boldsymbol{\Lambda}_1}=\abs{\boldsymbol{\Lambda}_2}=\abs{\boldsymbol{\Lambda}_3}=0} q^{\frac{p_{12}}{2}\lVert\boldsymbol{\Lambda}_1\rVert+\frac{p_{23}}{2}\lVert\boldsymbol{\Lambda}_2\rVert+\frac{p_{34}}{2}\lVert\boldsymbol{\Lambda}_3\rVert}\nonumber\\
&\qquad\qquad\qquad\qquad\qquad\qquad\qquad\qquad\times\Phi_{\boldsymbol{\Lambda}_1}^{m_1}(q)\Phi_{\boldsymbol{\Lambda}_1,\boldsymbol{\Lambda}_2}^{m_2}(q)\Phi_{\boldsymbol{\Lambda}_2,\boldsymbol{\Lambda}_3}^{m_3}(q)\Phi_{\boldsymbol{\Lambda}_3}^{m_4}(q)\nonumber\\
&= \prod_{n=1}^\infty  (1-q^n)^{m_1+m_2+m_3+m_4-4}\Bigg(\prod_{i=1}^4 (1-(m_i-1)q^n)\nonumber\\
&\qquad-\sum_{i=1}^3 (m_i-2)(m_{i+1}-2) q^{(p_{i,i+1}+2)n}\prod_{j=1,\, j \not\in\{ i,i+1\}}^4 (1-(m_j-1)q^n)\nonumber\\
&\qquad-(m_1-2)(m_3-2)q^{(p_{12}+p_{23}+2)n}(1+(m_2-3)q^n)(1-(m_4-1)q^n)\nonumber\\
&\qquad-(m_2-2)(m_4-2)q^{(p_{23}+p_{34}+2)n}(1+(m_3-3)q^n)(1-(m_1-1)q^n)\nonumber\\
&\qquad-(m_1-2)(m_2-2)(m_3-2)(m_4-2)q^{(p_{12}+p_{34}+4)n} \nonumber\\
&\qquad-(m_1-2)(m_4-2)q^{(p_{12}+p_{23}+p_{34}+2)n} (1+(m_2-3)q^n)(1+(m_3-3)q^n)\Bigg)\ .
\end{align}
We see the following structure emerging. Let $\mathcal{E}$ be the set of internal edges and $\mathcal{V}$ the set of vertices of the graph we are considering. All the internal edges of the internal edges are labeled by integers $p_e$. The vertices are labeled by integers $m_v$, which give the valence of these vertices.
We define
\begin{align}
D(q)=\sum_{ E \subset \mathcal{E}}(-1)^{|E|}\prod_{e \in E} q^{p_e} \prod_{v \in \mathcal{V}} \Big(1-m_v^E-(m_v-m_v^E-1)q\Big)\ .
\end{align} 
Here, the sum runs over all possible subsets of internal edges, which hence define internal subgraphs. $m_v^E$ denotes the valences of the vertices in this subgraph. As an example, for the linear quiver, the possible internal subgraphs are
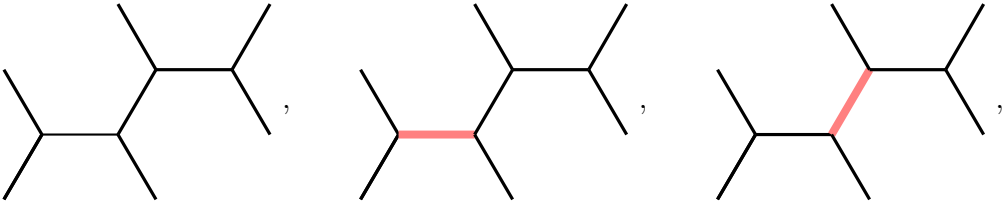
\begin{align}
&\begin{tikzpicture}[baseline={([yshift=-.5ex]current bounding box.center)}]
\draw[very thick] (-1,0) to (-1.5,.86);
\draw[very thick] (-1,0) to (-1.5,-.86);
\draw[very thick] (-1,0) to (-1.5,-.86);
\draw[thick] (-1,0) to (0,0);
\draw[very thick] (0,0) to (.5,-.86);
\draw[very thick] (0,0) to  (.5,.86);
\draw[very thick] (.5,.86) to  (0,1.73);
\draw[very thick] (.5,.86) to (1.5,.86);
\draw[very thick] (1.5,.86)  to (2,0);
\draw[very thick] (1.5,.86)  to (2,1.73);
\end{tikzpicture}\ , \qquad
\begin{tikzpicture}[baseline={([yshift=-.5ex]current bounding box.center)}]
\draw[very thick] (-1,0) to (-1.5,.86);
\draw[very thick] (-1,0) to (-1.5,-.86);
\draw[very thick] (-1,0) to (-1.5,-.86);
\draw[line width=3pt,white!50!red] (-1,0) to (0,0);
\draw[very thick] (0,0) to (.5,-.86);
\draw[very thick] (0,0) to  (.5,.86);
\draw[very thick] (.5,.86) to  (0,1.73);
\draw[very thick] (.5,.86) to (1.5,.86);
\draw[very thick] (1.5,.86)  to (2,0);
\draw[very thick] (1.5,.86)  to (2,1.73);
\end{tikzpicture}\ , \qquad
\begin{tikzpicture}[baseline={([yshift=-.5ex]current bounding box.center)}]
\draw[very thick] (-1,0) to (-1.5,.86);
\draw[very thick] (-1,0) to (-1.5,-.86);
\draw[very thick] (-1,0) to (-1.5,-.86);
\draw[very thick] (-1,0) to (0,0);
\draw[line width=3pt,white!50!red] (0,0) to (.5,.86);
\draw[very thick] (0,0) to  (.5,-.86);
\draw[very thick] (.5,.86) to  (0,1.73);
\draw[very thick] (.5,.86) to (1.5,.86);
\draw[very thick] (1.5,.86)  to (2,0);
\draw[very thick] (1.5,.86)  to (2,1.73);
\end{tikzpicture}\ , \nonumber\\
&\begin{tikzpicture}[baseline={([yshift=-.5ex]current bounding box.center)}]
\draw[very thick] (-1,0) to (-1.5,.86);
\draw[very thick] (-1,0) to (-1.5,-.86);
\draw[very thick] (-1,0) to (-1.5,-.86);
\draw[very thick] (-1,0) to (0,0);
\draw[very thick] (0,0) to (.5,-.86);
\draw[very thick] (0,0) to  (.5,.86);
\draw[very thick] (.5,.86) to  (0,1.73);
\draw[line width=3pt,white!50!red] (.5,.86) to (1.5,.86);
\draw[very thick] (1.5,.86)  to (2,0);
\draw[very thick] (1.5,.86)  to (2,1.73);
\end{tikzpicture}\ , \qquad
\begin{tikzpicture}[baseline={([yshift=-.5ex]current bounding box.center)}]
\draw[very thick] (-1,0) to (-1.5,.86);
\draw[very thick] (-1,0) to (-1.5,-.86);
\draw[very thick] (-1,0) to (-1.5,-.86);
\draw[line width=3pt,white!50!red] (-1,0) to (0,0);
\draw[very thick] (0,0) to (.5,-.86);
\draw[line width=3pt,white!50!red] (0,0) to  (.5,.86);
\draw[very thick] (.5,.86) to  (0,1.73);
\draw[very thick] (.5,.86) to (1.5,.86);
\draw[very thick] (1.5,.86)  to (2,0);
\draw[very thick] (1.5,.86)  to (2,1.73);
\end{tikzpicture}\ , \qquad
\begin{tikzpicture}[baseline={([yshift=-.5ex]current bounding box.center)}]
\draw[very thick] (-1,0) to (-1.5,.86);
\draw[very thick] (-1,0) to (-1.5,-.86);
\draw[very thick] (-1,0) to (-1.5,-.86);
\draw[very thick] (-1,0) to (0,0);
\draw[line width=3pt,white!50!red] (0,0) to (.5,.86);
\draw[very thick] (0,0) to  (.5,-.86);
\draw[very thick] (.5,.86) to  (0,1.73);
\draw[line width=3pt,white!50!red] (.5,.86) to (1.5,.86);
\draw[very thick] (1.5,.86)  to (2,0);
\draw[very thick] (1.5,.86)  to (2,1.73);
\end{tikzpicture}\ , \nonumber\\
&\begin{tikzpicture}[baseline={([yshift=-.5ex]current bounding box.center)}]
\draw[very thick] (-1,0) to (-1.5,.86);
\draw[very thick] (-1,0) to (-1.5,-.86);
\draw[very thick] (-1,0) to (-1.5,-.86);
\draw[line width=3pt,white!50!red] (-1,0) to (0,0);
\draw[very thick] (0,0) to (.5,-.86);
\draw[very thick] (0,0) to  (.5,.86);
\draw[very thick] (.5,.86) to  (0,1.73);
\draw[line width=3pt,white!50!red] (.5,.86) to (1.5,.86);
\draw[very thick] (1.5,.86)  to (2,0);
\draw[very thick] (1.5,.86)  to (2,1.73);
\end{tikzpicture}\ , \qquad\text{and}\qquad
\begin{tikzpicture}[baseline={([yshift=-.5ex]current bounding box.center)}]
\draw[very thick] (-1,0) to (-1.5,.86);
\draw[very thick] (-1,0) to (-1.5,-.86);
\draw[very thick] (-1,0) to (-1.5,-.86);
\draw[line width=3pt,white!50!red] (-1,0) to (0,0);
\draw[very thick] (0,0) to (.5,-.86);
\draw[line width=3pt,white!50!red] (0,0) to  (.5,.86);
\draw[very thick] (.5,.86) to  (0,1.73);
\draw[line width=3pt,white!50!red] (.5,.86) to (1.5,.86);
\draw[very thick] (1.5,.86)  to (2,0);
\draw[very thick] (1.5,.86)  to (2,1.73);
\end{tikzpicture}\ .
\end{align}
We observe that the various subgraphs correspond exactly to the various terms appearing in the denominator of the vacuum characters we have computed. We hence conjecture that the vacuum character is in general given by
\be
\label{eq:treegluingchar}
\prod_{n=1}^\infty \frac{(1-q^n)^{\sum_i(m_i-1)}}{D(q^n)}\ .
\ee
This simple formula seems to work for all tree graphs that we have tested.

Finally, we also briefly describe algebras whose gluing involves a loop. In this case, we cannot necessarily consistently choose $\varepsilon=1$ and each loop has an associated sign $\varepsilon$ to it.
Analogous formulae should exist for other gluings as well. We will see in section~\ref{subsec:N2 Grassmannian} an example where three Grassmannians are glued in a loop through a $p=1$ gluing, which again exhibits a very simple factorized vacuum character. 

\subsection{Comparison to \texorpdfstring{$\mathcal{W}_{1+\infty}$}{W1infty} gluing}
It would seem that the gluing procedure discussed here is different than the gluing studied in \cite{Prochazka:2017qum}, because here the basic gluing fields are in bi-adjoint representation of the Grassmannians that we are gluing while in \cite{Prochazka:2017qum} the basic gluing fields were generated by the bi-fundamental representations (fundamental with respect to both $\mathcal{W}_\infty$ algebras being glued). In the following we explain that this seeming difference only arises because of a different notation.

Consider the following gluing diagram that represents the decompositions of the numerator of \eqref{coset1} into the Grassmannian and the two affine Lie algebras in the denominator of \eqref{coset1}. We need to sum over all representations of the gluing fields which are labeled by $\boldsymbol{\Lambda}_\text{L}$ on the left and $\boldsymbol{\Lambda}_\text{R}$ on the right. If we were gluing only the left and middle sphere together, we would have $\boldsymbol{\Lambda}_\text{R} = \bullet$ and the Grassmannian selection rule would force us to have $|\boldsymbol{\Lambda}_\text{L}| = 0$, i.e. the representation $\boldsymbol{\Lambda}_\text{L}$ would have to be the same number of boxes and antiboxes, the simplest non-trivial representation being the adjoint representation.
\be 
\begin{tikzpicture}[baseline={([yshift=-.5ex]current bounding box.center)}]
\draw[thick] (0,0) circle (1.3);
\node[cross=3pt, thick] at (-.5,0) {};
\draw[thick] (4,0) circle (1.3);
\fill[black!50!white] (4,.8) circle (.1);
\fill[white] (1,.3) rectangle (3,-.3);
\draw[bend right=10, thick] (1,.2) to (3,.2);
\draw[bend left=10, thick] (1,-.2) to (3,-.2);
\draw[thick] (8,0) circle (1.3);
\node[cross=3pt, thick] at (8.5,0) {};
\fill[white] (5,.3) rectangle (7,-.3);
\draw[bend right=10, thick] (5,.2) to (7,.2);
\draw[bend left=10, thick] (5,-.2) to (7,-.2);
\fill[black!50!white] (1.1,0) circle (.1);
\fill[black!50!white] (2.9,0) circle (.1);
\fill[black!50!white] (5.1,0) circle (.1);
\fill[black!50!white] (6.9,0) circle (.1);
\end{tikzpicture}
\ee
But in the situation that we are considering we also have the right sphere so the selection rule (which only applies to Grassmannians and not to affine Lie algebras) is weaker, requiring only $|\boldsymbol{\Lambda}_\text{L}| + |\boldsymbol{\Lambda}_\text{R}| = 0$ so in particular from the point of view of left or right affine algebras we also sum over representations which do not have the same number of boxes as antiboxes. The simplest example would be the representation which is $\Box$ or its conjugate from the affine algebra point of view and which is the bifundamental representation of the Grassmannian (stretching between two punctures). These are exactly the representations of gluing fields that were considered in \cite{Prochazka:2017qum}.

\subsection[Characters of matrix \texorpdfstring{$\mathcal{W}_\infty$}{Winfty}]{Characters of matrix $\boldsymbol{\mathcal{W}_\infty}$}
As another application of the gluing procedure, we use it to compute the characters of $\mathcal{W}_\infty$ with the inclusion of matrix degrees of freedom. This also shows gluing at work in the presence of non-trivial external representations. The result turns out to be very simple. Representations of matrix $\mathcal{W}_{1+\infty}$ were already studied in \cite{Creutzig:2019qos}.

We recall that matrix $\mathcal{W}_{1+\infty}$ can be obtained from the Grassmannian by gluing an affine algebra. The resulting object has two punctures that can carry representation labels. Those representation labels have no selection rule, since the right number of boxes can always be absorbed by the gluing fields. For a non-trivial represention of matrix $\mathcal{W}_{1+\infty}$, \eqref{eq:Winf decomposition} reads
\be 
\big(\boldsymbol{\Lambda}_1,\boldsymbol{\Lambda}_2\big)= \bigoplus_{\boldsymbol{\Lambda},\, \abs{\boldsymbol{\Lambda}}+\abs{\boldsymbol{\Lambda}_1}+\abs{\boldsymbol{\Lambda}_2}=0} \big(\boldsymbol{\Lambda},\boldsymbol{\Lambda}_1,\boldsymbol{\Lambda}_2\big) \times \bar{\boldsymbol{\Lambda}}\tran\ ,
\ee
where the two labels on the LHS are the labels for the matrix $\mathcal{W}_{1+\infty}$ representation. Since the RHS only depends on the $\mathfrak{su}(\infty)$ tensor product $\boldsymbol{\Lambda}_1\otimes \boldsymbol{\Lambda}_2$, the same is true for the LHS and we can hence restrict our attention to only one non-zero label, say $\boldsymbol{\Lambda}_1$. Let in the following $m$ be the rank of the matrix.

We computed some of these characters explicitly (always up to $\mathcal{O}(q^{11})$). We find that they always split into a wedge character part times the vacuum character. 
The wedge characters take the form
\begin{subequations}
\begin{align}
\Phi_{\text{\tiny \ydiagram{1}},\, \bullet}(q)&=\frac{m}{1-q} \ ,\\
\Phi_{\text{\tiny \ydiagram{2}},\,\bullet}(q)&=\frac{m(m+1)+q m (m-1)}{2(1-q)(1-q^2)}\ , \\
\Phi_{\text{\tiny \ydiagram{1,1}},\,\bullet}(q)&=\frac{m(m-1)+q m (m+1)}{2(1-q)(1-q^2)} \ ,\\
\Phi_{\text{\tiny \ydiagram{3}},\,\bullet}(q)&=m\frac{(m+1)(m+2)+2(m-1)(m+1)(q+q^2)+(m-2) (m-1)  q^3}{6(1-q)(1-q^2)(1-q^3)} \ ,\\
\Phi_{\text{\tiny \ydiagram{2,1}},\,\bullet}(q)&=m\frac{2 (m-1) (m+1)+2 (m^2+2) q+2 (m-1) (m+1) q^2}{6(1-q)^2(1-q^3)} \ ,\\
\Phi_{\text{\tiny \ydiagram{1,1,1}},\,\bullet }(q)&=m\frac{(m-1)(m-2)+2(m-1)(m+1)(q+q^2)+(m+1) (m+2)  q^3}{6(1-q)(1-q^2)(1-q^3)} \ .
\end{align}
\end{subequations}
These results are compatible with the following structure. Defining $\Phi(q)=\frac{m}{1-q}$, the wedge character associated with the representation $\Phi_{\Lambda,\,\bullet}(q)$ can be obtained by the same formula as \eqref{eq:general wedge character}. We stress however that the symmetric group is embedded only in the box-part of the representation and is \emph{not} the diagonal symmetric group, as was the case in the Grassmannian. We have checked this up to five boxes in the wedge character. For non-trivial boxes and anti-boxes simultaneously, we find that the wedge characters simply factorize as follows,
\be 
\Phi_{\Lambda,\bar{\Lambda}}(q)=q^{\text{min}(\abs{\Lambda},\abs{\bar{\Lambda}})}\Phi_{\Lambda,\, \bullet}(q)\Phi_{\bar{\Lambda},\, \bullet}(q)\ .
\ee
This determines the character of any matrix $\mathcal{W}_{1+\infty}$ representation that is visible in the coset description.\footnote{There is reason to believe that representations outside of this class exist. In the case of $m=1$, the algebra is $\mathcal{W}_\infty$ and it enjoys also a triality symmetry \cite{Gaberdiel:2012ku}. Applying this triality to the coset representations leads to more representations. Even though this triality does not exist in the case of matrix $\mathcal{W}_{1+\infty}$, analogous representations might still exist. These conjectural `hidden' representations are analogous to the conjectural hidden representations for the Grassmannian that will be discussed in section~\ref{sec:4parameters}.}

\section{Variations and applications} \label{sec:variations}

\subsection{The orthosymplectic series}
We can analogously consider the orthogonal Grassmannian cosets
\begin{equation}
\frac{\mathfrak{so}(\mu_1+\mu_2)_k}{\mathfrak{so}(\mu_1)_k \times \mathfrak{so}(\mu_2)_k} \label{eq:orthogonal Grassmannian}
\end{equation}
and their level-rank duals
\begin{equation}
\frac{\mathfrak{so}(k)_{\mu_1} \times \mathfrak{so}(k)_{\mu_2}}{\mathfrak{so}(k)_{\mu_1+\mu_2}} \ .
\end{equation}
Analogously to the unitary case it is convenient to introduce a parametrization
\begin{equation}
k = \frac{1}{2}(4-\mu_1-\mu_2-\mu_3) \label{eq:orthogonal k}
\end{equation}
which brings the central charge to the form
\begin{equation}
c = -\frac{\mu_1 \mu_2 \mu_3 (\mu_1+\mu_2+\mu_3-2)(\mu_1+\mu_2+\mu_3-4)}{(\mu_1+\mu_2-\mu_3)(\mu_1-\mu_2+\mu_3)(-\mu_1+\mu_2+\mu_3)}\ ,
\end{equation}
symmetric under permutations of $\mu_j$. Introducing the parameters $\nu_j$ in the same way as in the unitary Grassmanian (\ref{eq:mutonu}) brings the central charge to the form
\begin{equation}
c = -\frac{(\nu_1+\nu_2)(\nu_1+\nu_3)(\nu_2+\nu_3)(\nu_1+\nu_2+\nu_3+1)(\nu_1+\nu_2+\nu_3+2)}{2\nu_1\nu_2\nu_3}\ .
\end{equation}
Unlike in the unitary case this central charge is not invariant under the simultaneous change of signs of $\mu_j$. Since $\mathfrak{so}(N)\leftrightarrow \mathfrak{sp}(-N)$, the coset gets instead mapped to the symplectic coset \cite{Cvitanovic:2008zz}
\begin{equation}
\frac{\mathfrak{sp}(\mu_1+\mu_2)_{k/2}}{\mathfrak{sp}(\mu_1)_{k/2} \times \mathfrak{sp}(\mu_2)_{k/2}}
\end{equation}
for $\mu_1$ and $\mu_2$ even or their level-rank dual cosets
\begin{equation}
\frac{\mathfrak{sp}(k)_{\mu_1/2}\times\mathfrak{sp}(k)_{\mu_2/2}}{\mathfrak{sp}(k)_{\mu_1/2+\mu_2/2}}
\end{equation}
for $k$ even. Parametrizing $k$ as
\begin{equation}
k = -\frac{1}{2}(4+\mu_1+\mu_2+\mu_3),
\end{equation}
the central charge of the symplectic Grassmannian can be written as
\begin{equation}
c = -\frac{\mu_1 \mu_2 \mu_3(\mu_1+\mu_2+\mu_3+2)(\mu_1+\mu_2+\mu_3+4)}{(\mu_1+\mu_2-\mu_3)(\mu_1-\mu_2+\mu_3)(-\mu_1+\mu_2+\mu_3)}
\end{equation}
which is the same as in the orthogonal coset if we simultaneously change signs of all $\mu_j$.

\paragraph{Representations}
One can develop a theory of representations as we did for the unitary coset. We work with the orthogonal realization of the algebra \eqref{eq:orthogonal Grassmannian} and hence $k$ is in the following given by \eqref{eq:orthogonal k}.
They are labeled by three representations of $\mathrm{SO}(\infty)$ that can be associated to three branes, or rather O-planes. We label $\mathrm{SO}(\infty)$ representation by a Young diagram $\Lambda$ with $\abs{\Lambda}$ boxes. There are no anti-boxes in this case.
The orthogonal coset has weaker selection rules; instead of box conservation, there is only a $\mathds{Z}_2$ selection rule on the number of boxes.
States in these representations are now characterized by \emph{non-orientable} string configurations that connect the three O-planes. Since the strings are non-orientable, it is clear that the number of endpoints only has to be an even integer and this corresponds to the selection rule mentioned above. Representations have conformal weight
\be 
h(\Lambda_1,\Lambda_2,\Lambda_3)=\sum_{i=1}^3 \frac{\mathcal{C}_k(\Lambda_i)}2({k+\mu_i-2)}+\text{integer}\ , 
\ee
where we recall that the $\mathfrak{so}(k)$ Casimir takes the form
\be 
\mathcal{C}_k(\Lambda)=(k-1)\abs{\Lambda}+\sum_i \text{row}_i(\Lambda)^2-\sum_i \text{column}_i(\Lambda)^2\ .
\ee
Thus, minimal representations have weight
\begin{subequations}
\begin{align}
h\big(\ydiagram{1},\ydiagram{1},\bullet\big)&=\frac{k-1}{2(\mu_1-2+k)}+\frac{k-1}{2(\mu_2-2+k)}\ , \\
h\big(\ydiagram{2},\bullet,\bullet\big)&=\frac{k}{\mu_1-2+k}+2\ , \\
h\big(\ydiagram{1,1},\bullet,\bullet\big)&=\frac{k-2}{\mu_1-2+k}+1\ .
\end{align}
\end{subequations}
We included the integer shifts in these formulas.

\paragraph{Vacuum character}
The vacuum character can be obtained from counting \emph{non-orientable} necklaces and some details of how to do this are spelled out in appendix~\ref{app:characters}. The result for the vacuum character reads
\be 
\text{ch}[\text{vac}](q)=\prod_{n=1}^\infty \frac{1-q^n}{(1-q^{2n})^{2^{n-2}}(1-2q^n)^{\frac{1}{2}}}\ .
\ee
The fact that this character produces integer degeneracies is not manifest, but one can easily convince oneself that this is true by expanding it to high order in $q$. The spin content of the non-composite primary fields agrees with $N_+(s)$ given in table \ref{tab:spin parity}.

\paragraph{Wedge characters}
The non-trivial representations can be dealt with as follows. From the combinatorial picture of counting non-orientable strings, it is again clear that the result for the characters factorize into the vacuum part and a wedge character. Let us compute the fundamental wedge character for one non-orientable string that connects one O-plane with itself. Under orientation reversal, the O-plane can be either even or odd (O$^+$ or O$^-$-plane). The two cases lead to different wedge characters. They can be again found via a recursion relation. Let us write
\be 
\Phi_{\text{\tiny \begin{ytableau}
\none[\pm] \\
\\
\end{ytableau}}}(q)=\sum_{n=1}^\infty c^\pm(n)q^n\ .
\ee
for the wedge character of a string starting and ending on the same O$^{\pm}$-plane. The reason why we introduce this notation will be explained below.
Clearly, we have $c^+(n)+c^-(n)=c(n)=2^{n-1}$, since the space of orientable strings can be decomposed into symmetric and antisymmetric sectors under the orientation-reversal. So let us focus on $c^+(n)$. It is convenient for the following to define also $c^+(0)=1$, even though it does not appear in the wedge character.

Consider now any numbers on the string with a fixed mode number sum $n$. Then there is either only one mode number or there are at least two. For this choice of parity of the primary state, a string with one number is odd under reversal and is projected out (remember that the mode operators have odd parity), hence $c^+(1)=0$. Thus, there have to be at least two mode numbers.
  We consider the left-most and right-most number and remove it. Call them $k$ and $\ell$. In case that they are equal, we obtain again a non-orientable string configuration with the same parity property. In case they are different, we have to sum the string configuration with the reversed configuration to obtain a symmetric non-orientable configuration. Thus, this possibility is counted by the known number of orientable strings. It follows that the coefficients satisfy the recursion relation\footnote{This is even true if terms like $c^+(0)$ appear in the sum by our convention $c^+(0)=1$.}
\be 
c^+(n)=\sum_{k=1}^{\frac{n}{2}} c^+(n-2k)+\sum_{k<\ell,\, k+\ell \le n} c(n-k-\ell), \quad\quad n \geq 1.
\ee
Here $c(n)$ are the degeneracies of the orientable string, see eq.~\eqref{eq:minimal wedge character} and we also put $c(0)=1$. The solution of the recursion relation is given by
\be 
c^\pm(n)=\begin{cases}
2^{n-2}\ , \qquad n\text{ even}\ , \\
2^{n-2}\mp 2^{\frac{n-3}{2}}\ , \qquad n\text{ odd}\ ,
\end{cases}
\ee
where we also gave the corresponding solution for $c^-(n)$.
The generating function is
\be 
\Phi_{\text{\tiny \begin{ytableau}
\none[\pm] \\
\\
\end{ytableau}}}(q)=\frac{1}{2}\left(\Phi_+(q)\pm \Phi_-(q)\right)=\frac{1}{2}\left(\frac{q}{1-2q}\mp \frac{q}{1-2q^2}\right)\ .
\ee
We denoted suggestively the two terms by $\Phi_+(q)$ (which is the original wedge character of the unitary model) and $\Phi_-(q)$.

To obtain the wedge characters of more complicated representations, we use the following algorithm. We first consider representations where all strings end on the same brane. Assume that there are $n$ strings ending on the brane, i.e.~there are $2n$ endpoints.
The boundary statistics of the non-orientable strings are captured by the Weyl group of $\mathrm{SO}(n)$, which is $\mathcal{S}_n \ltimes \mathds{Z}_2^n$.
%\footnote{For even $n$, the Weyl group would actually be slightly smaller because of the determinant condition. We are more precisely talking about the Weyl group of $\mathrm{O}(n)$.} 
The minimal representations $\Phi_{\text{\tiny \begin{ytableau}
\none[\pm] \\
\\
\end{ytableau}}}(q)$ are associated to the trivial and alternating representation of $\mathds{Z}_2$, respectively. For a configuration with $2n$ strings we proceed as follows. The $\mathrm{SO}(\infty)$ representation is naturally associated with a representation of $\mathcal{S}_{2n}$, which we branch down to $\mathcal{S}_n \ltimes \mathds{Z}_2^n$. This is the same process as in the orientable case, but the additional $\mathds{Z}_2$'s keep track of orientation reversals. For a single $\mathcal{S}_n \ltimes \mathds{Z}_2^n$ representation, the wedge character is obtained in the following manner. Conjugacy classes of this group are given by `generalized cycle types', which means that every cycle has additionally a sign attached, which is the product of all $\mathds{Z}_2$ elements contained in this cycle \cite[Chapter 4]{James:2009}. We say that an element $\sigma \in \mathcal{S}_n \ltimes \mathds{Z}_2^n$ has cycle type $(1,+)^{m_1^+}(1,-)^{m_1^-} (2,+)^{m_2^+}(2,-)^{m_2^-} (3,+)^{m_3^+}(3,-)^{m_3^-}\cdots$. For such a $\sigma$, we define
\be 
\Phi_\sigma(q)=\prod_{i \ge 1} \Phi^+(q^i)^{m_i^+}\Phi^-(q^i)^{m_i^-}\ .
\ee
The wedge character associated to an arbitrary representation of $\mathcal{S}_n \ltimes \mathds{Z}_2^n$ is thus given by (cf.~\eqref{eq:general wedge character})
\be 
\Phi_\lambda(q)=\frac{1}{2^n n!} \sum_{\sigma \in \mathcal{S}_n \ltimes \mathds{Z}_2^n} \chi_\lambda(\sigma) \Phi_\sigma(q)\ .
\ee
To finally obtain the most general wedge character, with different $\mathrm{SO}(\infty)$ representations associated to the different branes, one first takes the $\mathrm{SO}(\infty)$ tensor product. For each term in the tensor product decomposition, one applies the above steps. 

To exemplify this algorithm, we compute the wedge character of the representation $(\ydiagram{2},\ydiagram{1},\ydiagram{1})$. Here, we use Young diagrams for $\mathrm{SO}(\infty)$ and they specify the non-zero Dynkin labels. For instance
\be 
\ydiagram{2} \leftrightarrow [2,0,\dots]\ ,
\ee
as for $\mathrm{SU}(\infty)$. Taking the tensor product in $\mathrm{SO}(\infty)$ leads to
\be 
\ydiagram{2} \otimes \ydiagram{1} \otimes \ydiagram{1}=\bullet \oplus 2 \cdot \ydiagram{1,1} \oplus 3 \cdot \ydiagram{2} \oplus 	\ydiagram{4} \oplus \ydiagram{2,2} \oplus \ydiagram{2,1,1} \oplus 2 \cdot \ydiagram{3,1}\ .
\ee
To compute the wedge characters of the individual terms, we proceed as outlined above. The group $\mathcal{S}_2 \ltimes \mathds{Z}_2^2$ has five representations and conjugacy classes. They are conveniently denoted by Young diagrams with additional signs put on the columns, i.e.
\be 
{\begin{ytableau}
\none[+] &\none[+] \\
&
\end{ytableau}\ , \quad
\begin{ytableau}
\none[+] &\none[-] \\
&
\end{ytableau}\ , \quad
\begin{ytableau}
\none[-] &\none[-] \\
&
\end{ytableau}\ , \quad
\begin{ytableau}
\none[+]  \\
 \\
\\ 
\end{ytableau}\ , \quad
\begin{ytableau}
\none[-]  \\
\\
\\
\end{ytableau}\ .
}
\ee
The conjugacy classes have sizes 1, 2, 1, 2 and 2, respectively. The second representation is 2-dimensional and the other four are one-dimensional.
The have wedge characters
\begin{subequations}
\begin{align}
\Phi_{\text{\tiny \begin{ytableau}
\none[+] &\none[+] \\
&
\end{ytableau}}}(q)&=\frac{1}{8}\left(\Phi^+(q)^2+2\Phi^+(q)\Phi^-(q)+\Phi^-(q)^2+2\Phi^+(q^2)+2\Phi^-(q^2)\right) \ , \\
\Phi_{\text{\tiny \begin{ytableau}
\none[+]  \\
\\
\\
\end{ytableau}}}(q)&=\frac{1}{8}\left(\Phi^+(q)^2+2\Phi^+(q)\Phi^-(q)+\Phi^-(q)^2-2\Phi^+(q^2)-2\Phi^-(q^2)\right)\ , \\
\Phi_{\text{\tiny \begin{ytableau}
\none[-] &\none[-] \\
&
\end{ytableau}}}(q)&=\frac{1}{8}\left(\Phi^+(q)^2-2\Phi^+(q)\Phi^-(q)+\Phi^-(q)^2+2\Phi^+(q^2)-2\Phi^-(q^2)\right) \ , \\
\Phi_{\text{\tiny \begin{ytableau}
\none[-] \\
\\
\\
\end{ytableau}}}(q)&=\frac{1}{8}\left(\Phi^+(q)^2-2\Phi^+(q)\Phi^-(q)+\Phi^-(q)^2-2\Phi^+(q^2)+2\Phi^-(q^2)\right) \ , \\
\Phi_{\text{\tiny \begin{ytableau}
\none[+] &\none[-] \\
&
\end{ytableau}}}(q)&=\frac{1}{4}\left(\Phi^+(q)^2-\Phi^-(q)^2\right) \ .
\end{align}
\end{subequations}
The branching rules $\mathcal{S}_4 \mapsto \mathcal{S}_2 \ltimes \mathds{Z}_2^2$ are
\begin{subequations}
\begin{align}
\ydiagram{4} &\longmapsto {\begin{ytableau}
\none[+] &\none[+] \\
&
\end{ytableau}}\ , \\
\ydiagram{3,1} &\longmapsto {\begin{ytableau}
\none[+] &\none[-] \\
&
\end{ytableau}}\oplus  {\begin{ytableau}
\none[+] \\
\\
\\
\end{ytableau}}\ , \\
\ydiagram{2,2} &\longmapsto {\begin{ytableau}
\none[+] &\none[+] \\
&
\end{ytableau}}\oplus {\begin{ytableau}
\none[-] &\none[-] \\
&
\end{ytableau}}\ , \\
\ydiagram{2,1,1} &\longmapsto {\begin{ytableau}
\none[+] &\none[-] \\
&
\end{ytableau}}\oplus  {\begin{ytableau}
\none[-] \\
\\
\\
\end{ytableau}}\ , \\
\ydiagram{1,1,1,1} &\longmapsto {\begin{ytableau}
\none[-] &\none[-] \\
&
\end{ytableau}}\ .
\end{align}
\end{subequations}
Thus, after branching the wedge character becomes
\begin{align}
\Phi_{\text{\tiny \ydiagram{2}},\text{\tiny \ydiagram{1}},\text{\tiny \ydiagram{1}}}(q)&=1+3 \Phi_{\text{\tiny \begin{ytableau}
\none[+]  \\
\\
\end{ytableau}}}(q)+2 \Phi_{\text{\tiny \begin{ytableau}
\none[-]  \\
\\
\end{ytableau}}}(q)+2 \Phi_{\text{\tiny \begin{ytableau}
\none[+] & \none[+]  \\
 & \\
\end{ytableau}}}(q)+ \Phi_{\text{\tiny \begin{ytableau}
\none[-] & \none[-]  \\
 & \\
\end{ytableau}}}(q)\nonumber\\
&\qquad+3 \Phi_{\text{\tiny \begin{ytableau}
\none[+] & \none[-]  \\
 & \\
\end{ytableau}}}(q)+2 \Phi_{\text{\tiny \begin{ytableau}
\none[+] \\
 \\
 \\
\end{ytableau}}}(q)+\Phi_{\text{\tiny \begin{ytableau}
\none[-] \\
 \\
 \\
\end{ytableau}}}(q) \\
&=1+\frac{5}{2} \Phi^+(q)+\frac{1}{2}\Phi^-(q)+\frac{3}{2}\Phi^+(q)^2+\frac{1}{2} \Phi^+(q)\Phi^-(q)\\
&=\frac{(1-q)^3(1+q)}{(1-2 q)^2 \left(1-2 q^2\right)}\ .
\end{align}
We expect that one can develop a similar gluing theory for the orthogonal Grassmannian as we did in section~\ref{sec:gluing} for the unitary Grassmannian, although we have not tried to do so. Let us just mention that the analogue of the `top algebra', i.e.~the algebra one obtains by attaching three affine (orthogonal or symplectic) algebras to the orthogonal Grassmannian is 
\be 
\mathfrak{so}(\mu_1+\mu_2+\mu_3)_{-\frac{1}{2}(\mu_1+\mu_2+\mu_3-4)}\quad \text{or} \quad \mathfrak{sp}(\mu_1+\mu_2+\mu_3)_{-\frac{1}{2}(\mu_1+\mu_2+\mu_3+4)}\ .
\ee
The two possibilities are related by a sign change $\mu_i\to -\mu_i$.

\subsection{Lagrangian Grassmannian (orthounitary coset)}
We have now discussed three of the seven infinite families of symmetric space cosets. One might also wonder about the remaining four, which take the form
\be 
 \frac{\mathfrak{su}(\kappa)_n}{\mathfrak{so}(\kappa)_{2n}}\ , \qquad \frac{\mathfrak{so}(2n)_\kappa}{\mathfrak{u}(n)_\kappa}\ , \qquad \frac{\mathfrak{su}(\kappa)_n}{\mathfrak{sp}(\kappa)_{2n}}\ , \qquad\text{and}\qquad  \frac{\mathfrak{sp}(2n)_\kappa}{\mathfrak{u}(n)_\kappa}\ .
\ee
Of those, the first and second as well as the third and fourth are related by level-rank duality. This statement seems to be less-known in the literature, so let us review the argument. We need the fact that the embeddings
\begin{subequations}
\begin{align} 
\mathfrak{u}(n)_\kappa \times \mathfrak{su}(\kappa)_n &\subset \mathfrak{u}(n\kappa)_1 \subset \mathfrak{so}(2n\kappa)_1\ , \\
\mathfrak{so}(2n)_\kappa \times \mathfrak{so}(\kappa)_{2n} &\subset \mathfrak{so}(2n\kappa)_1
\end{align}
\end{subequations}
are conformal. We can then write
\be 
\frac{\mathfrak{so}(2n)_\kappa}{\mathfrak{u}(n)_\kappa} \cong \frac{\mathfrak{so}(2n\kappa)_1}{\mathfrak{u}(n)_\kappa \times \mathfrak{so}(\kappa)_{2n}} \cong \frac{\mathfrak{u}(n\kappa)_1}{\mathfrak{u}(n)_\kappa \times \mathfrak{so}(\kappa)_{2n}}\cong \frac{\mathfrak{su}(\kappa)_n}{\mathfrak{so}(\kappa)_{2n}}\ .
\ee
A similar argument shows equivalence for the symplectic version. Furthermore, the orthogonal and symplectic versions are related by a $\mathds{Z}_2$ duality that sends $(n,\kappa)\to (-n,-\kappa)$. Thus, there is really only one independent coset and we may restrict our attention to the first one. The central charge is easily computable and equals
\be 
c=\frac{(\kappa-1)(\kappa+2)(n-1)n}{(\kappa+n)(2n+\kappa-2)}\ .
\ee
There are no obvious symmetries in this formula and we do not expect any dualities. These cosets have one orthogonal and one unitary factor and in a brane picture correspond to having one D-brane and one O-plane. They can hence be thought of as being an interface between the unitary and the orthogonal Grassmannian.

The vacuum character can again be computed by taking the large-level limit, where the system reduces to a free boson in the symmetric representation of $\mathfrak{so}(\kappa)$ with a singlet-constraint, which is counted by non-orientable necklaces without any signs for the reflection. The result is
\be 
\text{ch}[\text{vac}](q)=\prod_{n=1}^\infty \frac{(1-q^n)^2}{(1-q^{2n})^{3 \cdot 2^{n-2}} (1-q^{2n-1})^{2^{n-1}}(1-2q^n)^{\frac{1}{2}}}\ .
\ee
This character deviates from the unitary Grassmannian vacuum character only at spin 6.

Representations are labeled by one $\mathfrak{su}(\kappa)$ representation $\boldsymbol{\Lambda}_1$ and one $\mathfrak{so}(\kappa)$ representation $\Lambda_2$, subject to a $\mathds{Z}_2$ selection rule $\abs{\boldsymbol{\Lambda}_1}+\abs{\Lambda_2} \equiv 0 \bmod 2$. There are hence several possible `minimal' representations, where the string either connects the O-plane with itself, and the O-plane can be either an $\text{O}^+$ or $\text{O}^-$ plane, or it can connect the D-brane with itself. Finally, it can also stretch from the O-plane to the D-brane.

\subsection[The \texorpdfstring{$\mathcal{N}=2$}{N=2}  supersymmetric Grassmannian]{The $\mathcal{N}=2$ supersymmetric Grassmannian}
\label{subsec:N2 Grassmannian}
In the literature, the $\mathcal{N}=2$ of the Grassmannian is probably the most widely studied version of the Grassmannian. Since the complex Grassmannian is a hermitian symmetric space, the $\mathcal{N}=1$ version of the coset has actually $\mathcal{N}=2$ supersymmetry \cite{Kazama:1988qp, Kazama:1988uz}. It takes the form
\be
\label{eq:n2grasscoset}
\frac{\mathfrak{u}(\rho_1+\rho_2)_k^{(1)}}{\mathfrak{u}(\rho_1)_k^{(1)} \times \mathfrak{u}(\rho_2)_k^{(1)}} \cong \frac{\mathfrak{u}(\rho_1+\rho_2)_{k-\rho_1-\rho_2} \times \textrm{$\rho_1 \rho_2$ complex free fermions}}{\mathfrak{u}(\rho_1)_{k-\rho_1} \times \mathfrak{u}(\rho_2)_{k-\rho_2}}\ .
\ee
Here, $\mathfrak{u}(N)^{(1)}_k$ denotes an $\mathcal{N}=1$ affine Lie algebra at (supersymmetric) level $k$ which is equivalent to \cite{Kazama:1988qp,Kazama:1988uz}
\begin{equation}
\mathfrak{u}(N)_{k-N} \times \textrm{$N^2$ real free fermions}.
\end{equation}
We introduce the parameter
\be 
k=\rho_1+\rho_2+\rho_3
\ee
and in this parametrization the central charge simply takes the form
\be 
c=\frac{3\rho_1\rho_2\rho_3}{\rho_1+\rho_2+\rho_3}\ . \label{eq:N2 central charge}
\ee

\paragraph{Triality and supersymmetric top algebra} This suggests the existence of a triality symmetry that permutes $\rho_1$, $\rho_2$ and $\rho_3$. This triality symmetry was already discussed in \cite{Kazama:1988qp, Blau:1995np, Naculich:1997ic, Ali:2002vd}.
 Moreover, we also expect a $\mathds{Z}_2$ duality symmetry that exchanges all three signs of $\rho_i$ simultaneously.

We can make the triality symmetry manifest by finding a supersymmetric analogue of the coset \eqref{eq:grassfromtop} as follows. First notice that
 \be
 \frac{\text{$k^2$ real free fermions} \times \mathfrak{u}(1)}{\mathfrak{u}(\rho_1+\rho_2)^{(1)}_k \times \mathfrak{u}(\rho_3)^{(1)}_k}
 \ee
is unitary and has vanishing central charge. It is hence trivial. The embedding of the denominator into the numerator works as follows. We can write
\begin{multline}
\text{$k^2$ real free fermions}  \times \mathfrak{u}(1) \supset \\
\supset \text{$((\rho_1+\rho_2)^2+\rho_3^2)$ real free fermions} \times \mathfrak{u}((\rho_1+\rho_2)\rho_3)_1 \times \mathfrak{u}(1)
\end{multline}
Next, we use the conformal embedding $\mathfrak{su}(m)_n \times \mathfrak{su}(n)_m \subset \mathfrak{su}(mn)_1$ to construct the denominator algebra. It follows that we can write
\be 
\mathfrak{u}(\rho_1+\rho_2)^{(1)}_k \cong  \frac{\text{$k^2$ real free fermions} \times \mathfrak{u}(1)}{\mathfrak{u}(\rho_3)^{(1)}_k}
\ee
Plugging this into \eqref{eq:n2grasscoset}, we finally conclude that we can write the $\mathcal{N}=2$ Grassmannian as
 \be 
 \text{Gr}_{\mathcal{N}=2}(\rho_1,\rho_2,\rho_3) \cong \frac{\text{$k^2$ real free fermions} \times \mathfrak{u}(1)}{\mathfrak{u}(\rho_1)_k^{(1)}\times \mathfrak{u}(\rho_2)_k^{(1)} \times \mathfrak{u}(\rho_3)_k^{(1)}}\ ,
 \ee
 which makes the triality manifest.
  
\paragraph{Decomposition in bosonic Grassmannians} We now argue that the $\mathcal{N}=2$ Grassmannian can be obtained by gluing cyclically \emph{three} bosonic Grassmannians together, together with an additional $\mathfrak{u}(1)$ factor.
 These three bosonic Grassmannians are permuted under the above-mentioned triality symmetry of the $\mathcal{N}=2$ coset. The existence of such a decomposition can be seen directly at the level of the coset. Writing $\supset$ for conformal embedding, we have
 \begin{align}
 \text{Gr}&_{\mathcal{N}=2}(\rho_1,\rho_2,\rho_3)\nonumber\\
  & \supset \frac{\mathfrak{u}(\rho_1+\rho_2)_{\rho_3} \times \mathfrak{u}(\rho_1\rho_2)_1}{\mathfrak{u}(\rho_1)_{\rho_2+\rho_3} \times \mathfrak{u}(\rho_2)_{\rho_1+\rho_3}}\\
 &\supset \frac{\mathfrak{u}(\rho_1+\rho_2)_{\rho_3}}{\mathfrak{u}(\rho_1)_{\rho_3} \times \mathfrak{u}(\rho_2)_{\rho_3}} \times \frac{\mathfrak{u}(\rho_1)_{\rho_3} \times \mathfrak{u}(\rho_2)_{\rho_3} \times \mathfrak{su}(\rho_1)_{\rho_2} \times \mathfrak{su}(\rho_2)_{\rho_1} \times \mathfrak{u}(1)}{\mathfrak{u}(\rho_1)_{\rho_2+\rho_3} \times \mathfrak{u}(\rho_2)_{\rho_1+\rho_3}} \\
& \supset \frac{\mathfrak{u}(\rho_1+\rho_2)_{\rho_3}}{\mathfrak{u}(\rho_1)_{\rho_3} \times \mathfrak{u}(\rho_2)_{\rho_3}} \times \frac{\mathfrak{su}(\rho_1)_{\rho_2} \times \mathfrak{su}(\rho_1)_{\rho_3}}{\mathfrak{su}(\rho_1)_{\rho_2+\rho_3}} \times \frac{\mathfrak{su}(\rho_2)_{\rho_1} \times \mathfrak{su}(\rho_2)_{\rho_3}}{\mathfrak{su}(\rho_2)_{\rho_1+\rho_3}} \times \mathfrak{u}(1)  \\
 &=\text{Gr}(-\rho_1-\rho_2-\rho_3,\rho_1+\rho_2,\rho_1+\rho_3)\times \text{Gr}(\rho_1+\rho_2,-\rho_1-\rho_2-\rho_3,\rho_1+\rho_3)\nonumber\\
 &\qquad\times \text{Gr}(\rho_1+\rho_3,\rho_2+\rho_3,-\rho_1-\rho_2-\rho_3)\times \mathfrak{u}(1)\ .
 \end{align}
 Here, the arguments of the bosonic Grassmannians refer to the $\nu$-parameters.
 If one tracks more carefully the decomposition of the algebra under the three Grassmannian subalgebras, one finds the decomposition
 \begin{align}
  \text{Gr}_{\mathcal{N}=2}(\rho_1,\rho_2,\rho_3)=\!\!\!\bigoplus_{\genfrac{}{}{0pt}{}{\boldsymbol{\Lambda}_1,\,\boldsymbol{\Lambda}_2,\, \boldsymbol{\Lambda}_3}{
  \abs{\boldsymbol{\Lambda}_1}=\abs{\boldsymbol{\Lambda}_2}=\abs{\boldsymbol{\Lambda}_3}}} \!\!\! (\bullet,\boldsymbol{\Lambda}_3,\bar{\boldsymbol{\Lambda}}_2\tran)\times (\bar{\boldsymbol{\Lambda}}_3\tran,\bullet,\boldsymbol{\Lambda}_1)\times (\boldsymbol{\Lambda}_2,\bar{\boldsymbol{\Lambda}}_1\tran,\bullet)\times \abs{\boldsymbol{\Lambda}_1}\ . \label{eq:decomposition N2 Grassmannian}
 \end{align}
 Here, the three factors label the Grassmannian representations as usual and the last factor captures the $\mathfrak{u}(1)$ charge. As a consistency check, we compute the conformal dimension of the additional gluing fields appearing in this decomposition,
 \begin{align} 
 h_0(\boldsymbol{\Lambda}_1,\boldsymbol{\Lambda}_2,\boldsymbol{\Lambda}_3)&=\frac{(\rho_1+\rho_2+\rho_3)\abs{\boldsymbol{\Lambda}_1}^2}{2\rho_1\rho_2\rho_3}+\sum_{i=1}^3 \frac{\mathcal{C}_{\rho_i}(\boldsymbol{\Lambda}_{i+1})}{2(\rho_i+\rho_{i+2})}+\frac{\mathcal{C}_{\rho_i}(\bar{\boldsymbol{\Lambda}}_{i+2}\tran)}{2(\rho_i+\rho_{i+1})}\\
 &= \frac{(\rho_1+\rho_2+\rho_3)\abs{\boldsymbol{\Lambda}_1}^2}{2\rho_1\rho_2\rho_3}+\sum_{i=1}^3 \frac{\mathcal{C}_{\rho_{i+2}}(\boldsymbol{\Lambda}_{i})+\mathcal{C}_{\rho_{i+1}}(\bar{\boldsymbol{\Lambda}}_{i}\tran)}{2(\rho_{i+1}+\rho_{i+2})} \\
 &= \frac{(\rho_1+\rho_2+\rho_3)\abs{\boldsymbol{\Lambda}_1}^2}{2\rho_1\rho_2\rho_3}+\sum_{i=1}^3 \left( \frac{\lVert \boldsymbol{\Lambda}_i \rVert}{2}-\frac{\abs{\boldsymbol{\Lambda}_i}}{2\rho_{i+1}\rho_{i+2}}\right)\\
 &=\frac{1}{2} \sum_{i=1}^3 \lVert \boldsymbol{\Lambda}_i \rVert\ ,
 \end{align}
 which is indeed half-integer as expected. Here and in the following, all indices are understood to be mod 3. We used that the level of the $\mathfrak{u}(1)$ current in the $\mathcal{N}=2$ superconformal algebra is $\frac{c}{3}$, which fixes the normalization of the first term. The lowest-lying gluing fields are the two dimension $\frac{3}{2}$ supercharges of the $\mathcal{N}=2$ algebra. They appear in the representations $\boldsymbol{\Lambda}_i=(\ydiagram{1},\bullet)$ and $\boldsymbol{\Lambda}_i=(\bullet,\ydiagram{1})$ for $i=1,\,2,\, 3$, respectively. In the language of section~\ref{subsec:more gluings}, this gluing corresponds to $\varepsilon=-1$ and $p=1$. 

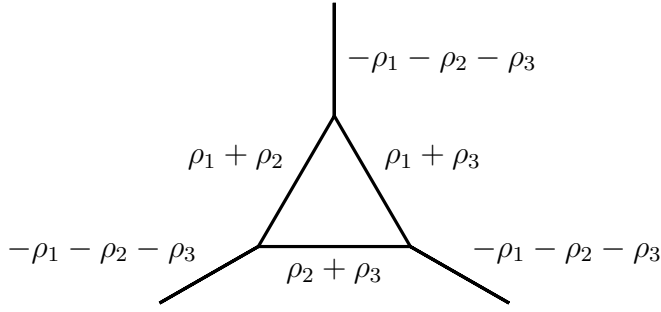
\begin{figure}
\centering
\begin{tikzpicture}[baseline={([yshift=-.5ex]current bounding box.center)}]
\draw[very thick] (0,0) -- node[below] {$\rho_2+\rho_3$} ++(0:2) -- node [above right] {$-\rho_1-\rho_2-\rho_3$} ++(330:1.5) -- ++(150:1.5) -- node[above right] {$\rho_1+\rho_3$} ++(120:2) -- node [right] {$-\rho_1-\rho_2-\rho_3$} ++(90:1.5) -- ++(270:1.5) -- node[above left] {$\rho_1+\rho_2$} ++(240:2) -- node [above left] {$-\rho_1-\rho_2-\rho_3$} ++(210:1.5) -- ++(30:1.5);
\end{tikzpicture}
\caption{Gluing diagram for $\mathcal{N}=2$ supersymmetric Grassmannian. The parameters are the values of $\nu_j$ parameters associated to the corresponding legs. The gluing parameter $p=1$ and $\varepsilon=1$ for all three edges.}
\label{fig:n2grassgluing}
\end{figure}

This decomposition is the generalization of the well-known fact that the $\mathcal{N}=2$ $\mathcal{W}_\infty$ algebra can be constructed out of two bosonic $\mathcal{W}_\infty$ algebras \cite{Gaberdiel:2017hcn, Prochazka:2017qum, Gaberdiel:2018nbs}. This statement is recovered by setting $\rho_3=1$, thus reducing $\text{Gr}_{\mathcal{N}=2}$ to $\mathcal{N}=2$ $\mathcal{W}_\infty$. The three bosonic Grassmannians also simplify. Since $\rho_3$ appears in two of these Grassmannians as a parameter, these two bosonic Grassmannians reduce to usual $\mathcal{W}_\infty$ algebras. The third Grassmannian can be seen to become trivial upon using its definition in terms of a coset
 \be 
 \text{Gr}(\rho_1+1,\rho_2+1,-\rho_1-\rho_2-1)=\frac{\mathfrak{su}(1)_{\rho_1} \times \mathfrak{su}(1)_{\rho_2}}{\mathfrak{su}(1)_{\rho_1+\rho_2}}\ .
 \ee
 
% For the $\mathcal{N}=2$ $\mathcal{W}_\infty$ algebra (that is the special case of the Grassmannian with $N=1$), it was established in a series of papers \cite{}
% Decomposition:
%\begin{equation}
%c=c_A+c_B+1
%\end{equation}
%where
%\begin{align}
%c_A & = -\frac{MN(M+N-2k)(M+N-k-1)(M+N-k+1)}{k(M-k)(N-k)} \\
%c_B & = \frac{(M+N-k)(M^3N+MN^3-M^2N^2-MN+k^2-MNk^2)}{k(M-k)(N-k)}
%\end{align}
%The bifundamental fermionic fields of spin $\frac{3}{2}$ have dimensions
%\begin{equation}
%(h_A,h_B) = -\frac{k}{2MN(M+N-k)} (c_A,c_B)
%\end{equation}
%The first factor can be identified with Grassmannian with parameters
%\begin{equation}
%\mu_1 = \frac{M}{2}, \quad\quad \mu_2 = \frac{N}{2}, \quad\quad \mu_3=\frac{M+N-2k}{2}
%\end{equation}
%
%This model has other virtues over the bosonic one. For one, it is manifestly unitary in all triality frames. This would make it very interesting to study protected quantities like the BPS spectrum and the elliptic genus. We comment more on this below.

\paragraph{Representations and characters} Representations are labeled by three Young diagrams that are associated to three branes. In terms of the bosonic decomposition, they correspond to the three empty labels in \eqref{eq:decomposition N2 Grassmannian}, i.e. to external legs in figure \ref{fig:n2grassgluing}.
Additionally, representations can be in the NS-sector or the R-sector. We will only consider NS-sector representations, since R-sector representations can be obtained by spectral flow of the $\mathcal{N}=2$ algebra.
The theory can be viewed as describing supersymmetric string configurations. Let us use the same methods as in the bosonic case to derive the vacuum character, refined by the $\mathfrak{u}(1)$ charge that is present in the model. We normalize the supercharges to have charge $\pm 1$.

Consider the large level limit, where the model becomes free. The coset \eqref{eq:n2grasscoset} reduces to $\rho_1\rho_2$ free complex bosons and fermions in the bifundamental representation of $\mathrm{U}(\rho_1) \times \mathrm{U}(\rho_2)$, subject to a singlet condition. Let us denote the bosons by $X$ (viewed as an $\rho_1 \times \rho_2$ matrix) and the fermions by $\psi$ (viewed as an $\rho_1 \times \rho_2$ matrix). $\mathrm{U}(\rho_1)\times \mathrm{U}(\rho_2)$ invariant combinations are given by a trace (or products thereof)
\be 
\tr(X_{-m_1} \bar{X}_{-m_2} \cdots X_{-m_{\ell-1}} \bar{X}_{-m_\ell})\ ,
\ee
where some of the $X$'s ($\bar{X}$'s) can be replaced by $\psi$'s ($\bar{\psi}$'s). Consider then $X_{-m_1}\bar{X}_{-m_2}$, $X_{-m_1}\bar{\psi}_{-m_2}$, $\psi_{-m_1}\bar{X}_{-m_2}$ and  $\psi_{-m_1}\bar{\psi}_{-m_2}$ as the fundamental letters of which the necklaces are build out. They have generating function \eqref{eq:polyacolorfun}
\be 
f(q,y)=\frac{q-q^{\frac{3}{2}}(y+y^{-1})+q^2}{(1-q)^2}\ .
\ee
We recall that the fermions are half-integer moded and the bosons are integer-moded. $y$ keeps track of the $\mathfrak{u}(1)$ charge and we counted fermions with a sign. Necklaces with these letters are counted by \eqref{eq:necklace counting} \cite{Flajolet}
\be 
\sum_{d=1}^\infty \frac{\phi(d)}{d} \log \left(\frac{1}{1-f(q^d,y^d)}\right)\ .
\ee
Finally, we obtain the full vacuum character by applying the plethystic exponential, leading to
\be 
\prod_{n=1}^\infty \frac{1}{1-f(q^n,y^n)}=\prod_{n=1}^\infty \frac{(1-q^n)^2}{1-3q^n+q^{\frac{3n}{2}}(y^n+y^{-n})}\ .
\ee
Thus, the formula has a similar structure to the bosonic version and to the the general gluing formula for tree gluings \eqref{eq:treegluingchar}. If we are interested in the vacuum character without $(-1)^\text{F}$ insertions, we simply replace $y \to -y$.

The character counting works in exactly the same way as in the bosonic coset. The minimal wedge character takes the form
\be 
\Phi_{\text{\tiny \ydiagram{1}}}(q)=\frac{(1-q)(-y q^{\frac{1}{2}}+q)}{1-3q+q^{3/2} \left(y+y^{-1}\right)}\ .
\ee
Curiously, the denominator of this wedge character is equal to the denominator of the vacuum character (for $n=1$), which was also true in the bosonic coset.

The decomposition of the supersymmetric Grassmannian in terms of bosonic Grassmannian \eqref{eq:decomposition N2 Grassmannian} leads to the following character identity
\begin{multline}
\prod_{n=1}^\infty \frac{(1-q^n)^2}{1-3q^n+(-1)^nq^{\frac{3n}{2}}(y^n+y^{-n})}\\
=\prod_{n=1}^\infty \frac{(1-q^n)^5}{(1-2q^n)^3}\sum_{\genfrac{}{}{0pt}{}{\boldsymbol{\Lambda}_1,\,\boldsymbol{\Lambda}_2,\, \boldsymbol{\Lambda}_3}{
  \abs{\boldsymbol{\Lambda}_1}=\abs{\boldsymbol{\Lambda}_2}=\abs{\boldsymbol{\Lambda}_3}}} \!\!\! q^{\frac{1}{2}\sum_j \lVert \boldsymbol{\Lambda}_j \rVert} y^{\abs{\boldsymbol{\Lambda}_1}} \prod_{i=1}^3\Phi_{\boldsymbol{\Lambda}_i,\bar{\boldsymbol{\Lambda}}_{i+1}\tran}(q)\ ,
\end{multline}
which we have checked up to $\mathcal{O}(q^{\frac{15}{2}})$.
\subsection[The \texorpdfstring{$\mathcal{N}=4$}{N=4} algebra]{The $\boldsymbol{\mathcal{N}=4}$ algebra}\label{subsec:N4 algebra}
Here, we discuss another example where the Grassmannian appears, namely in the large $\mathcal{N}=4$ superconformal algebra $\tilde{A}_\gamma$ \cite{Sevrin:1988ew}. This algebra has played an important role in holography on $\text{AdS}_3 \times \text{S}^3\times \text{S}^3 \times \text{S}^1$ \cite{deBoer:1999gea,Gukov:2004ym, Eberhardt:2017fsi,Eberhardt:2017pty, Eberhardt:2019niq} and we show here that it can be understood by a simple gluing of a Grassmannian (or rather a specific truncation) with two affine $\mathfrak{su}(2)$'s that yield the R-symmetry of the algebra. There are two varieties of this algebra in the literature, the so-called linear $\mathcal{N}=4$ algebra $A_\gamma$ and the non-linear $\mathcal{N}=4$ algebra $\tilde{A}_\gamma$. The former algebra has additionally four free fermions and a free boson and hence the latter can be obtained from the former by dividing out these free fermions and the free boson. By doing so, quadratic terms in the R-symmetry currents appear in the OPEs, hence the name. Since it has less fields, we will only discuss the non-linear version of the algebra. The algebra depends on two parameters $k^+$ and $k^-$, which appear as the levels of the two R-symmetry currents $\mathfrak{su}(2)_{k^+-1}$ and $\mathfrak{su}(2)_{k^--1}$. The parameter $\gamma$ that appears in the algebra is given by $\gamma=\frac{k^+}{k^++k^-}$. The central charge of the algebra equals
\be 
c=\frac{6k^+k^-}{k^++k^-}-3\ .
\ee
The algebra enjoys a duality symmetry that exchanges the two R-symmetry currents and sends $k^+ \leftrightarrow k^-$.

By definition, we can realize the algebra as a conformal extension of the coset algebra
\be 
\frac{\tilde{A}_\gamma}{\mathfrak{su}(2)_{k^+-1}\times \mathfrak{su}(2)_{k^--1}} \times \mathfrak{su}(2)_{k^+-1} \times \mathfrak{su}(2)_{k^--1}\ .
\ee
We will argue in the following that this coset coincides with the Grassmannian with parameters
\be 
\nu_1=\frac{1}{k^+}+1\ , \qquad \nu_2=\frac{1}{k^-}+1\ , \qquad \nu_3=-\frac{1}{k^+}-\frac{1}{k^-}\ .
\ee
It has central charge
\be 
c=\frac{3(k^+-1)(k^--1)(2k^+k^-+k^++k^-)}{(k^++1)(k^-+1)(k^++k^-)}\ ,
\ee
which indeed matches with the central charge of the Grassmannian with the above parameters. We can also make further consistency checks. The lowest-lying gluing fields are the supercharges of the $\tilde{A}_\gamma$ algebra. Since they are in the representation $(\mathbf{2},\mathbf{2})$ of the $\mathfrak{su}(2)$'s, they lead to fields of coset conformal dimension
\be 
h=\frac{3}{2}-\frac{3}{4(k^++1)}-\frac{3}{4(k^-+1)}=\frac{3(2k^+k^-+k^++k^-)}{4(k^++1)(k^-+1)}\ .
\ee
This coincides with the conformal dimension of the minimal representation of the Grassmannian \eqref{eq:conformal dimension minimal representation}. We can also determine the vacuum character of this coset, which reads explicitly
\be 
1+q^2+q^3+3q^4+3q^5+8q^6+9q^7+19q^8+25q^9+45q^{10}+\mathcal{O}(q^{11})\ .
\ee
It indicates the existence of one spin 4 field and two spin 6 fields. Hence the field content is too big to fit into $\mathcal{W}_\infty$ and we really have to turn to the Grassmannian to identify it. We also see that we are dealing with a level 3 truncation of the Grassmannian, since the spin 3 field is absent. Comparing with table~\ref{tab:low-lying truncations}, we see that the relevant truncation curve is
\be 
\nu_1+\nu_2+\nu_3-2=0\ ,
\ee
which provides another consistency condition. Since $k=\nu_1+\nu_2+\nu_3=2$ is a positive integer, we can give a simple coset realization of this particular Grassmannian,
\be 
\frac{\mathfrak{su}(2)_{\frac{1}{k^+}-1}\times \mathfrak{su}(2)_{\frac{1}{k^-}-1}}{\mathfrak{su}(2)_{\frac{1}{k^+}+\frac{1}{k^-}-2}}\ .
\ee
This gives us a way to compute the character of this particular truncation by directly using eq.~\eqref{eq:character projection formula}. We computed the characters up to $\mathcal{O}(q^{21})$, obtaining perfect agreement and thus strong support for the claim.

Next, we analyze the relevant gluing of the coset with the two $\mathfrak{su}(2)_{k^\pm-1}$'s. Because $k=2$, all the representations are just given in terms of $\mathfrak{su}(2)$ spins $\ell^\pm \in \frac{1}{2}\mathds{Z}_{\ge 0}$. We claim that the decomposition of the $\mathcal{N}=4$ algebra as the Grassmannian times the $\mathfrak{su}(2)$'s is given by the formula
\be 
\tilde{A}_\gamma=\bigoplus_{\genfrac{}{}{0pt}{}{\ell^+,\,\ell^-\in \frac{1}{2}\mathds{Z}_{\ge 0}}{\ell^++\ell^- \in \mathds{Z}}} (\ell^+,\ell^-,\bullet) \times \ell^+ \times \ell^-\ ,
\ee
where as usual the first factor specifies the representation of the Grassmannian and the second and third factor specify the two $\mathfrak{su}(2)$ representations.\footnote{Since $k=2$, the usual labels of boxes and anti-boxes degenerate to $\mathfrak{su}(2)$ spins. Similarly the selection rule of the Grassmannian degenerates to the condition that the sum of all spins should be an integer.} The gluing fields have dimension
\begin{align} 
h&=\frac{\ell^+(\ell^++1)}{\frac{1}{k^+}+1}+\frac{\ell^-(\ell^-+1)}{\frac{1}{k^-}+1}+\frac{\ell^+(\ell^++1)}{k^++1}+\frac{\ell^-(\ell^-+1)}{k^-+1}\\
&=\ell^+(\ell^++1)+\ell^-(\ell^-+1)\in \frac{1}{2} \mathds{Z}\ .
\end{align}
We have checked that this leads to the correct vacuum character of the $\mathcal{N}=4$ algebra up to $\mathcal{O}(q^{21})$.

We should mention that this gluing is of a different type than discussed in section~\ref{sec:gluing}. The gluing fields manage to be half-integer only because $k=2$.

\section{Operator product expansions}
\label{sec:OPEs}

In this section we will explicitly determine the operator product expansions of fields of low spin in the Grassmannian algebra. There are several reasons for doing this: first of all, having access to the structure constants lets us test the dualities of algebra which are not manifest in the coset description. We can also use the OPE coefficients to study the truncations of the universal algebra. Finally, the OPE bootstrap is also a way to see if there are any deformations of the algebra, i.e. whether a given algebra sits in a larger family of algebras.

Determination of the structure constants of the Grassmannian algebra is much more complicated than in the case of $\mathcal{W}_\infty$ \cite{Gaberdiel:2012ku}, especially due to large number of primary fields. In the table \ref{tab:spin parity} we can see the number of primary fields of spin $\geq 3$ present in the Grassmannian algebra. Starting from spin $4$, there are multiple primary generators of the same spin. For this reason, in order to use the Jacobi identities (associativity conditions) to fix the form of the operator product expansions, we need to fix the resulting large freedom of redefinition of fields. Unfortunately there is no clear canonical choice, so we have to make certain ad hoc choices which leads to complicated expressions for the structure constants.

There is another difference compared to $\mathcal{W}_\infty$: analogously to the situation there, there is a charge conjugation $\mathds{Z}_2$ automorphism flipping the sign of spin $3$ generator. In the Grassmannian case for higher spins (starting with spin $6$) there is no relation between the $\mathds{Z}_2$ conjugation parity of a field and its spin being even or odd. See table~\ref{tab:spin parity}.

We will label fields by their parity and their spin. If there is still degeneracy, the fields receive an additional label. Thus, the higher spin fields are given by $W_{3^-}$, $W_{4^+_1}$, $W_{4^+_2}$, $W_{5^+_1}$, $W_{5^-_1}$, $W_{6^-}$, \dots\ .

We start with the unique spin $3$ primary field $W_{3^-}$ (which has odd parity under the charge conjugation symmetry). We are free to rescale it arbitrarily and it is convenient to fix the normalization in such a way that the normalization respects the triality symmetry. The operator product expansion of $W_{3^-}$ with itself is of the form
\begin{equation}
W_{3^-} W_{3^-} \sim {C_{3^- 3^-}^0} \mathbbm{1} + {C_{3^- 3^-}^{4^+_1}} W_{4^+_1} + \mathit{C_{3^- 3^-}^{4^+_2}} W_{4^+_2} + \ldots
\end{equation}
We follow the convention that we do not explicitly list all the Virasoro descendants because they are uniquely fixed by the Virasoro subalgebra. No composite primary fields have dimension less than $6$ so these cannot appear in the singular part of the OPE. The dimension $5$ primaries are not allowed due to $\mathds{Z}_2$ parity. Any linear combination of two spin $4$ primaries can appear in this OPE but we fix the first of these, $W_{4^+_1}$, (up to rescaling) by requiring the coefficient $\mathit{C_{3^- 3^-}^{4^+_2}}$ to be zero. We use the italics to indicate OPE coefficients that we put to zero by fixing part of the field redefinition freedom.

At the next order we have the pair of OPEs
\begin{subequations}
\begin{align}
W_{3^-} W_{4^+_1} & \sim {C_{3^- 4^+_1}^{3^-}} W_{3^-} + {C_{3^- 4^+_1}^{5^-_1}} W_{5^-_1} + \mathit{C_{3^- 4^+_1}^{5^-_2}} W_{5^-_2} + {C_{3^- 4^+_1}^{6^-}} W_{6^-} + \ldots \\
W_{3^-} W_{4^+_2} & \sim \mathit{C_{3^- 4^+_2}^{3^-}} W_{3^-} + {C_{3^- 4^+_2}^{5^-_1}} W_{5^-_1} + {C_{3^- 4^+_2}^{5^-_2}} W_{5^-_2} + {C_{3^- 4^+_2}^{6^-}} W_{6^-} + \ldots
\end{align}
\end{subequations}
We can use these to first of all fix $W_{4^+_2}$ up to rescaling: requiring the orthogonality of $W_{4^+_1}$ and $W_{4^+_2}$ in the sense of two-point function, i.e. $C_{4^+_1 4^+_2}^0 = 0$, is equivalent to the requirement
\begin{equation}
C_{3^- 3^-}^{4^+_2} C_{4^+_2 4^+_2}^0 = C_{3^- 4^+_2}^{3^-} C_{3^- 3^-}^0.
\end{equation}
But we defined $W_{4^+_1}$ in such a way that $\mathit{C_{3^- 3^-}^{4^+_2}} = 0$ so the orthogonality of $W_{4^+_1}$ and $W_{4^+_2}$ can be imposed if we put $\mathit{C_{3^- 4^+_2}^{3_-}} = 0$. Now both spin $4$ primaries are fixed up to rescaling freedom. The redefinition freedom is given by $\mathrm{GL}(2)$ group and we reduced it to rescaling symmetry $\mathrm{GL}(1) \times \mathrm{GL}(1)$ by imposing two conditions. For the higher spins we will make analogous choices.

Returning to OPE of $3$ with spin $4$, we have three new fields in the singular part of the OPE: $W_{5^-_1}, W_{5^-_2}$ and $W_{6^-}$. We can fix $W_{5^-_1}$ up to normalization by requiring $\mathit{C_{3^- 4^+_1}^{5^-_2}} = 0$ and $W_{5^-_2}$ up to normalization by $\mathit{C_{3^- 5^-_2}^{4^+_1}} = 0$. The field $W_{6^-}$ is the only odd parity spin $6$ primary so it does not mix with other spin $6$ fields and therefore is uniquely determined up to rescaling freedom. The appearance of this field in OPE of spin $3$ and spin $4$ fields is a new feature not present in $\mathcal{W}_\infty$.

Turning next to OPEs with sum of spins $8$, we have the following:
\begin{subequations}
\label{opespins8}
\begin{align}
W_{3^-}W_{ 5^-_1} & \sim C_{3^- 5^-_1}^{4^+_1} W_{4^+_1} + C_{3^- 5^-_1}^{4^+_2} W_{4^+_2} + C_{3^- 5^-_1}^{6^+_1} W_{6^+_1} + \mathit{C_{3^- 5^-_1}^{6^+_2}} W_{6^+_2} + \ldots + \mathit{C_{3^- 5^-_1}^{6^+_5}} W_{6^+_5} \nonumber\\
& \qquad + C_{3^- 5^-_1}^{[3^- 3^-]} [W_{3^-} W_{3^-}] + C_{3^- 5^-_1}^{7^+} W_{7^+} + \ldots \\
W_{3^-}W_{ 5^-_2} & \sim \mathit{C_{3^- 5^-_2}^{4^+_1}} W_{4^+_1} + C_{3^- 5^-_2}^{4^+_2} W_{4^+_2} + C_{3^- 5^-_2}^{6^+_1} W_{6^+_1} + \ldots + \mathit{C_{3^- 5^-_2}^{6^+_5}} W_{6^+_5} \nonumber\\
& \qquad+ C_{3^- 5^-_2}^{[3^- 3^-]} [W_{3^-} W_{3^-}] + C_{3^- 5^-_2}^{7^+} W_{7^+} + \ldots \\
W_{4^+_1}W_{ 4^+_1} & \sim C_{4^+_1 4^+_1}^0 \mathbbm{1} + C_{4^+_1 4^+_1}^{4^+_1} W_{4^+_1} + C_{4^+_1 4^+_1}^{4^+_2} W_{4^+_2} + C_{4^+_1 4^+_1}^{6^+_1} W_{6^+_1} + \ldots \nonumber\\
& \qquad+ \mathit{C_{4^+_1 4^+_1}^{6^+_5}} W_{6^+_5} + C_{4^+_1 4^+_1}^{[3^- 3^-]} [W_{3^-} W_{3^-}] + C_{4^+_1 4^+_1}^{7^+} W_{7^+} + \ldots \\
\nonumber
W_{4^+_1} W_{4^+_2} & \sim C_{4^+_1 4^+_2}^0 \mathbbm{1} + C_{4^+_1 4^+_2}^{4^+_1} W_{4^+_1} + C_{4^+_1 4^+_2}^{4^+_2} W_{4^+_2} + C_{4^+_1 4^+_2}^{6^+_1} W_{6^+_1} + \ldots \\
&\qquad + \mathit{C_{4^+_1 4^+_2}^{6^+_5}} W_{6^+_5} + C_{4^+_1 4^+_2}^{[3^- 3^-]} [W_{3^-} W_{3^-}] + C_{4^+_1 4^+_2}^{7^+} W_{7^+} + \ldots \\
\nonumber
W_{4^+_2 }W_{4^+_2} & \sim C_{4^+_2 4^+_2}^0 \mathbbm{1} + C_{4^+_2 4^+_2}^{4^+_1} W_{4^+_1} + C_{4^+_2 4^+_2}^{4^+_2} W_{4^+_2} + C_{4^+_2 4^+_2}^{6^+_1} W_{6^+_1} + \ldots \\
&\qquad + \mathit{C_{4^+_2 4^+_2}^{6^+_5} W_{6^+_5}} + C_{4^+_2 4^+_2}^{[3^- 3^-]} [W_{3^-} W_{3^-}] + C_{4^+_2 4^+_2}^{7^+} W_{7^+} + \ldots
\end{align}
\end{subequations}
At this order the composite primaries like $[W_{3^-} W_{3^-}]$ appear in the singular part.

The field redefinitions are harder to fix at this order: the possibility of setting certain OPE coefficients to zero that we used so far assumes that there is a new primary that is linearly independent from the previous ones. Assuming that all five spin $6$ primaries appearing in (\ref{opespins8}) are linearly independent (and orthogonal to the composite field $[W_{3^-} W_{3^-}]$) leads to inconsistency in Jacobi identities. One can evaluate explicitly the OPEs using matrix $\mathcal{W}_{1+\infty}$ at $M=3$ and the following pattern of linear dependence emerges: we can choose the first spin $6$ field $W_{6^+_1}$ to be the one in $W_{3^- 5^-_1}$ OPE. The spin $6$ field in $W_{3^- 5^-_2}$ OPE is linearly independent so we can choose it to be a linear combination $W_{6^+_1}$ and $W_{6^+_2}$ and on top of that require the orthogonality of these two primaries. Next the OPE $W_{4^+_1} W_{4^+_1}$ is a linear combination of $W_{6^+_1}, W_{6^+_2}$ and $W_{6^+_3}$. So far everything followed the naive expectations, but it turns out that the spin $6$ field appearing in $W_{4^+_1} W_{4^+_2}$ OPE is not linearly independent from those defined so far so we cannot use it to define $W_{6^+_4}$. This field appears in $W_{4^+_2} W_{4^+_2}$ OPE together with $W_{6^+_1}, W_{6^+_2}, W_{6^+_3}$. The last spin $6$ field, $W_{6^+_5}$ does not appear at this order of OPE but we can define it (up to rescaling) by requiring its orthogonality to all spin $6$ fields that we discussed so far.

To summarize, we have $7$ primary spin $6$ fields in the algebra. One of them has negative parity so it doesn't mix in the OPE with the remaining fields. Another field is the composite primary $[W_{3^-} W_{3^-}]$ which is determined uniquely and there are $5$ even parity spin $6$ primaries which are orthogonal to it. To fix the redefinition freedom of these fields up to rescaling, we need to determine $6^2-6 = 30$ additional conditions. $25$ of these can be chosen to be
\begin{enumerate}
\item orthogonality to $[W_{3^-} W_{3^-}]$ $\leftrightarrow$ vanishing of $C_{3^- 6^+_1}^{3^-}$, $\ldots$, $C_{3^- 6^+_5}^{3^-}$ (5 conditions)
\item vanishing of $C_{3^- 5^-_1}^{6^+_2}$, $C_{3^- 5^-_1}^{6^+_3}$, $C_{3^- 5^-_1}^{6^+_4}$, $C_{3^- 5^-_1}^{6^+_5}$, $C_{3^- 5^-_2}^{6^+_3}$, $C_{3^- 5^-_2}^{6^+_4}$, $C_{3^- 5^-_2}^{6^+_5}$, $C_{4^+_1 4^+_1}^{6^+_4}$, $C_{4^+_1 4^+_1}^{6^+_5}$ and $C_{4^+_2 4^+_2}^{6^+_5}$ (10 conditions)
\item vanishing of $C_{3^- 6^+_2}^{5^-_1}$, $C_{3^- 6^+_3}^{5^-_1}$, $C_{3^- 6^+_4}^{5^-_1}$, $C_{3^- 6^+_5}^{5^-_1}$, $C_{3^- 6^+_3}^{5^-_2}$, $C_{3^- 6^+_4}^{5^-_2}$, $C_{3^- 6^+_5}^{5^-_2}$, $C_{4^+_1 6^+_4}^{4^+_1}$, $C_{4^+_1 6^+_5}^{4^+_1}$ and $C_{4^+_2 6^+_5}^{4^+_2}$ (10 conditions)
\end{enumerate}
These conditions determine spin $6$ fields uniquely up to rescaling ($5$ remaining degrees of freedom) but of course there is nothing canonical about this choice of fields.

At this point one can study what the Jacobi identities impose on the structure constants that have been independent so far. The fact that the Grassmannian depends on three parameters implies that all scaling invariant combinations of structure constants should be expressible in terms of the central charge and two additional independent constants. Surprisingly, up to the order we were able to solve the bootstrap equations, there seem to be four independent structure constants. This seems to indicate that there could possibly exist a four-parametric family of algebras with the same fields of low spin as in the Grassmannian family. The family of Lagrangian Grassmannian cosets has the same spin content at low spins. In section \ref{sec:4parameters} we will conjecture the existence of a four-parametric family of algebras which has specializations to both of these families.

We solved the bootstrap equations using \texttt{OPEdefs} by Kris Thielemans \cite{Thielemans:1991uw} up to fields of spin $7-8$ appearing in OPE of spins with sum up to $9-10$ (depending on the conjugation parity of the fields). In this way ca. $60$ structure constants were determined in terms of the following four structure constants:
\begin{equation}
c\ , \quad C_{4^+_1 4^+_1}^0\ , \quad C_{4^+_1 4^+_1}^{4^+_1}\ , \quad C_{4^+_1 4^+_2}^{4^+_2}\ .
\end{equation}
Since the expressions are extremely long, we list these only for lower spins, see appendix \ref{sec:appOPE}.

The solution of Jacobi identities does not come with any convenient parametrization of the structure constants in terms of level/rank parameters such as $\mu_j$ or $\nu_j$. One needs to find this parametrization by either studying families of simple truncations, by using coset representations or by using the minimal representations of the algebra. In our case we need to determine the values of three structure constants (including the central charge) in terms of these rank-like parameters (because we expect a three-parametric family of algebras). We performed a direct calculation using the coset representation \eqref{coset1}. This calculation was simplified by dividing it in two steps: first we divided by only one factor in the denominator which reduced the algebra to matrix $\mathcal{W}_{1+\infty}$. We studied studied this algebra in detail in \cite{Eberhardt:2019xmf} and in particular we found a closed-form expression for all OPEs in quadratic basis of the algebra. This significantly simplified the analysis of the coset \eqref{coset1}. As a second step we studied the OPEs of the commutant of the spin $1$ subsector in matrix $\mathcal{W}_{1+\infty}$.

The result is given in appendix \ref{sec:appOPE}. In the first section we list examples of structure constant up to spin $5$ in terms of four independent constants, $c$, $C_{4^+_1 4^+_1}^0$, $C_{4^+_1 4^+_1}^{4^+_1}$ and $C_{4^+_1 4^+_2}^{4^+_2}$ (as well as constants such as $C_{3^- 3^-}^0$ which can be chosen arbitrarily by rescaling the generators). The second section of the appendix lists expressions of the structure constants in terms of symmetric polynomials in five parameters $\nu_j$ where $\nu_1,\nu_2$ and $\nu_3$ are the parameters of the Grassmannian algebra and $\nu_4 = 1 = -\nu_5$. The reason for choosing this parametrization will be discussed in section \ref{sec:4parameters}.

\paragraph{Lagrangian Grassmannian family}
Since the field content of the Lagrangian Grassmannian family agrees with that of the Grassmannian family, we expect the bootstrap equations at lower spins to be the same. This is indeed the case so the formulas given in the appendix \ref{sec:appOPE} also determine the structure constants of the Lagrangian Grassmannian family. All we need to do is to correctly identify the parameters. The identification is given in \eqref{eq:apporthounitarymap} and with this identification, all the OPE structure constants that we determined directly from the coset agree with the bootstrap calculation (the comparison was done up to spin $5$).

\section{Truncations} \label{sec:truncations}
In this section, we combine known information about the truncations and guess the structure of all truncations for the algebra. This leads to the analogous larger cousins to the $Y_{M,N,L}$ algebras \cite{Gaiotto:2017euk}. Understanding these truncations is quite important, since they carry a large amount of information about the algebra. These truncations are also crucial if we want to consider the particular rational models corresponding to specializations of the universal algebra. In particular, the minimal models are typically found at the intersection of several such truncation curves. It turns out to be simplest if we describe the truncations in terms of the parameters $\nu_i$. There are various sources of known truncations, which we shall describe in detail. We will see that there is a very simple pattern that governs these truncations and we conjecture that this is the complete list.

\subsection{Explicit truncations}
\paragraph{From matrix $\mathcal{W}_{1+\infty}$} We can realize the Grassmannian as a coset of the matrix $\mathcal{W}_{1+\infty}$ algebra and the truncations of matrix $\mathcal{W}_{1+\infty}$ are expected to carry through to the coset. The level at which they first appear can however be different. This realization is valid as long as  the matrix rank ($=-\nu_2-\nu_3$) of matrix $\mathcal{W}_{1+\infty}$ is a positive integer. We shall take it to be very large such that this restriction does not matter. These truncations were (conjecturally) classified in \cite{Eberhardt:2019xmf}. 
The truncations take the form
\be 
\nu_1+(1-k_X)\nu_2+(1-k_X\tran)\nu_3 =N\in \mathds{Z}\ .
\ee
with $k_X$, $k_X\tran\in \mathds{Z}_{\ge 0}$. These truncations appear in matrix $\mathcal{W}_{1+\infty}$ at level $(k_X+1)(k_X\tran+1)(\abs{N}+1)$. Moreover, experimentally, one observes that the truncation always has a state transforming in the trivial representation of $\mathfrak{u}(\mu_1)$.\footnote{In \cite{Eberhardt:2019xmf} we focused for each null state on the highest spin representation of the global subalgebra in which these null states states transform. But there was also always a null state transforming as a singlet and this state is expected to survive in the quotient.} Thus, it is natural to assume that the truncation appears at the same level in the Grassmannian.
\paragraph{From the coset realization \eqref{coset2}} From the level-rank dual coset, we get another set of truncations. The coset truncates if one of the factors $\mathfrak{su}(k)_{\mu_1}$ or $\mathfrak{su}(k)_{\mu_2}$ in the numerator truncates. This realization of the Grassmannian is valid as long as $k=\nu_1+\nu_2+\nu_3 \in \mathds{Z}_{\ge 1}$. A large class of such truncations is given by affine algebras of admissible level \cite{Kac:1988qc}
\be 
\nu_i=k+\mu_i =\frac{p}{q}\ , \qquad p \ge k\ ,
\ee
where $i=1$ or $2$ and $p$ and $q$ are positive integers. This can be rewritten as
\be 
(1-q) \nu_1+\nu_2+\nu_3 =N=k-p\in \mathds{Z}_{\le 0}\ .
\ee

\paragraph{From $\mathcal{W}_\infty$} For $\mu_3=1$, the Grassmannian algebra truncates to $\mathcal{W}_\infty$, whose truncations are well-studied \cite{Prochazka:2014gqa,Gaiotto:2017euk,Prochazka:2018tlo,Creutzig:2020zaj}. Assuming that the $\mathcal{W}_\infty$ truncations are the intersection of the $\mu_3$ truncation curve with another truncation in the Grassmannian, this gives us further data. We hence have $\nu_1+\nu_2=-1$. The $\lambda$-parameters of $\mathcal{W}_\infty$ become
\begin{subequations}
\label{eq:winftonu}
\begin{align}
\lambda_1&=\frac{\nu_1+\nu_2+\nu_3}{\nu_1}\ , \\
\lambda_2&=\frac{\nu_1+\nu_2+\nu_3}{\nu_2}\ , \\
\lambda_3&=\nu_1+\nu_2+\nu_3\ .
\end{align}
\end{subequations}
Truncations of $\mathcal{W}_\infty$ occur whenever
\be 
\frac{N_1}{\lambda_1}+\frac{N_2}{\lambda_2}+\frac{N_3}{\lambda_3}=1\ ,
\ee
for three integers $N_1$, $N_2$ and $N_3 \in \mathds{Z}_{\ge 1}$. 
In terms of Grassmannian parameters, the truncation curve can be rewritten as\footnote{We used the relation $\nu_1+\nu_2+1=0$ in order to bring the result to a suggestive form.}
\be 
(1-N_1)\nu_1+(1-N_2)\nu_2+\nu_3=N_3 \in \mathds{Z}_{\ge 0}\ .
\ee
The level of the truncation in $\mathcal{W}_\infty$ and hence in the Grassmannian is $(N_1+1)(N_2+1)(N_3+1)$.\footnote{We should also mention that the truncations with shifted $(N_1,N_2,N_3) \to (N_1+a,N_2+a,N_3+a)$ for some integer $a$ define the same truncation curve, but the truncation appears at different levels \cite{Prochazka:2017qum}.}
\paragraph{Direct computation} We used the explicit knowledge of the OPEs at low levels to compute some low-lying truncation explicitly. They are listed in table~\ref{tab:low-lying truncations}.

\begin{table}
\begin{center}
\begin{tabular}{|c|c|c|c|}
\hline
level & curve & coset parameters \\
\hline
$2$ & $\nu_1+\nu_2+\nu_3 \pm 1$ & $k=1$  \\
$2$ & $\nu_1+\nu_2$ & $\mu_3=0$ \\
\hline
$3$ & $\nu_1+\nu_2+\nu_3 \pm 2$ & $k=2$ \\
$3$ & $\nu_1+\nu_2-\nu_3$ & $k+\mu_3=\frac{k}{2}$ \\
\hline
$4$ & $\nu_1$ & $N_1=N_2=1$ in $\mathcal{W}_\infty$  \\
$4$ & $\nu_1+\nu_2+\nu_3 \pm 3$ & $k=3$  \\
$4$ & $\nu_1+\nu_2-2\nu_3$ & $k+\mu_3=\frac{k}{3}$ \\
$4$ & $\nu_1+\nu_2 \pm 1$ & $\mu_3=1$ \\
\hline
$5$ & $\nu_1+\nu_2+\nu_3 \pm 4$ & $k=5$  \\
$5$ & $\nu_1+\nu_2-3\nu_3$ & $k+\mu_3=\frac{k}{4}$ \\
\hline
\end{tabular}
\end{center}
\caption{Low-lying truncation curves. These are directly extracted from the OPEs. It is understood that all the triality images of the truncation also appear. We added comments on the truncation curves which indicate how they can be understood from the different inputs we have given above.} \label{tab:low-lying truncations}
\end{table}

\subsection{All truncation curves}
Based on this data, together with triality symmetry and the symmetry $\nu_i \to -\nu_i$, we can guess the form of all truncation curves in the Grassmannian. We conjecture that they all take the form
\be
\label{eq:grasstruncfull1}
(1-N_1)\nu_1+(1-N_2)\nu_2+(1-N_3)\nu_3+ N\ ,
\ee
where $N_1$, $N_2$ and $N_3 \in \mathds{Z}_{\ge 0}$ and $N \in \mathds{Z}$. The first null-vector appears at level
\be 
\label{eq:grasstruncfull2}
(N_1+1)(N_2+1)(N_3+1)(\abs{N}+1)\ .
\ee
Of course, some truncation curves like $(N_1,N_2,N_3)=(1,1,1)$ should be discarded, since they can never be satisfied.\footnote{We should also add that we have no evidence that all three of the $N_i$'s can be non-zero at the same time. It might be that there are additional restrictions on the $N_i$'s that are not visible from our analysis.}

\subsection{Lagrangian Grassmannian cosets}
\label{sec:orthounitarytrunc}
We can also analyze the truncation curves of the Lagrangian Grassmannian family of cosets. Since we have calculated the OPE coefficients up to spin $5$, we can extract from their zeros and poles candidates for truncation curves. The list is given in the following table. Except for three exceptional cases which we will discuss and where we understand the origin of the discrepancy, there is a good evidence that the truncation curves can be parametrized by quintuples of non-negative integers $N_j$ such that a truncation of the algebra at level
\begin{equation}
(N_1+1)(N_2+1)(N_3+1)(N_4+1)(N_5+1)
\end{equation}
correspond to parameters $n$ and $\kappa$ satisfying
\begin{equation}
(N_1-2N_2+N_3)n + (N_1-N_2+N_3)\kappa + (2N_2-N_3+N_4-2N_5) = 0.
\end{equation}
We will give a conjectural explanation for this in section \ref{sec:4parameters}. The exceptions from this pattern are as follows:
\begin{enumerate}
\item The level $4$ truncation $(0,1,1,0,0)$ does not appear in the list because it corresponds to a same curve as level $2$ truncation $(0,1,0,0,0)$. Similarly, the level $5$ truncation $(0,0,4,0,0)$ corresponds to the same curve as level $3$ truncation $(0,2,0,0,0)$.
\item The extraction of level $3$ curves depends on normalization of spin $3$ field which cannot be fixed canonically. For this reason, with our ad hoc normalization we saw the truncation curve $n+\kappa$ corresponding to parameters $(0,1,0,0,1)$ already at level $3$ while the curve $2n+\kappa-2$ corresponding to $(0,0,2,0,0)$ only showed up at level $4$.
\item The curve $n+\kappa-1$ corresponding to $(0,1,0,0,0)$ or $(0,1,1,0,0)$ would be expected at level $2$ but is not visible at the level of OPE coefficients due to cancellations: at level $2$ it should appear as a zero of the central charge, but here it is canceled by $\nu_3$ in the denominator of the central charge formula.
\item The previous three discrepancies exactly correspond to one of the $\nu_j$ parameters vanishing (where parameters $\nu_j$ are introduced in section \ref{sec:4parameters}).
\end{enumerate}

\begin{table}[ht]
\centering
\begin{tabular}{|c|c|c|}
\hline
level & truncation curve & $N_j$ \\
\hline
$2$ & $n$ & $(1,0,0,0,0)$ \\
$2$ & $n-1$ & $(0,0,1,0,0)$ \\
$2$ & $\kappa-1$ & $(0,0,0,1,0)$ \\
$2$ & $\kappa+2$ & $(0,0,0,0,1)$ \\
\hline
$3$ & $2n+\kappa$ & $(2,0,0,0,0)$ \\
$3$ & $4n+3\kappa-4$ & $(0,2,0,0,0)$ \\
$3$ & $2n+\kappa-2$ & $(0,0,2,0,0)$ \\
$3$ & $\kappa-2$ & $(0,0,0,2,0)$ \\
$3$ & $\kappa+4$ & $(0,0,0,0,2)$ \\
\hline
$4$ & $3n+2\kappa$ & $(3,0,0,0,0)$ \\
$4$ & $3n+2\kappa-3$ & $(0,3,0,0,0), (0,0,3,0,0)$ \\
$4$ & $\kappa-3$ & $(0,0,0,3,0)$ \\
$4$ & $\kappa+6$ & $(0,0,0,0,3)$ \\
$4$ & $n+\kappa-2$ & $(1,1,0,0,0)$ \\
$4$ & $2n+\kappa-1$ & $(1,0,1,0,0)$ \\
$4$ & $n+1$ & $(1,0,0,1,0)$ \\
$4$ & $n-2$ & $(1,0,0,0,1)$  \\
$4$ & $2n+2\kappa-3$ & $(0,1,0,1,0)$ \\
$4$ & $n+\kappa$ & $(0,1,0,0,1)$ \\
$4$ & $\kappa$ & $(0,0,1,1,0)$ \\
$4$ & $n-3$ & $(0,0,1,0,1)$ \\
$4$ & $\kappa+1$ & $(0,0,0,1,1)$ \\
\hline
$5$ & $4n+3\kappa$ & $(4,0,0,0,0)$ \\
$5$ & $8n+5\kappa-8$ & $(0,4,0,0,0)$ \\
$5$ & $\kappa-4$ & $(0,0,0,4,0)$ \\
$5$ & $\kappa+8$ & $(0,0,0,0,4)$ \\
\hline
\end{tabular}
\caption{Truncation curves of Lagrangian Grassmannian family of algebras up to level $5$.}
\end{table}

\subsection{Orthosymplectic Grassmannians}

\paragraph{Reduction to even spin $\mathcal{W}_\infty$}
Similarly as in the unitary case, when we choose $\mu_1 = 1$ or $\mu_2 = 1$, the orthosymplectic Grassmannian coset reduces to a coset associated to even spin $\mathcal{W}_\infty$. The identification of parameters is very analogous to the one in the unitary case \eqref{eq:winftonu}:
\begin{subequations}
\begin{align}
\tilde{\mu}_1&=\frac{\nu_1+\nu_2+\nu_3+1}{\nu_1}\ , \\
\tilde{\mu}_2&=\frac{\nu_1+\nu_2+\nu_3+1}{\nu_2}\ , \\
\tilde{\mu}_3&=\nu_1+\nu_2+\nu_3+1\ .
\end{align}
\end{subequations}
with $\nu_1+\nu_2+1 = 0$. The parameters $\tilde{\mu}_j$ correspond to the parameters $\mu_j$ used in \cite{Prochazka:2019yrm}.

\paragraph{Truncation curves at levels $2$ and $4$}
We can list the truncation curves appearing at levels $2$ and $4$. We will not try to conjecture the general expression as in the unitary case, the purpose of this is to illustrate the richness of the truncations in orthosymplectic algebras which was in the even spin $\mathcal{W}_\infty$ case observed in \cite{Prochazka:2019yrm}.
\begin{table}[ht]
\centering
\begin{tabular}{|c|c|c|}
\hline
level & $\nu_1+\nu_2+\nu_3+\nu_4$ \\
\hline
$2$ & $-3$ \\
$2$ & $\nu_1-1$ \\
$2$ & $\nu_4-1$ \\
\hline
$4$ & $-1$ \\
$4$ & $\nu_1+1$ \\
$4$ & $\nu_4+1$ \\
$4$ & $2\nu_1-3$ \\
$4$ & $2\nu_4-3$ \\
$4$ & $\nu_1+\nu_2-1$ \\
$4$ & $\nu_1+\nu_4-1$ \\
$4$ & $3\nu_1-1$ \\
$4$ & $3\nu_4-1$ \\
\hline
\end{tabular}
\caption{Truncation curves of orthosymplectic family of algebras at levels $2$ and $4$. Out of all curves related by the triality symmetry permuting $\nu_1, \nu_2$ and $\nu_3$ we list only one. $\nu_4=-1$}
\end{table}
When listing the curves it was convenient to introduce an auxiliary parameter $\nu_4 = -1$. After doing that, the list of truncation curves at levels $2$ and $4$ is invariant under permutations of all four $\nu_j$ parameters. It would be interesting if this pattern survives also at higher levels. The table shows the richness of the orthosymplectic case, in the unitary situation we had only $4$ types of truncation curves at level $4$, now we have $9$ different types, but already at this level a pattern resembling the structures encountered in even spin $\mathcal{W}_\infty$ \cite{Prochazka:2019yrm} is emerging, with the parity of $N_j$ parameters used as labels of truncation curves (such as in \eqref{eq:grasstruncfull1}) playing a role.

\section{Four-parametric family of algebras}
\label{sec:4parameters}

Recall that our original goal was to find a generalization of $\mathcal{W}_\infty$ algebra that would serve as a building block from which as many VOAs as possible could be constructed by the gluing procedure. The unitary Grassmannian family of algebras is a three-parametric family of algebras with a group of duality symmetries $\mathcal{S}_3 \times \mathds{Z}_2$. There are however few problems with this family of algebras that give a hint that there should be a generalization of these as well:
\begin{enumerate}
\item The triality symmetry of $\mathcal{W}_\infty$ that appears when we impose the condition $\mu_3 = 1$ is not a subgroup of the triality symmetries of the Grassmannian. Only the duality symmetry $\nu_1 \leftrightarrow \nu_2$ is manifest.
\item $\mathcal{W}_\infty$ has three minimal representations (and their charge conjugates) and out of these only two of them are visible from the coset construction. To find the third representation, one has to use the OPE bootstrap or apply the $\mathcal{W}_\infty$ triality symmetry. Due to the fact that we constructed the Grassmannian algebras using cosets, we do not see the potential analogue of the third representation. This discussion suggests that there could possibly  exist other minimal representations of the algebra generalizing the third minimal representation of $\mathcal{W}_\infty$.
\item The truncation curves (\ref{eq:grasstruncfull1}) and (\ref{eq:grasstruncfull2}) exhaust all the truncations of the algebra that we are aware of, but the formulas (\ref{eq:grasstruncfull1}) and (\ref{eq:grasstruncfull2}) do not take as nice form as one knows from the discussion of $\mathcal{W}_\infty$ \cite{Prochazka:2014gqa} or its even spin extension \cite{Prochazka:2019yrm}. In particular the absolute value and the way the $\mathds{Z}_2$ symmetry acts is not as esthetically nice as one could expect.
\item The bootstrap equations for the algebra actually seem to have a four-parametric family of solutions (to the order we were able to reach). The Lagrangian Grassmannian series of coset algebras has the same spin content up to spin $6$ as the Grassmannian family but cannot be identified with it. If one believes that all possible branches of $\mathcal{W}$-algebras with the same spin content at lower spins should live in the same universal family of algebras, one is tempted to look for a unified four-parametric family of algebras of which the unitary Grassmannian family and the Lagrangian Grassmannian family are specializations. This is somewhat analogous to the situation encountered in \cite{Hornfeck:1992tm} where the OPE bootstrap approach finds different branches of algebras such as $\mathcal{W}(2,3,4,5)$, but these are unified when viewed as truncations of $\mathcal{W}_\infty$ \cite{Prochazka:2018tlo}.
\end{enumerate}

Surprisingly enough, all these small problems seem to have a simple resolution if we conjecture the existence of a four-parametric family of algebras. Our starting point is the formula (\ref{eq:grasscnu}) for the central charge. We could introduce another variable $\nu_4$ equal to $1$ and write the central charge as
\begin{multline}
c=-\frac{(\nu_1+\nu_2+\nu_3-1)(\nu_1+\nu_2+\nu_4-1)(\nu_1+\nu_3+\nu_4-1)(\nu_2+\nu_3+\nu_4-1)}{\nu_1\nu_2\nu_3\nu_4}\\
\times (\nu_1+\nu_2+\nu_3+\nu_4)\ .
\end{multline}
If we would allow $\nu_4$ to be arbitrary, the formula would now exhibit quadrality symmetry instead of triality. There is still a problem with this parametrization, as it is not compatible with $\mathds{Z}_2$ symmetry mapping $\nu_j$ parameters to their negatives. In fact $\mathcal{S}_3 \times \mathds{Z}_2$ cannot be a subgroup of $\mathcal{S}_4$. We can however make one more step, introducing $\nu_5$ such that we have now five parameters
\begin{equation}
\label{eq:grassid}
\nu_1, \quad \nu_2, \quad \nu_3, \quad \nu_4 = 1, \quad \nu_5 = -1
\end{equation}
The central charge can now be written as
\begin{equation}
c = \frac{(k-\nu_1)(k-\nu_2)(k-\nu_3)(k-\nu_4)(k-\nu_5)}{\nu_1\nu_2\nu_3\nu_4\nu_5}
\end{equation}
where
\begin{equation}
k = \nu_1 + \nu_2 + \nu_3 + \nu_4 + \nu_5.
\end{equation}
Although there are five $\nu_j$ parameters entering this equation, both numerator and denominator are homogeneous functions of $\nu_j$ of degree $5$ so the central charge actually only depends on a point in the complex projective space $\mathbb{CP}^4$, i.e. we have four continuous parameters as suggested by the OPE bootstrap. In fact, we can write the central charge in form analogous to the central charge of $\mathcal{W}_\infty$ \cite{Prochazka:2014gqa}:
\begin{equation}
\label{eq:bigcentralcharge}
c = (\lambda_1-1)(\lambda_2-1)(\lambda_3-1)(\lambda_4-1)(\lambda_5-1)
\end{equation}
if we introduce
\begin{equation}
\lambda_j = \frac{k}{\nu_j}
\end{equation}
These parameters are constrained by
\begin{equation}
\label{eq:biglambdaconstraint}
\frac{1}{\lambda_1} + \frac{1}{\lambda_2} + \frac{1}{\lambda_3} + \frac{1}{\lambda_4} + \frac{1}{\lambda_5} = 1.
\end{equation}
The unbroken symmetry is now the pentality symmetry $\mathcal{S}_5$ permuting all five $\nu_j$ parameters and also the $\mathds{Z}_2$ duality symmetry of the Grassmannian family is easy to realize: changing the sign of $\nu_1, \nu_2$ and $\nu_3$ accompanied with the exchange $\nu_4 \leftrightarrow \nu_5$ is projectively equivalent to the identity transformation.

\paragraph{Truncation curves}
Perhaps the strongest argument for this conjecture is the form of the truncation curves. We conjecture that there is a singular vector at level
\begin{equation}
(N_1+1)(N_2+1)(N_3+1)(N_4+1)(N_5+1)
\end{equation}
if the parameters $\nu_j$ satisfy the equation
\begin{equation}
\nu_1+\nu_2+\nu_3+\nu_4+\nu_5 = N_1 \nu_1 + N_2 \nu_2 + N_3 \nu_3 + N_4 \nu_4 + N_5 \nu_5.
\end{equation}
or
\begin{equation}
\label{eq:bigtruncations}
1 = \frac{N_1}{\lambda_1} + \frac{N_2}{\lambda_2} + \frac{N_3}{\lambda_3} + \frac{N_4}{\lambda_4} + \frac{N_5}{\lambda_5}
\end{equation}
Choosing here $N_4 = 0$ and $N_5 = N$ or $N_4 = -N$ and $N_5 = 0$ reproduces uniformly the curves found previously \eqref{eq:grasstruncfull1} and \eqref{eq:grasstruncfull2}. This form of the truncation curves is very similar to the one found in $\mathcal{W}_\infty$ \cite{Prochazka:2014gqa,Prochazka:2017qum}.

\paragraph{Lagrangian Grassmannian family}
Additionally, the Lagrangian Grassmannian series fits nicely in this picture. As mentioned in the context of truncation curves, we can identify the parameters as
\begin{align}
\label{eq:orthounitid}
\nu_1 = n+\kappa, \quad \nu_2 = -\kappa-2n+2, \quad \nu_3 =  n + \kappa - 1, \quad \nu_4 = 1, \quad \nu_5 = -2.
\end{align}
With this identification, the central charge as well as all the truncation curves discussed in section \ref{sec:orthounitarytrunc} and the structure constants discussed in section \ref{sec:OPEs} nicely fit into this picture.

\paragraph{OPEs for the four-parametric family}
The Grassmannian structure constants are symmetric functions of $\nu_1, \nu_2$ and $\nu_3$. We can assume that they can be written as symmetric homogeneous functions of degree zero in five parameters $\nu_1, \ldots, \nu_5$. For the structure constants we have at hand (up to spin $6$) this is indeed possible, in some cases uniquely and in some cases with a certain ambiguity. Requiring that the specialization of the parameters (\ref{eq:orthounitid}) reproduces the Lagrangian Grassmannian structure constants uniquely determines the structure constants of the algebra (to the order we were able to check). In particular, since the OPE bootstrap conjecturally determines all the structure constants of the algebra in terms of four independent parameters, it is enough to find a parametrization of four independent structure constants in terms of $\nu_1, \ldots, \nu_5$ in such a way that the specializations (\ref{eq:grassid}) and (\ref{eq:orthounitid}) reproduce the OPEs of the Grassmannian family and of the Lagrangian Grassmannian family. One can check that this is indeed the case. The expressions for the structure constants are given in appendix \ref{sec:appOPE}.

Unfortunately we do not have so far any alternative description of the conjectural algebra, so getting information even about simple properties of the algebra can be hard. It is not clear for example, what is the vacuum character of the algebra. Based on the truncation curves we can however conjecture that it differs from the Grassmannian character at level $8$. This is because the truncation curve associated to special values of $\nu_4$ and $\nu_5$ in the Grassmannian, $\nu_4 + \nu_5 = 0$, corresponds to for $N = (1,1,1,0,0)$ and the associated null state is therefore at level $8$. As a consistency check, the analogous truncation to Lagrangian Grassmannian algebra is at level $6$ as we can see from the null state parametrized by $N = (0,0,0,2,1)$. Level $6$ is indeed the level where the Lagrangian Grassmannian algebra has less states than the Grassmannian family.

\paragraph{Minimal representations}
Another check of this proposal is related to the fate of the missing representations. Remember that for the Grassmannian family the minimal representations are the bifundamental representations, each associated to a pair of $\nu_j$ parameters. The conformal dimension is given by (\ref{eq:conformal dimension minimal representation}). To conjecture generalization of this formula to the four-parameter algebra, we need to rewrite it in form which respects the symmetries and is a homogeneous function of $\nu_j$ of degree zero. Fortunately we can simply write
\begin{align}
\label{eq:bifundbig}
h_\text{bif} & = \frac{(\nu_1+\nu_2+\nu_3+\nu_4)(\nu_1+\nu_2+\nu_3+\nu_5)(\nu_1+\nu_2+\nu_4+\nu_5)}{2\nu_1\nu_2(\nu_1+\nu_2+\nu_3+\nu_4+\nu_5)} \\
& = \frac{\lambda_1\lambda_2(\lambda_3-1)(\lambda_4-1)(\lambda_5-1)}{2\lambda_3\lambda_4\lambda_5}.
\end{align}
This is a direct generalization of a formula for the minimal representation of $\mathcal{W}_\infty$ \cite{Prochazka:2015deb,Prochazka:2017qum} which is given by the same formula without the terms containing $\lambda_4$ and $\lambda_5$. The $\mathcal{S}_5$ symmetry is broken to $\mathcal{S}_2 \times \mathcal{S}_3$ because the bifundamental representation picks two out of five $\nu_j$ parameters (in the formula written these are $\nu_1$ and $\nu_2$). In total there should be 10 representations of this kind (as well as their charge conjugates). Specializing $\nu_4 \to 1$ and $\nu_5 \to -1$ gives us $3$ conformal dimensions of bifundamental representations that we started with, one additional orbit under $\mathcal{S}_3 \times \mathds{Z}_2$ of $6$ conformal dimensions of the form
\begin{equation}
\frac{(\nu_1+\nu_3)(\nu_2+\nu_3)(\nu_1+\nu_2+\nu_3+1)}{2\nu_3(\nu_1+\nu_2+\nu_3)}
\end{equation}
and finally a representation of conformal dimension
\begin{equation}
-\frac{(\nu_1+\nu_2)(\nu_1+\nu_3)(\nu_2+\nu_3)}{2(\nu_1+\nu_2+\nu_3)}
\end{equation}
(associated to $45$ directions) invariant under $\mathcal{S}_3 \times \mathds{Z}_2$ duality symmetries. It would be nice to see if these representations can be realized in the Grassmannian algebra, but we do not expect this to be possible using the coset description.

Let us verify that the newly conjectured minimal representations of the algebra include all three minimal representations known in $\mathcal{W}_\infty$. We restrict the parameters as in (\ref{eq:winftonu}). The representations (\ref{eq:bifundbig}) with $(\nu_1,\nu_3)$ and $(\nu_1,\nu_5)$ correspond to the minimal representation of $\mathcal{W}_\infty$ associated to $\lambda_1$. Analogously for the second minimal representation if we exchange $\nu_1 \leftrightarrow \nu_2$. The minimal representation of $\mathcal{W}_\infty$ associated to $\lambda_3$ which is not visible at the level of the coset corresponds to parameters $(\nu_4,\nu_3)$ or $(\nu_4,\nu_5)$ so indeed we reproduce all minimal representations of $\mathcal{W}_\infty$.

We can not only compare the conformal dimensions of the minimal primaries, but also higher spin charges. The $w_3$ charge (rescaled to be independent of the normalization) of the minimal representation is
\begin{equation}
\frac{w_3^2}{C_{33}^0} = -\frac{\nu_3(\nu_1+\nu_2)(\nu_1+\nu_2-\nu_3)(k-1)(k+1)(k-2)(k+2)}{6k^3\nu_1\nu_2(\nu_1+\nu_3)(\nu_2+\nu_3)(\nu_1-\nu_2+\nu_3)(-\nu_1+\nu_2+\nu_3)}.
\end{equation}
Introducing $\nu_4$ and $\nu_5$ we can again write it in form which is homogeneous and invariant under $\mathcal{S}_2 \times \mathcal{S}_3$\footnote{We use the notation $\mathcal{S}_2$ for the symmetry $\nu_1 \leftrightarrow \nu_2$, not to be confused with $\mathcal{Z}_2$ duality symmetry of Grassmannian which was acting as $\nu_4 \leftrightarrow \nu_5$ together with the change of sign of the remaining parameters.},
\begin{align}
\label{universalw3min}
\frac{w_{3,\text{bif}}^2}{C_{33}^0} & = \frac{\nu_3\nu_4\nu_5(k-\nu_3)(k-\nu_4)(k-\nu_5)(k-2\nu_3)(k-2\nu_4)(k-2\nu_5)}{6k^3\nu_1\nu_2(k-\nu_1)(k-\nu_2)(k-2\nu_1)(k-2\nu_2)} \\
& = \frac{\lambda_1^3\lambda_2^3(\lambda_3-1)(\lambda_4-1)(\lambda_5-1)(\lambda_3-2)(\lambda_4-2)(\lambda_5-2)}{6\lambda_3^3\lambda_4^3\lambda_5^3(\lambda_1-1)(\lambda_2-1)(\lambda_1-2)(\lambda_2-2)}.
\end{align}
The fact that we were able to write it in this form, symmetric in $\lambda_3, \lambda_4$ and $\lambda_5$ is not entirely obvious because a priori nothing guarantees that $\nu_3$ which is a genuine parameter of the Grassmannian algebra would appear in exactly the same way as the parameters $\nu_4$ and $\nu_5$ that we introduced on symmetry grounds. As a last example, let us have a look at the charge $w_{4^+_2}$ of the minimal representation. It can again be written in the universal form
\begin{align}
\nonumber
\label{universalw42min}
\frac{w_{4^+_2,\text{bif}}^2}{C_{4^+_2 4^+_2}^0} & = \frac{\nu_3\nu_4\nu_5}{4k\Delta_8\nu_1\nu_2} \frac{(k-\nu_3)(k-\nu_4)(k-\nu_5)}{(k-\nu_1)(k-\nu_2)} \frac{(k-\nu_3-\nu_4)(k-\nu_3-\nu_5)(k-\nu_4-\nu_5)}{(k-\nu_1-\nu_2)}  \\
\nonumber
& \times \frac{(k-3\nu_3)(k-3\nu_4)(k-3\nu_5)}{(k-3\nu_1)(k-3\nu_2)} \times (k-\nu_1-\nu_3)(k-\nu_1-\nu_4)(k-\nu_1-\nu_5)  \\
& \times (k-\nu_2-\nu_3)(k-\nu_2-\nu_4)(k-\nu_2-\nu_5)
\end{align}
where $\Delta_8$ is given in appendix \ref{sec:appOPE} and is a consequence of our choice of normalization of the charge. We again see the nice symmetry between the parameters $\nu_3, \nu_4$ and $\nu_5$.

The universal parametrization of the conformal dimension and $w_3$ and $w_{4^+_2}$ charges of the minimal primary also gives a correct result for the minimal bifundamental representation of the Lagrangian Grassmannian series. In fact, using \eqref{eq:orthounitid} in \eqref{eq:bifundbig}, \eqref{universalw3min} and \eqref{universalw42min} we obtain the correct parameters of the representation (that we verified directly from the operator product expansions):
\begin{equation}
h_\text{bif} = \frac{(n-1)(\kappa-1)(\kappa+2)}{2\kappa(n+\kappa)(2n+\kappa-2)}
\end{equation}
for the conformal dimension,
\begin{equation}
\frac{w_{3,\text{bif}}^2}{C_{3^- 3^-}^0} = \frac{(n-1)(\kappa+2)(\kappa+4)(\kappa-1)(\kappa-2)}{6n\kappa^3(n+\kappa)(2n+\kappa)(4n+3\kappa-4)}\ .
\end{equation}
for spin $3$ charge and finally
\begin{multline}
\frac{w_{4^+_2,\text{bif}}^2}{C_{4^+_2 4^+_2}^0} = -\frac{(n-3)(n-2)(n-1)(n+1)(\kappa-3)(\kappa-1)(\kappa+1)(\kappa+2)(\kappa+6)}{4\Delta_8 \kappa(n+\kappa-2)(2n+\kappa-2)(3n+2\kappa)} \times \\
\times (n+\kappa-1)(2n+\kappa-1)(2n+2\kappa-3)\ ,
\end{multline}
where $\Delta_8$ is given in appendix \ref{sec:appOPE}.

\paragraph{Specialization to $\mathcal{W}_\infty$}
It is interesting to see how the specialization of the four-parametric algebra to $\mathcal{W}_\infty$ looks like at the level of parameters. Comparing the expression for the central charge \eqref{eq:bigcentralcharge} to the central charge of $\mathcal{W}_\infty$ \cite{Prochazka:2014gqa}
\begin{equation}
c_{\mathcal{W}_\infty} = (\lambda_1^\prime-1)(\lambda_2^\prime-1)(\lambda_3^\prime-1)
\end{equation}
we see that these two agree if
\begin{equation}
\label{eq:winfred}
(\lambda_4-1)(\lambda_5-1) = 1 \quad \quad \leftrightarrow \quad \quad \frac{1}{\lambda_4} + \frac{1}{\lambda_5} = 1
\end{equation}
and if we identify $\lambda_j=\lambda_j^\prime, j=1,2,3$. In $\mathcal{W}_\infty$ the $\lambda^\prime$-parameters are restricted to satisfy
\begin{equation}
\label{eq:winflambdaconstraint}
\frac{1}{\lambda_1^\prime} + \frac{1}{\lambda_2^\prime} + \frac{1}{\lambda_3^\prime} = 0
\end{equation}
while in the four-parametric algebra we have instead \eqref{eq:biglambdaconstraint}. But it is immediate that these are equivalent if \eqref{eq:winfred} is satisfied. We can use this argument also in the opposite direction: assuming that $\lambda_1, \lambda_2$ and $\lambda_3$ are constrained by \eqref{eq:winflambdaconstraint}, \eqref{eq:biglambdaconstraint} implies \eqref{eq:winfred} and as a consequence of this both expressions for the central charge agree. The condition \eqref{eq:winflambdaconstraint} corresponds to the truncation curve parametrized by $N=(0,0,0,1,1)$. This means that the first singular vector corresponding to truncation is at level $4$ which is indeed the case because there is only one spin $4$ primary field in $\mathcal{W}_\infty$.

The same condition \eqref{eq:winfred} also guarantees that the conformal dimension of the minimal primary reduces to the $\mathcal{W}_\infty$ result. Take for instance the representation associated to parameters $(\nu_1,\nu_4)$. Its conformal dimension from \eqref{eq:bifundbig} is
\begin{equation}
\frac{\lambda_1(\lambda_2-1)(\lambda_3-1)}{2\lambda_2\lambda_3} \times \frac{\lambda_4(\lambda_5-1)}{\lambda_5}.
\end{equation}
The first term is the correct minimal dimension in $\mathcal{W}_\infty$ while the second term is equal to $1$ by virtue of \eqref{eq:winfred} (independently of the value of $\lambda_4$).

It might seem from this discussion that to restrict the four-parametric algebra to $\mathcal{W}_\infty$ we need to impose only one independent condition on the parameters which is in tension with the fact that $\mathcal{W}_\infty$ has only two independent parameters. But at the level of structure constants one can check that if \eqref{eq:winfred} is satisfied, the OPEs of fields that survive in $\mathcal{W}_\infty$ are independent of $\lambda_4$ (and $\lambda_5$) which is another constraint that the structure constants of the algebra satisfy. The mechanism of this is the same as in discussion of the minimal dimension.

\section{Discussion and Conclusions}
\label{sec:discussion}
In this paper, we have discussed in detail the VOA of the Grassmannian coset \eqref{coset1}. In the following, we outline interesting applications and possible avenues for future research.

\paragraph{Holography}
The Grassmannian or rather its $\mathcal{N}=2$ version discussed in section~\ref{sec:variations} is expected to have a holographic dual. In many ways, the Grassmannian should be considered to be the analogous theory to ABJM in two dimensions \cite{Aharony:2008ug}. Correspondingly, although its full holographic dual is currently unknown, there should be a version of the ABJ triality for the Grassmannian \cite{Chang:2012kt}. Similar ideas were already put forward in \cite{Gaberdiel:2013vva, Gaberdiel:2014cha}.

Let us recall that the central charge of the supersymmetric Grassmannian coset with $\text{U}(\rho_1)$ and $\text{U}(\rho_2)$ groups in the denominator is \eqref{eq:N2 central charge}
\be 
c=\frac{3\rho_1\rho_2\rho_3}{\rho_1+\rho_2+\rho_3}
\ee
%There are various limits of the model that are expected to lead to different physics. 
To obtain a holographic dual, we need to consider a large $N$ limit, in which the central charge becomes large. There are essentially two ways (up to permutation of parameters) to do so:\footnote{Of course various refined ways of scaling are possible.}
\begin{enumerate}
\item $\rho_2 \to \infty$ and $\rho_3 \to \infty$ with $\rho_1$ kept finite and $\lambda=\rho_2/(\rho_2+\rho_3)$ is kept finite.
\item All three parameters tend to infinity with their ratios kept fixed.
\end{enumerate}
In the first case, the central charge diverges vector-like,
whereas it diverges matrix-like in the second case. In the first parameter regime, one should hence expect that the holographic dual is described by a higher spin theory and this was discussed in \cite{Creutzig:2013tja, Candu:2013fta, Creutzig:2018pts}. The 't Hooft parameter $\lambda$  is identified with the coupling $\lambda$ of the higher spin theory. 
%Notice in particular that in the free limit $\lambda \to 0$, the Grassmannian degenerates to one of the free-field cases discussed in section~\ref{sec:grassmannian} (or rather its $\mathcal{N}=2$ analogue), namely free bosons and fermions in the bifundamental representation of $\mathrm{U}(M) \times \mathrm{U}(N)$, which is a two-dimensional vector-model for $M$ finite and $N \to \infty$.
  
In second parameter regime, the dual should be instead stringy. One can understand this transition as follows. In the free-field limit, the model reduces to $2\rho_1\rho_2$ complex bosons and fermions in the bifundamental representation of $\mathrm{U}(\rho_1) \times \mathrm{U}(\rho_2)$. Singlet operators are of the form
\be 
\tr \left(X_1 \bar{X}_1 \cdots X_n \bar{X}_n \right)\ ,
\ee
where $X_i$ are $\rho_1 \times \rho_2$ matrices and $\bar{X}_i$ are $\rho_2 \times \rho_1$ matrices (that can be either bosonic or fermionic). Let us assume that $\rho_1 \le \rho_2$. For finite $\rho_1$, the maximum length of such a trace is $2\rho_1$ operators, since $\tr \left(X_{1} \bar{X}_{1} \cdots X_{\rho_1+1} \bar{X}_{\rho_1+1} \right)$ could be written as product of shorter traces.
Identifying single-trace operators as usual with single-trace states, this implies that for finite $\rho_1$, the strings break up into smaller constituents that can be identified with the higher spin fields in the bulk. For small 't Hooft coupling, we can think of these constituents as being $X_i \bar{X}_j$, transforming in the adjoint representation of $\mathrm{U}(\rho_1)$ and weakly interacting in the bulk.

\paragraph{Stringy dual} For large parameters $\rho_1$, $\rho_2$, $\rho_3 \in \mathds{Z}_{\ge 0}$, the supersymmetric Grassmannian coset is unitary and is expected to have a good stringy dual. As we saw in the discussion of the characters, their Hilbert space has naturally a Fock space structure, with single-trace and multi-trace states. Moreover, in this regime, the vacuum character has a Hagedorn growth with Hagedorn temperature
\be 
T_\text{Hagedorn}=\frac{1}{\log(2)}\ .
\ee
We expect that this statement is not qualitatively changed when including also all the non-trivial characters in the partition function.

We currently do not have a good idea what the dual string worldsheet should look like and whether there is a region in moduli space, where it is described by a supergravity background. As a first step in this direction, one should analyze the BPS spectrum and the elliptic genus of the theory, along the lines of \cite{Belin:2020nmp}. Elliptic genera of minimal models and Kazama-Suzuki models take very simple forms \cite{Witten:1993jg, DiFrancesco:1993dg}. This is related to the fact that they have a simple Landau-Ginzburg description. It is unknown to us whether such a description also exists for the Grassmannian coset. It would provide a useful tool to study the BPS sector of the coset.

\paragraph{Four-parametric family of algebras}
Although we collected some evidence for the existence of the four-parametric family of algebras unifying the Grassmannian cosets with the orthosymplectic cosets, the four-parametric algebra itself (if it exists) has not been constructed. One possibility would be to use the bootstrap approach as in \cite{Linshaw:2017tvv,Kanade:2018qut}, but due to large number of primary fields in Grassmannian this approach seems to be much more complicated. The $\mathcal{W}_{1+\infty}$ can be also constructed from its free field representations and coproduct, but the coproduct structure is not manifest in the coset description and for this reason it is not clear at this moment if such a structure exists in the Grassmannian or in the hypothetical four-parametric family. The transition from matrix $\mathcal{W}_{1+\infty}$ to the Grassmannian by decoupling the spin $1$ sector is analogous to transition from $\mathcal{W}_{1+\infty}$ to $\mathcal{W}_\infty$. The basic Miura factor is an object naturally associated to $\mathcal{W}_{1+\infty}$ rather than to $\mathcal{W}_\infty$ so in analogy it is possible that the Miura transformation which is very useful feature of matrix $\mathcal{W}_{1+\infty}$ does not have a simple analogue in the Grassmannian algebra.

Parallel to the rational case of $\mathcal{W}_\infty$ there is an ongoing research in the $q$-deformed setting \cite{Feigin:2013fga,Mironov:2016yue,Awata:2016riz,Awata:2016mxc,Awata:2016bdm,Negut:2016dxr,Fukuda:2017qki,Awata:2018svb,Bourgine:2017jsi,Bourgine:2018uod}. Since the $q$-deformed theory is currently undoubtedly more developed, it could perhaps be easier to search for the conjectural four-parametric algebra in this setting first.

\paragraph{Vacuum character and Hall algebras}
It is interesting to notice that some of the vacuum characters have a nice combinatorial interpretation. The most famous example is the vacuum character of $\mathcal{W}_{1+\infty}$ counting the plane partitions. The same function also counts the types of Jordan forms of matrices (where by type we mean including the information whether two eigenvalues are the same or different). Similarly, the Grassmannian vacuum character
\begin{equation}
\prod_{n=1}^\infty \frac{1-q^n}{1-2q^n}
\end{equation}
(where we did not remove the $\mathfrak{u}(1)$ factor) is counting the conjugacy classes of invertible $n \times n$ matrices with values in $\mathds{F}_2$, the field of two elements and the formula generalizes to $\mathds{F}_p$ for any $p$ prime (and $(p+1)$-punctured algebra). One can associate to this counting problem a Hall algebra and it would be interesting to see if there is any connection between these algebraic structures and the structures that we were studying. Another analogous natural algebraic structure is the cohomological Hall algebra \cite{Kontsevich:2010px} which in the context of $\mathcal{W}_{1+\infty}$ and its truncations was discussed in \cite{Rapcak:2018nsl}.

\paragraph{Product cosets} There is another class of interesting cosets which we have not discussed, namely the cosets of the form
\begin{equation}
\frac{\mathfrak{su}(N_1 N_2)_{N_3}}{\mathfrak{su}(N_1)_{N_2 N_3} \times \mathfrak{su}(N_2)_{N_1 N_3}}\cong \frac{\mathfrak{su}(N_1N_2N_3)_1}{\mathfrak{su}(N_1)_{N_2 N_3} \times \mathfrak{su}(N_2)_{N_1 N_3}\times \mathfrak{su}(N_3)_{N_1 N_1}}\ . \label{eq:product coset}
\end{equation}
Their central charge is
\begin{equation}
\frac{(N_1-1)N_1(N_1+1)(N_2-1)N_2(N_2+1)(N_3-1)N_3(N_3+1)}{(N_1+N_2 N_3)(N_2+N_1 N_3)(N_3+N_1 N_2)}\ .
\end{equation}
The second description makes it manifest that also this class of cosets has a triality symmetry exchanging $(N_1,N_2,N_3)$.
The vacuum character of this class of algebras seems to be much larger than in those algebras that we considered. 
We computed their vacuum character up to $\mathcal{O}(q^8)$ using the technology outlined in appendix~\ref{subapp:alternative counting} and found that the strong generators are given by one spin-2 field, two spin-3 fields, 9 spin-4 fields, 26 spin-5 fields, 213 spin-6 fields and 1315 spin-7 fields. In eq.~\eqref{eq:product algebra vacuum character}, we give a closed form expression for the vacuum character of the gluing of this coset with two Grassmannians as studied in this paper. We find that it has a super Hagedorn growth which suggests that the same is also true for \eqref{eq:product coset}.
However, this coset has still only one energy-momentum tensor and thus cannot be obtained from gluing smaller algebras. Thus, it seems that $\mathcal{W}_{1+\infty}$ and the Grassmannian are at the beginning of a vast hierarchy of more and more complicated VOAs.\footnote{It is easy to see that one can in principle have VOAs without spin-1 fields, one energy-momentum tensor, but arbitrarily many higher spin fields. One example is given by the singlet sector of a free boson in the $m$-adjoint representation of the group $\mathrm{SU}(n_1) \times\mathrm{SU}(n_2) \times \cdots \mathrm{SU}(n_m)$. This VOA has one energy-momentum tensor, $2^{m-1}$ spin-3 fields and $\frac{1}{24}(9^m+3 \cdot 5^m+6 \cdot 3^m+6)$ spin-4 fields.}

\paragraph{General gluing, dualities}
The list of possible gluings that we studied is surely not exhaustive. In particular, we have not attempted to study the possible gluings of the four-parametric family of algebras. Even if we restrict our attention to Grassmannian gluings, the vacuum character \eqref{eq:treegluingchar} for special values of parameters shows interesting coincidences between algebras glued in different ways. It would be interesting if there is any unifying picture explaining these coincidences. We were also able to give a coset realization of only a particular subclass of these gluings, so it would be nice to find alternative ways of constructing these algebras.

\paragraph{Higher spin square}
In the $\text{AdS}_3/\text{CFT}_2$ holography there is an interesting algebra constructed by Gaberdiel and Gopakumar, the higher spin square \cite{Gaberdiel:2014cha, Gaberdiel:2015mra, Gaberdiel:2015wpo}. The bosonic toy model consists of a copy of $\mathcal{W}_{1+\infty}$ and even spin $\mathcal{W}_\infty$ intertwined together. It can be realized as symmetric product orbifold of the free boson theory. The field content of this algebra at low spins agrees with our Grassmannian family up to spin $6$ so one might hope that there could be a specialization of the Grassmannian algebra or the four-parametric family which reduces to the higher spin square. Unfortunately, we were not able to find such a truncation.

\paragraph{Integrability and Bethe ansatz}
The underlying integrable structure of $\mathcal{W}_{1+\infty}$ and the description of this algebra as a Yangian of $\widehat{\mathfrak{u}}(1)$ is a source of many recent developments. The matrix $\mathcal{W}_{1+\infty}$ is expected to be the Yangian of $\widehat{\mathfrak{u}}(M)$. It would be interesting to see if this has any generalization to the Grassmannian algebra or to the four-parametric family of algebras. On the level of Bethe ansatz equations \cite{Litvinov:2013zda,Alfimov:2014qua} the main difference between the usual Yangians of finite algebras used in spin chains and the Yangians of affine algebras is that the structure constant (scattering phase) is a ratio of cubic polynomials rather than of linear factors. The expressions for the central charge given in section \ref{sec:4parameters} seem to indicate that the possible generalization could be using the quintic polynomials instead. If such a naive approach was successful, it could pave the way to an integrable approach to the description of the algebras that we discussed.

\paragraph{Exceptional algebras}
The even spin $\mathcal{W}_\infty$ associated to orthosymplectic cosets or principal Drinfe\v{l}d-Sokolov reductions of orthosymplectic algebras is a subalgebra of $\mathcal{W}_{1+\infty}$ \cite{Prochazka:2019yrm}. The principal Drinfe\v{l}d-Sokolov reductions of the exceptional Lie algebras of $\mathrm{E}$- and $\mathrm{G}$-type are also subalgebras of truncations of $\mathcal{W}_{1+\infty}$ so in a sense $\mathcal{W}_{1+\infty}$ can be thought of as a unifying algebra for all principal DS reductions (except possibly for $\mathrm{F}_4$). It would be interesting to see if such a unifying picture also applies to the cosets considered here, namely if the orthogonal Grassmannians can be embedded into unitary ones or in the four-parametric family of algebras. In this sense the unifying algebra could be a generalization of the Vogel plane. It is amusing to notice that our parameters $\nu_4$ and $\nu_5$ have a similar role as the $\alpha$ and $\beta$ parameters of \cite{mkrtchyan2012casimir}.

\paragraph{Relation to 4d $\mathcal{N}=2$ chiral algebras}
In \cite{Beem:2013sza}, a program was started that relates the Schur operators of 4d $\mathcal{N}=2$ theories and their correlation functions to vertex operator algebras. While the relevant vertex operator algebras are non-unitary, they often have special null-vectors. It is interesting to note that our gluing constructions reproduce many of these `special' vertex operator algebras. For instance, the `top' algebra that we considered in section~\ref{sec:gluing} can be obtained from gluing three affine algebras to the Grassmannian. The resulting algebra is simply $\mathfrak{u}(N)_{-N/2}$. At this level, the affine algebra has a null-vector at level 2 and as a consequence saturates the unitarity bound for vertex operator algebras coming from 4d $\mathcal{N}=2$ theories \cite{Beem:2013sza}. The associated Higgs branch in the $\mathcal{N}=2$ theory has hence a quadratic relation \cite{Beem:2017ooy}. This is also true for the orthogonal and symplectic versions of the top algebra.
As such, the top algebra shows up in a number of physically interesting theories \cite{Chacaltana:2010ks}, as discussed in \cite[section 4.3]{Beem:2013sza}. 
We find it probable that by following the gluing construction of section~\ref{sec:gluing} and attaching affine algebras to all the external legs of the gluing graph, one can recover many more chiral algebras of $\mathcal{N}=2$ theories.
There are however some 4d $\mathcal{N}=2$ theories, whose chiral algebras clearly \emph{cannot} be constructed in this way, such as the $T_N$ theory \cite{Gaiotto:2009we, Lemos:2014lua}, since its central charge grows cubically with $N$, whereas the Grassmannian central charge grows only quadratically. However, since the putative four-parametric algebra of section~\ref{sec:4parameters} can have cubic growth in the central charge, we find it intriguing to speculate that it could lead to a direct construction of more general chiral algebras.

\paragraph{Relation to AGT} The AGT correspondence relates partition functions of class $\mathcal{S}$ theories to correlation functions of Toda theory on the compactification surface. There are various generalizations of the correspondence and in \cite{Nishioka:2011jk, Wyllard:2011mn, Belavin:2011sw, Alfimov:2011ju, Foda:2019msm, Manabe:2020etw} evidence was collected that $\mathrm{SU}(N)$ $\mathcal{N}=2$ gauge theory on the ALE space $\mathds{C}^2/\mathds{Z}_n$ is related to the Grassmannian coset with $\mu_1=n$, $\mu_2$ related to the parameter of the $\Omega$-deformation and $N=-\frac{1}{2}(\mu_1+\mu_2+\mu_3)$. We expect that the techniques developed in this paper will be useful for a further study of the correspondence.

\section*{Acknowledgements}
We thank to Bruno Le Floch, Madalena Lemos, Andy Linshaw, Miroslav Rap\v{c}\'{a}k and Menika Sharma for interesting discussions. We thank Thomas Creutzig and Matthias Gaberdiel for useful comments on the manuscript. We thank Alessandro Sfondrini for organizing the workshop ``A fresh look at AdS$_3$/CFT$_2$'' in Castasegna, where this work started. LE gratefully acknowledges support from the Della Pietra family at IAS.
%the Swiss National Science Foundation, and by the NCCR SwissMAP which is also funded by the Swiss National Science Foundation. He also acknowledges support from the Della Pietra Family at IAS.
The research of TP was supported, in parts, by the DFG cluster of excellence ORIGINS.

\appendix

\section{Character background}
\label{app:characters}
In this appendix, we review some useful combinatorics that we use to determine characters. Similar discussions can be found in \cite{Aharony:2003sx, Geloun:2013kta, Bae:2017fcs}.
\subsection{The cycle index and the P\'olya enumeration theorem}
For a permutation action of a group $\mathrm{G}$ on a set $X$, we introduce the cycle index
\be 
Z(t_1,t_2,\dots)=\frac{1}{|\mathrm{G}|}\sum_{g \in \mathrm{G}} \prod_{i \ge 1} t_i^{m_i(g)}\ .
\ee
Here, the $t_i$'s are infinity many formal variables, but for finite groups $\mathrm{G}$, only finitely many are needed. For a group element $g$, we denoted by $m_i(g)$ the number of cycles of length $i$ in its cycle decomposition.

We shall mainly need the cycle index of the cyclic group, which takes the form
\be 
Z(\mathds{Z}_n)=\frac{1}{n} \sum_{d|n} \phi(d) t_d^{\frac{n}{d}}\ ,
\ee
where $\phi(d)$ is the Euler totient function. For the dihedral group, we have instead
\begin{align}
Z(\mathrm{D}_n)&=\frac{1}{2} Z(\mathds{Z}_n)+\begin{cases}
\tfrac{1}{2}t_1t_2^{\frac{n-1}{2}}\ , & n\text{ odd}\, , \\
\tfrac{1}{4}\left(t_1^2t_2^{\frac{n-2}{2}}+t_2^{\frac{n}{2}}\right)\ , & n\text{ even}\, .
\end{cases}
\end{align}

\subsection{P\'olya enumeration theorem}
\label{app:combinatorics}
We now apply the cycle index to counting problems. Suppose that we want to count the number of orbits of the permutation group on a given set of colors. For example, say we want to count how many necklaces there are, where each bead can have three color -- red, blue or green. Then the P\'olya enumeration theorem states that the answer is obtained from replacing every $t_i$ in the cycle index by the number of colors. In our example, the relevant permutation group would be the cyclic group. So there are
\be 
\frac{1}{n} \sum_{d|n} \phi(d) m^{\frac{n}{d}}
\ee
necklaces with $m$ colors. If two necklaces that are obtained by reflection are considered equivalent, one should use the cycle index of the dihedral group $\mathrm{D}_n$ instead. 

For the applications that are relevant to this paper, we need a small generalization, in which the colors are also allowed to have some weight (which will be the mode number in our application). Suppose for example that the colors are given by all the nonnegative integers and their weight is the respective integer. Then we would like to know how many necklaces of length $n$ exist with a fixed total weight. We introduce the generating function
\be
\label{eq:polyacolorfun}
f(q)=\sum_{m=0}^\infty c_m q^m
\ee
that keeps track of the colors -- there are $c_m$ colors of weight $m$. Then the generating function for the number of orbits we want to count is obtained from the cycle index by replacing $t_i$ by $f(q^i)$. So for the example of counting necklaces with nonnegative integers of length $n$ with a fixed total weight, the generating function is
\be 
\frac{1}{n} \sum_{d|n} \phi(d) f(q^d)^{\frac{n}{d}}=\frac{1}{n} \sum_{d|n} \frac{ \phi(d)}{(1-q^d)^{\frac{n}{d}}}\ .
\ee
%For the symmetric group $\mathcal{S}_n$, it is easier to write down the generating function of the cycle index in $n$. It is given by
%\be 
%\sum_{\ell=0}^\infty x^n\, Z(\mathcal{S}_n)= \exp\left(\sum_{j=1}^\infty \frac{t_j x^j}{j}\right)\ . \label{eq:cycle index symmetric group}
%\ee
\subsection{The vacuum character of the Grassmannian} \label{subapp:vac char}
We are now prepared to derive the vacuum character of the Grassmannian. We address first the unitary version. As we have argued in section~\ref{subsec:vacuum character}, the vacuum character can be obtained by taking the plethystic exponential of a single-particle contribution. The problem of determining the single-particle contribution is equivalent to determining the number of necklaces of arbitrary length $n \ge 2$ with positive integers associated to the beads and given total weight. The relevant group in the P\'olya enumeration theorem is $\mathrm{G}=\mathds{Z}_n$ and the relevant generating function of the colors is
\be
f(q)=\frac{q}{1-q}\ .
\ee
Then the generating function for the number of necklaces is given by
\be 
Z(q)=\sum_{n=2}^\infty Z(\mathds{Z}_n,t_i=f(q^i))=\sum_{n=2}^\infty \frac{1}{n} \sum_{d|n} \phi(d) f (q^d)^{\frac{n}{d}}
\ee
This expression can be simplified swapping the summation and setting $n=r d$ and summing over $r \ge 1$ instead. We deal with the $n=1$ term later. This yields
\be 
Z(q)=\sum_{d=1}^\infty\sum_{r=1}^\infty \frac{\phi(d)}{rd} f(q^d)^r=\sum_{d=1}^\infty \frac{\phi(d)}{d} \log \frac{1}{1-f(q^d)}=\sum_{d=1}^\infty \frac{\phi(d)}{d} \log \frac{1-q^d}{1-2q^d}\ . \label{eq:necklace counting}
\ee
To deal with the $n=1$ term, we notice that it equals
\be 
\frac{q}{1-q}=\sum_{m=1}^\infty  \frac{q^m}{m}\sum_{d|m}\phi(d)=\sum_{d=1}^\infty \frac{\phi(d)}{d} \log \left(\frac{1}{1-q^d}\right)\ .
\ee
In total, we hence obtain \eqref{eq:single necklaces generating function}.

Let us perform also the same analysis for necklaces where reflections are allowed. These are relevant for the orthogonal cosets and for the parity operation. Here, we want to count necklaces of length $n$ with the positive integers as weights. Under reflection, the necklace is even (odd) if the length of the necklace is even (odd). We introduce another variable $x$ that satisfies $x^2=1$ and is the $\mathds{Z}_2$ fugacity of reflections. 

Let us begin with an even length necklace. In this case, the relevant generating function is
\begin{align} 
Z_n^\text{even}(q,x)&=Z(\mathrm{D}_n,t_i=f(q^i))+x \big(Z(\mathds{Z}_n,t_i=f(q^i))-Z(\mathrm{D}_n,t_i=f(q^i))\big)\\
&=\frac{1+x}{2n} \sum_{d|n} \frac{ \phi(d)q^n}{(1-q^d)^{\frac{n}{d}}}+\frac{1-x}{4} \left(\frac{q^n}{(1-q^2)^{\frac{n}{2}}} +\frac{q^n}{(1-q)^2(1-q^2)^{\frac{n-2}{2}}}\right)\ .
\end{align}
For odd length, the situation is reversed and we have
\begin{align} 
Z_n^\text{odd}(q,x)&=\big(Z(\mathds{Z}_n,t_i=f(q^i))-Z(\mathrm{D}_n,t_i=f(q^i))\big)+x Z(\mathrm{D}_n,t_i=f(q^i))\\
&=\frac{1+x}{2n} \sum_{d|n} \frac{ \phi(d)q^n}{(1-q^d)^{\frac{n}{d}}}+\frac{x-1}{2} \frac{q^n}{(1-q)(1-q^2)^{\frac{n-1}{2}}}\ .
\end{align}
To obtain the full answer, we should sum over all $n \ge 2$.  We discussed the first term already above, it is the same as for the cyclic group. The second term is straightforward to add and we obtain
\be 
Z(q,x)=\frac{1+x}{2}\sum_{m=1}^\infty \frac{\phi(m)}{m} \log \frac{(1-q^m)^2}{1-2q^m}+\frac{(1-x)q^2}{2(1-2q^2)}\ .
\ee
\subsection{Alternative character counting} \label{subapp:alternative counting}
Let us present an alternative way to arrive at the vacuum character of the Grassmannian. It is more direct, but hides the string interpretation that we developed in section~\ref{sec:characters}. We again count the vacuum character of the coset realization \eqref{coset2} in the limit $\mu_2 \to \infty$. For simplicity, we also add one $\mathfrak{u}(1)$ current algebra, which we can always easily remove in the end. As discussed in section~\ref{subsec:free fields}, the algebra reduces in this limit to the singlet sector of one free boson in the adjoint representation. Let us denote this boson by $\tensor{J}{^a_b}$.

So far, our discussion has been analogous to what has been discussed in section~\ref{sec:characters}. A general state in the vacuum representation can be written as
\be 
\tensor{J}{^{a_1}_{\sigma(a_1),-m_1}}\tensor{J}{^{a_2}_{\sigma(a_2),-m_2}} \cdots \tensor{J}{^{a_\ell}_{\sigma(a_\ell),-m_\ell}} \ket{0}\ .
\ee
Here, $(m_1,\dots,m_\ell)$ can be viewed as an unordered partition (a composition) of $n$ with $\ell$ elements. $\sigma \in \mathcal{S}_\ell$ is a permutation that specifies how the indices are contracted. Thus, any state in the vacuum representation at level $n$ can be viewed as a pair $(\sigma,p)$, where $\sigma$ is a permutation in $\mathcal{S}_\ell$ and $p$ is an unordered partition of $n$ with $\ell$ elements. We can reorder the currents, which means that 
\be 
(\sigma_1,p_1) \sim (\sigma_2,p_2) \quad \Longleftrightarrow \quad \sigma_2=\tau \sigma_1 \tau^{-1}\ , \quad p_2=\tau \cdot p_1\ .
\ee
Thus, we need to count the number of equivalence classes and then sum over all possible $\ell \in \mathds{Z}_{\ge 1}$. We do this by using Burnside's lemma, which states that the number of orbits of a group $\mathrm{G}$ acting on a set $X$ can be written as
\be 
|X/\mathrm{G}|=\frac{1}{|\mathrm{G}|} \sum_{g \in \mathrm{G}} |X^g|\ ,
\ee
where $X^g \subset X$ is the set of elements in $X$ fixed under the action of the group element $g$. In our context, 
\be 
\mathrm{G}=\mathcal{S}_\ell\ , \qquad X_{n,\ell}=\mathcal{S}_\ell \times \Big\{ p=(m_1,\dots ,m_\ell) \in \mathds{Z}_{\ge 1}^\ell \, \Big| \, \sum_{i=1}^\ell m_i=n \Big\}\ .
\ee
Rather than working with the set $X_{n,\ell}$, it is more convenient to just work with the set $X_\ell=\bigcup_{n=0}^\infty X_{n,\ell}$ and endow it with a weight.
For a permutation $\sigma \in \mathcal{S}_\ell$, the fixed point set $X^\sigma_n$ factorizes into the centralizer of $\sigma$  $\text{Cen}(\sigma)$ and the collection of unordered partitions invariant under the permutation $\sigma$. Let us the denote the number of such partitions by $f_\sigma(q)$ (endowed with the appropriate weight). Then $f_\sigma(q)$ is obtained analogously to what we saw in \eqref{eq:Phi permutation}, and we have
\be 
f_\sigma(q)=\prod_{j \ge 1}f(q^j)^{m_j}\ ,
\ee
where $\sigma$ has cycle type $(1)^{m_1}(2)^{m_2} (3)^{m_3} \cdots$. Here, $f(q)=q/(1-q)$ as in appendix~\ref{subapp:vac char}. Thus, 
\be 
|X^\sigma_n|=|\text{Cen}(\sigma)| f_\sigma(q)=\frac{\ell!}{|\text{Conj}(\sigma)|} f_\sigma(q)\ ,
\ee
where $\text{Conj}(\sigma)$ is the conjugacy class of $\sigma$.
Employing Burnside's lemma, the number of orbits is given by
\be 
|X_\ell/\mathcal{S}_\ell|=\frac{1}{\ell!} \sum_{\sigma \in \mathcal{S}_\ell} \frac{\ell!}{|\text{Conj}(\sigma)|} f_\sigma(q)=\sum_{\genfrac{}{}{0pt}{}{m_1\ge 0,m_2\ge 0,\dots }{ \sum_j j m_j=\ell}} \prod_{j \ge 1} \frac{q^{j m_j}}{(1-q^j)^{m_j}}\ .
\ee
As a final step, we should sum over $\ell$, which gives the vacuum character with an additional $\mathfrak{u}(1)$ factor. We obtain
\be 
\sum_{\ell=0}^\infty |X_\ell/\mathcal{S}_\ell| =\!\!\! \sum_{m_1 \ge 0, m_2 \ge 0, \dots} \prod_{j=1}^\infty \frac{q^{j m_j}}{(1-q^j)^{m_j}} 
=\prod_{j=1}^\infty \sum_{m=0}^\infty \frac{q^{j m}}{(1-q^j)^m} 
=\prod_{j=1}^\infty \frac{1-q^j}{1-2q^j} \ .
\ee 
This indeed coincides with \eqref{eq:Grassmannian vacuum character} (up to the $\mathfrak{u}(1)$ factor that we added).

Let us mention that this method of counting also works for also for the product coset that is mentioned in the conclusion~\ref{sec:discussion}. In that case, the large level limit leads to the singlet sector of a free boson that transforms in the biadjoint representation of the denominator groups. The character counting becomes again a lot simpler once we add back in the traces. In this case, this is not simply adding a $\mathfrak{u}(1)$ current, but rather modifying the coset to
\be 
\frac{\mathfrak{u}(N_1N_2)_{N_3} \times \mathfrak{su}(N_1)_{k_1} \times \mathfrak{su}(N_2)_{k_2}}{\mathfrak{su}(N_1)_{N_2N_3+k_1}\times \mathfrak{su}(N_2)_{N_1N_3+k_2}}\ .
\ee
Let us denote the biadjoint free boson by $\tensor{J}{^a_b^\alpha_\beta}$. Here, $a$, $b$ are $\mathfrak{su}(N_1)$ indices and $\alpha$, $\beta$ are $\mathfrak{su}(N_2)$ indices. A general state in the vacuum state can be written as
\be 
\tensor{J}{^{a_1}_{\sigma(a_1)}^{\alpha_1}_{\tau(\alpha_1),-m_1}}\tensor{J}{^{a_2}_{\sigma(a_2)}^{\alpha_2}_{\tau(\alpha_2),-m_2}} \cdots \tensor{J}{^{a_\ell}_{\sigma(a_\ell)}^{\alpha_\ell}_{\tau(\alpha_\ell),-m_\ell}} \ket{0}\ ,
\ee
where $\sigma$, $\tau \in \mathcal{S}_\ell$ are two permutations. Thus, nothing changes in the previous discussion, except that there are two permutations and thus
\be 
|X_\ell/\mathcal{S}_\ell|=\frac{1}{\ell!} \sum_{\sigma \in \mathcal{S}_\ell} \left(\frac{\ell!}{|\text{Conj}(\sigma)|}\right)^2 Z_\sigma(q)\ .
\ee
The vacuum character evaluates to
\be 
\sum_{\ell=0}^\infty |X_\ell/\mathcal{S}_\ell|=\prod_{j=1}^\infty \sum_{m=0}^\infty m! \left(\frac{j q^j}{1-q^j}\right)^m\ . \label{eq:product algebra vacuum character}
\ee

\section{OPE structure constants}
\label{sec:appOPE}

\subsection{Equations from OPE bootstrap}

For illustration, the following are the expressions for some structure constants of the Grassmannian algebra obtained via the OPE bootstrap as explained in section \ref{sec:OPEs}.
\begin{subequations}
\begin{align}
C_{5^-_1 5^-_1}^0 & = -\frac{12(c-130)}{c(5c+22)} \frac{C_{3^- 3^-}^0 C_{4^+_1 4^+_1}^0}{(C_{3^- 4^+_1}^{5^-_1})^2} + \frac{4(c-32)}{7c+114} \frac{(C_{3^- 3^-}^{4^+_1})^2 (C_{4^+_1 4^+_1}^0)^2}{C_{3^- 3^-}^0 (C_{3^- 4^+_1}^{5^-_1})^2} \nonumber\\
&+ \frac{C_{3^- 3^-}^{4^+_1} C_{4^+_1 4^+_1}^0 C_{4^+_1 4^+_1}^{4^+_1}}{(C_{3^- 4^+_1}^{5^-_1})^2} \ ,\\
C_{4^+_2 4^+_2}^0 & = \frac{(82944(c+10)^2}{c^2(5c+22)^2} \frac{(C_{3^- 3^-}^0)^2 C_{4^+_1 4^+_1}^0}{(C_{3^- 3^-}^{4^+_1})^2 (C_{4^+_1 4^+_1}^{4^+_2})^2} -\frac{864(c+3)(c+10)}{c(c+2)(5c+22)} \frac{(C_{4^+_1 4^+_1}^0)^2}{(C_{4^+_1 4^+_1}^{4^+_2})^2} \nonumber \\
& -\frac{3(c+3)}{(c+2)} \frac{C_{3^- 3^-}^{4^+_1} (C_{4^+_1 4^+_1}^0)^2 C_{4^+_1 4^+_2}^{4^+_2}}{C_{3^- 3^-}^0 (C_{4^+_1 4^+_1}^{4^+_2})^2} +\frac{C_{4^+_1 4^+_1}^0 C_{4^+_1 4^+_1}^{4^+_1} C_{4^+_1 4^+_2}^{4^+_2}}{(C_{4^+_1 4^+_1}^{4^+_2})^2} \nonumber\\
& +\frac{288(c+10)}{c(5c+22)} \frac{C_{3^- 3^-}^0 C_{4^+_1 4^+_1}^0 (C_{4^+_1 4^+_1}^{4^+_1} + C_{4^+_1 4^+_2}^{4^+_2})}{C_{3^- 3^-}^{4^+_1} (C_{4^+_1 4^+_1}^{4^+_2})^2} \ ,\\
C_{6^- 6^-}^0 & = \frac{576 (c+10)}{5c(5c+22)} \frac{C_{3^- 3^-}^0 C_{4^+_1 4^+_1}^0}{(C_{3^- 4^+_1}^{6^-})^2} -\frac{6(c+3)}{5(c+2)} \frac{(C_{3^- 3^-}^{4^+_1})^2 (C_{4^+_1 4^+_1}^0)^2}{C_{3^- 3^-}^0 (C_{3^- 4^+_1}^{6^-})^2} \nonumber\\
&+\frac{2}{5} \frac{C_{3^- 3^-}^{4^+_1} C_{4^+_1 4^+_1}^0 C_{4^+_1 4^+_1}^{4^+_1}}{(C_{3^- 4^+_1}^{6^-})^2} \ ,\\
C_{3^- 4^+_2}^{6^-} & = \frac{288(c+10)}{c(5c+22)} \frac{C_{3^- 3^-}^0 C_{3^- 4^+_1}^{6^-}}{C_{3^- 3^-}^{4^+_1} C_{4^+_1 4^+_1}^{4^+_2})} + \frac{C_{3^- 4^+_1}^{6^-} C_{4^+_1 4^+_2}^{4^+_2}}{C_{4^+_1 4^+_1}^{4^+_2}} \ ,\\
C_{4^+_2 4^+_2}^{4^+_2} & = \frac{82944(c+10)^2}{c^2(5c+22)^2} \frac{(C_{3^- 3^-}^0)^2}{(C_{3^- 3^-}^{4^+_1})^2 C_{4^+_1 4^+_1}^{4^+_2}} -\frac{48(11c^2+248c+596)}{c(c+2)(5c+22)} \frac{C_{4^+_1 4^+_1}^0}{C_{4^+_1 4^+_1}^{4^+_2}} +\frac{(C_{4^+_1 4^+_2}^{4^+_2})^2}{C_{4^+_1 4^+_1}^{4^+_2}} \nonumber\\
& -\frac{3(c+3)}{c+2} \frac{C_{3^- 3^-}^{4^+_1} C_{4^+_1 4^+_1}^0 C_{4^+_1 4^+_2}^{4^+_2}}{C_{3^- 3^-}^0 C_{4^+_1 4^+_1}^{4^+_2}} +\frac{288(c+10)}{c(5c+22)} \frac{C_{3^- 3^-}^0 (C_{4^+_1 4^+_1}^{4^+_1} + C_{4^+_1 4^+_2}^{4^+_2})}{C_{3^- 3^-}^{4^+_1} C_{4^+_1 4^+_1}^{4^+_2}} \ ,\\
C_{3^- 4^+_2}^{5^-_1} & = \frac{1}{\Delta_1}\Big[ -\frac{82944(c+10)^2(7c+114)}{c(5c+22)} \frac{(C_{3^- 3^-}^0)^3 C_{3^- 4^+_1}^{5^-_1}}{C_{3^- 3^-}^{4^+_1} C_{4^+_1 4^+_1}^{4^+_2}} \ ,\nonumber\\
& +\frac{3c(c+3)(5c+22)(7c+114)}{c+2} \frac{(C_{3^- 3^-}^{4^+_1})^2 C_{3^- 4^+_1}^{5^-_1} C_{4^+_1 4^+_1}^0 C_{4^+_1 4^+_2}^{4^+_2}}{C_{4^+_1 4^+_1}^{4^+_2}} \nonumber\\
& -288(c+10)(7c+114) \frac{(C_{3^- 3^-}^0)^2 C_{3^- 4^+_1}^{5^-_1} (C_{4^+_1 4^+_1}^{4^+_1} + C_{4^+_1 4^+_2}^{4^+_2})}{C_{4^+_1 4^+_1}^{4^+_2}} \nonumber\\
& +\frac{864(c+3)(c+10)(7c+114)}{(c+2)} \frac{C_{3^- 3^-}^0 C_{3^- 3^-}^{4^+_1} C_{3^- 4^+_1}^{5^-_1} C_{4^+_1 4^+_1}^0}{C_{4^+_1 4^+_1}^{4^+_2}} \nonumber\\
& -c(5c+22)(7c+114)\frac{C_{3^- 3^-}^0 C_{3^- 3^-}^{4^+_1} C_{3^- 4^+_1}^{5^-_1} C_{4^+_1 4^+_1}^{4^+_1} C_{4^+_1 4^+_2}^{4^+_2}}{C_{4^+_1 4^+_1}^{4^+_2}} \Big] \ ,\\
C_{5^-_2 5^-_2}^0 & = \frac{1}{c^3(c+2)^2(5c+22)^2 C_{3^- 3^-}^0 (C_{3^- 3^-}^{4^+_1})^2 (C_{3 4^+_2}^{5^-_2})^2 (C_{4^+_1 4^+_1}^{4^+_2})^2\Delta_1}  \nonumber\\
& \times \Big[ 59719680(c+2)^2(c+10)^2(7c+114)(23c+370) (C_{3^- 3^-}^0)^6 C_{4^+_1 4^+_1}^0 \nonumber\\
& +15 c^4 (c+3)(c+7)(5c+22)^4 (C_{3^- 3^-}^{4^+_1})^6 (C_{4^+_1 4^+_1}^0)^3 (C_{4^+_1 4^+_2}^{4^+_2})^2\nonumber \\
& +207360 c(c+2)^2(c+10)(5c+22)(7c+114)(47c+610)  \nonumber\\
& \qquad\times (C_{3^- 3^-}^0)^5 C_{3^- 3^-}^{4^+_1} C_{4^+_1 4^+_1}^0 (C_{4^+_1 4^+_1}^{4^+_1} + C_{4^+_1 4^+_2}^{4^+_2}) \nonumber\\
& +720 c^3(c+3)(5c+22)^3(11c^2+234c+904) C_{3^- 3^-}^0 (C_{3^- 3^-}^{4^+_1})^5 C_{4^+_1 4^+_2}^{4^+_2} (C_{4^+_1 4^+_1}^0)^3 \nonumber\\
& -5c^4(c+2)(c+7)(5c+22)^4 C_{3^- 3^-}^0 (C_{3^- 3^-}^{4^+_1})^5 (C_{4^+_1 4^+_1}^0)^2 C_{4^+_1 4^+_1}^{4^+_1} (C_{4^+_1 4^+_2}^{4^+_2})^2 \nonumber\\
& +207360c^2(c+3)(c+10)(c+22)(5c+22)^3 (C_{3^- 3^-}^0)^2 (C_{3^- 3^-}^{4^+_1})^4 (C_{4^+_1 4^+_1}^0)^3 \nonumber\\
& -60c^3(c+2)(5c+22)^4(13c+211) (C_{3^- 3^-}^0)^2 (C_{3^- 3^-}^{4^+_1})^4 (C_{4^+_1 4^+_1}^0)^2 C_{4^+_1 4^+_1}^{4^+_1} C_{4^+_1 4^+_2}^{4^+_2} \nonumber\\
& -180c^3(c+2)(3c+41)(5c+22)^4 (C_{3^- 3^-}^0)^2 (C_{3^- 3^-}^{4^+_1})^4 (C_{4^+_1 4^+_1}^0)^2 (C_{4^+_1 4^+_2}^{4^+_2})^2 \nonumber\\
& -622080c(c+2)(c+10)(5c+22) (321 c^3+11519 c^2+115544c+281420)  \nonumber\\
& \qquad\times (C_{3^- 3^-}^0)^4 (C_{3^- 3^-}^{4^+_1})^2 (C_{4^+_1 4^+_1}^0)^2 \nonumber\\
& +17280c^2(c+2)^2(c+10)(5c+22)^2(7c+114)\nonumber\\
&\qquad\times (C_{3^- 3^-}^0)^4 (C_{3^- 3^-}^{4^+_1})^2 C_{4^+_1 4^+_1}^0 (C_{4^+_1 4^+_1}^{4^+_1})^2 \nonumber\\
& +720c^2(c+2)^2(5c+22)^2(7c+114)(71c+850) \nonumber \\
& \qquad\times (C_{3^- 3^-}^0)^4 (C_{3^- 3^-}^{4^+_1})^2 C_{4^+_1 4^+_1}^0 C_{4^+_1 4^+_1}^{4^+_1} C_{4^+_1 4^+_2}^{4^+_2} \nonumber \\
& +17280c^2(c+2)^2(c+10)(5c+22)^2(7c+114) \nonumber\\
&\qquad\times(C_{3^- 3^-}^0)^4 (C_{3^- 3^-}^{4^+_1})^2 C_{4^+_1 4^+_1}^0 (C_{4^+_1 4^+_2}^{4^+_2})^2 \nonumber \\
& +60c^3(c+2)^2(5c+22)^3(7c+114) (C_{3^- 3^-}^0)^3 (C_{3^- 3^-}^{4^+_1})^3 C_{4^+_1 4^+_1}^0 C_{4^+_1 4^+_2}^{4^+_2} (C_{4^+_1 4^+_1}^{4^+_1})^2 \nonumber \\
& +60c^3(c+2)^2(5c+22)^3(7c+114) (C_{3^- 3^-}^0)^3 (C_{3^- 3^-}^{4^+_1})^3 C_{4^+_1 4^+_1}^0 C_{4^+_1 4^+_1}^{4^+_1} (C_{4^+_1 4^+_2}^{4^+_2})^2 \nonumber \\
& -2160c^2(c+2)(5c+22)^2(681c^3+21623c^2+202232c+497900) \nonumber  \\
& \qquad\times (C_{3^- 3^-}^0)^3 (C_{3^- 3^-}^{4^+_1})^3 (C_{4^+_1 4^+_1}^0)^2 C_{4^+_1 4^+_2}^{4^+_2} \nonumber \\
& -17280c^2(c+2)(c+10)(5c+22)^2(41c^2+933c+2962)  \nonumber\\
& \qquad\times (C_{3^- 3^-}^0)^3 (C_{3^- 3^-}^{4^+_1})^3 (C_{4^+_1 4^+_1}^0)^2 C_{4^+_1 4^+_1}^{4^+_1} \Big]\ ,
\end{align}
\end{subequations}
where
\begin{multline}
\Delta_1 = - c(5c+22)(7c+114) C_{3^- 3^-}^0 C_{3^- 3^-}^{4^+_1} C_{4^+_1 4^+_1}^{4^+_1} \\
+ 12(7c+114)(c-130) (C_{3^- 3^-}^0)^2 - 4(c-32)c(5c+22) (C_{3^- 3^-}^{4^+_1})^2 C_{4^+_1 4^+_1}^0\ .
\end{multline}

\subsection{Universal structure constants}

The bootstrap equations determine OPE coefficients of higher spin fields in terms of four independent coefficients. These in turn can be written in form that is a homogeneous symmetric function of five variables $\nu_1, \ldots, \nu_5$ in such a way that it reduces to both the Grassmannian algebra and the orthosymplectic algebra under the appropriate specialization. These four independent OPE coefficients are
\begin{subequations}
\begin{align}
c & = \frac{e_1^3 e_2 - e_1^2 e_3 + e_1 e_4 - e_5}{e_5} \\
C_{4^+_1 4^+_1}^0 & = \frac{\Delta_2}{\Delta_3} \frac{(C_{3^- 3^-}^0)^2}{(C_{3^- 3^-}^{4^+_1})^2} \\
C_{4^+_1 4^+_1}^{4^+_1} & = \frac{\Delta_4}{\Delta_5} \frac{C_{3^- 3^-}^0}{C_{3^- 3^-}^{4^+_1}} \\
C_{4^+_1 4^+_2}^{4^+_2} & = \frac{\Delta_6}{\Delta_7 \Delta_8} \frac{C_{3^- 3^-}^0}{C_{3^- 3^-}^{4^+_1}}
\end{align}
\end{subequations}
with 
\begin{subequations}
\begin{align}
\Delta_2 & = 144e_5(e_1^8e_2-4e_1^6e_2^2-e_1^7e_3+17e_1^5e_2e_3-13e_1^4e_3^2+e_1^6e_4-40e_1^4e_2e_4+49e_1^3e_3e_4\nonumber\\
& -36e_1^2e_4^2-3e_1^5e_5+104e_1^3e_2e_5-99e_1^2e_3e_5+72e_1e_4e_5+108e_5^2) \ ,\\
\Delta_3 & = (e_1^3e_2-e_1^2e_3+e_1e_4-e_5)(5e_1^3e_2-5e_1^2e_3+5e_1e_4+17e_5) \nonumber \\
&\qquad \times (e_1^5-4e_1^3e_2+8e_1^2e_3-16e_1e_4+32e_5) \ , \\
\Delta_4 & = 48e_5(3e_1^16e_2^2-24e_1^{14}e_2^3+48e_1^{12}e_2^4-6e_1^{15}e_2e_3+156e_1^{13}e_2^2e_3-553e_1^{11}e_2^3e_3 \nonumber\\
& +3e_1^{14}e_3^2-240e_1^{12}e_2e_3^2+1884e_1^{10}e_2^2e_3^2+108e_1^{11}e_3^3-2301e_1^9e_2e_3^3+922e_1^8e_3^4 \nonumber\\
& +6e_1^{14}e_2e_4-359e_1^{12}e_2^2e_4+1440e_1^{10}e_2^3e_4-6e_1^{13}e_3e_4+886e_1^{11}e_2e_3e_4\nonumber \\
& -9146e_1^9e_2^2e_3e_4-527e_1^{10}e_3^2e_4+14998e_1^8e_2e_3^2e_4-7292e_1^7e_3^3e_4+3e_1^{12}e_4^2 \nonumber\\
& -646e_1^{10}e_2e_4^2+10944e_1^8e_2^2e_4^2+730e_1^9e_3e_4^2-30457e_1^7e_2e_3e_4^2+20026e_1^6e_3^2e_4^2 \nonumber\\
& -311e_1^8e_4^3+17760e_1^6e_2e_4^3-21864e_1^5e_3e_4^3+8208e_1^4e_4^4-106e_1^{13}e_2e_5 \nonumber\\
& +1585e_1^{11}e_2^2e_5-4944e_1^9e_2^3e_5+106e_1^{12}e_3e_5-4114e_1^{10}e_2e_3e_5+29206e_1^8e_2^2e_3e_5 \nonumber\\
& +2529e_1^9e_3^2e_5-44130e_1^7e_2e_3^2e_5+19868e_1^6e_3^3e_5-106e_1^{11}e_4e_5+6268e_1^9e_2e_4e_5 \nonumber\\
& -66768e_1^7e_2^2e_4e_5-7212e_1^8e_3e_4e_5+163078e_1^6e_2e_3e_4e_5-96860e_1^5e_3^2e_4e_5 \nonumber\\
& +4683e_1^7e_4^2e_5-114096e_1^5e_2e_4^2e_5+129264e_1^4e_3e_4^2e_5-52272e_1^3e_4^3e_5+35e_1^{10}e_5^2 \nonumber\\
& -9406e_1^8e_2e_5^2+93936e_1^6e_2^2e_5^2+8666e_1^7e_3e_5^2-184657e_1^5e_2e_3e_5^2+89802e_1^4e_3^2e_5^2 \nonumber\\
& -5877e_1^6e_4e_5^2+135552e_1^4e_2e_4e_5^2-127656e_1^3e_3e_4e_5^2+39312e_1^2e_4^2e_5^2-12023e_1^5e_5^3 \nonumber\\
& +122256e_1^3e_2e_5^3-134784e_1^2e_3e_5^3+123120e_1e_4e_5^3+23328e_5^4)\ ,\\
\Delta_5 & = (e_1^3e_2-e_1^2e_3+e_1e_4-e_5)(5e_1^3e_2-5e_1^2e_3+5e_1e_4+17e_5) \nonumber \\
&\qquad \times (e_1^5-4e_1^3e_2+8e_1^2e_3-16e_1e_4+32e_5) (e_1^8e_2-4e_1^6e_2^2-e_1^7e_3+17e_1^5e_2e_3 \nonumber\\
&\qquad -13e_1^4e_3^2+e_1^6e_4-40e_1^4e_2e_4+49e_1^3e_3e_4-36e_1^2e_4^2-3e_1^5e_5+104e_1^3e_2e_5 \nonumber\\
&\qquad -99e_1^2e_3e_5+72e_1e_4e_5+108e_5^2) \ ,\\
\Delta_6 & = 48e_5(4e_1^{11}e_2^2-21e_1^9e_2^3-8e_1^10e_2e_3+99e_1^8e_2^2e_3+4e_1^9e_3^2-135e_1^7e_2e_3^2+57e_1^6e_3^3 \nonumber\\
& +8e_1^9e_2e_4-231e_1^7e_2^2e_4-8e_1^8e_3e_4+534e_1^6e_2e_3e_4-303e_1^5e_3^2e_4+4e_1^7e_4^2 -399e_1^5e_2e_4^2\nonumber\\
& +435e_1^4e_3e_4^2-189e_1^3e_4^3+8e_1^8e_2e_5+609e_1^6e_2^2e_5-8e_1^7e_3e_5 -1338e_1^5e_2e_3e_5\nonumber\\
& +729e_1^4e_3^2e_5+8e_1^6e_4e_5+1554e_1^4e_2e_4e_5-1674e_1^3e_3e_4e_5 +945e_1^2e_4^2e_5\nonumber\\
& +196e_1^5e_5^2-1071e_1^3e_2e_5^2+1107e_1^2e_3e_5^2-567e_1e_4e_5^2-1701e_5^3) \ ,\\
\Delta_7 & = (e_1^3e_2-e_1^2e_3+e_1e_4-e_5)(5e_1^3e_2-5e_1^2e_3+5e_1e_4+17e_5) \ ,\\
\Delta_8 & = (e_1^8e_2-4e_1^6e_2^2-e_1^7e_3+17e_1^5e_2e_3-13e_1^4e_3^2+e_1^6e_4-40e_1^4e_2e_4+49e_1^3e_3e_4 \nonumber\\
& -36e_1^2e_4^2-3e_1^5e_5+104e_1^3e_2e_5-99e_1^2e_3e_5+72e_1e_4e_5+108e_5^2)
\ .
\end{align}
\end{subequations}
We use the notation $e_j$ for elementary symmetric polynomials of variables $\nu_1,\ldots,\nu_5$.

\paragraph{Structure constants for unitary Grassmannian}
The structure constants for the Grassmannian can be obtained from the universal structure constants by replacing $\nu_4 \to 1, \nu_5 \to -1$.

\paragraph{Structure constants for Lagrangian Grassmannian algebras}
The structure constants for the Lagrangian Grassmannian cosets are obtained from the universal structure constants by replacing
\begin{equation*}
\label{eq:apporthounitarymap}
\nu_1 \to n+\kappa, \quad \nu_2 \to -2n-\kappa+2, \quad \nu_3 \to n+\kappa-1, \quad \nu_4 \to 1, \quad \nu_5 \to -2.
\end{equation*}

\bibliographystyle{JHEP}
\bibliography{bib}

\end{document}